\documentclass[longauth]{aa} 

\usepackage{graphicx}
\usepackage{epstopdf}
\usepackage{natbib}

\begin{document}

\title{WISH  VI. Constraints on UV and X-ray irradiation from a survey of hydrides in low- to high-mass YSOs \thanks{{\it Herschel} is an ESA space observatory with science instruments provided by a European-led Principal Investigator consortia and with important participation from NASA.}}
\author{A.O. Benz\inst{\ref{inst1}}
\and S. Bruderer\inst{\ref{inst1},\ref{inst2}}
\and E.F. van Dishoeck\inst{\ref{inst2},\ref{inst3}}
\and M. Melchior\inst{\ref{inst1},\ref{inst4}}
\and S.F. Wampfler\inst{\ref{inst1},\ref{inst5}}
\and F. van der Tak\inst{\ref{inst6},\ref{inst7}}
\and J.R. Goicoechea\inst{\ref{inst8}}
\and N. Indriolo\inst{\ref{inst9}}
\and L.E. Kristensen\inst{\ref{inst10}}
\and D.C. Lis\inst{\ref{inst11},\ref{inst12}}
\and J.C. Mottram\inst{\ref{inst3}}
\and E.A. Bergin\inst{\ref{inst9}}
\and P. Caselli\inst{\ref{inst2}}
\and F. Herpin\inst{\ref{inst13},\ref{inst14}}
\and M.R. Hogerheijde\inst{\ref{inst3}}
\and D. Johnstone\inst{\ref{inst15},\ref{inst16}}
\and R. Liseau\inst{\ref{inst17}}
\and B. Nisini\inst{\ref{inst18}}
\and M. Tafalla\inst{\ref{inst19}}
\and R. Visser\inst{\ref{inst20}}
\and F. Wyrowski\inst{\ref{inst21}}}

\institute{
Institute for Astronomy, ETH Zurich, 8093 Zurich, Switzerland\label{inst1}
\and
Max-Planck-Institut f\"{u}r extraterrestrische Physik, Giessenbachstrasse 1, 85748 Garching, Germany\label{inst2}
\and
Leiden Observatory, Leiden University, PO Box 9513, 2300 RA Leiden, The Netherlands\label{inst3}
\and
Institute of 4D Technologies, University of Applied Sciences FHNW, CH-5210 Windisch, Switzerland\label{inst4}
\and
Centre for Star and Planet Formation, Natural History Museum of Denmark, and Niels Bohr Institute,
{\O}ster Voldgade 5-7, DK-1350 Copenhagen K., Denmark\label{inst5}
\and
SRON Netherlands Institute for Space Research, PO Box 800, 9700 AV, Groningen, The Netherlands\label{inst6}
\and
Kapteyn Astronomical Institute, University of Groningen, PO Box 800, 9700 AV, Groningen, The Netherlands\label{inst7}
\and
Grupo de Astrofisica Molecular, Instituto de Ciencia de Materiales de Madrid (ICMM), Consejo Superior de Investigaciones Cientificas (CSIC), Calle Sor Juana Ines de la Cruz, 3, E-28049 Cantoblanco, Madrid, Spain\label{inst8}
\and
Department of Astronomy, The University of Michigan, 1085 S. University Ave., Ann Arbor, MI 48109-1107, USA\label{inst9}
\and
Harvard-Smithsonian Center for Astrophysics, 60 Garden Street, Cambridge, MA, 02138, USA\label{inst10}
\and
LERMA, Observatoire de Paris, PSL Research University, CNRS, Sorbonne Universit\'es, UPMC Univ. Paris 06, F-75014, Paris, France\label{inst11}
\and
California Institute of Technology, Cahill Center for Astronomy and Astrophysics, MS 301-17, Pasadena, CA 91125, USA\label{inst12}
\and
Univ. Bordeaux, LAB, UMR 5804, F-33270, Floirac, France\label{inst13}
\and
CNRS, LAB, UMR 5804, F-33270, Floirac, France\label{inst14}
\and
National Research Council Canada, Herzberg Astronomy and Astrophysics, 5071 West Saanich Rd, Victoria, BC, V9E 2E7, Canada\label{inst15}
\and
Department of Physics \& Astronomy, University of Victoria, Victoria, BC, V8P 1A1, Canada\label{inst16}
\and
Department of Radio and Space Science, Chalmers University of Technology, Onsala Space Observatory, 439 92 Onsala, Sweden\label{inst17}
\and
INAF - Osservatorio Astronomico di Roma, 00040 Monte Porzio catone, Italy\label{inst18}
\and
Observatorio Astron\'{o}mico Nacional (IGN), Calle Alfonso XII,3. 28014, Madrid, Spain\label{inst19}
\and
European Southern Observatory, Karl-Schwarzschild-Strasse 2, D-85748, Garching, Germany\label{inst20}
\and
Max-Planck-Institut f\"{u}r Radioastronomie, Auf dem H\"{u}gel 69, 53121 Bonn, Germany\label{inst21}}

\authorrunning{A.O. Benz et al.}
\titlerunning{Radiation tracers in YSOs}
   \date{Received: February 6, 2015}

\abstract
{Hydrides are simple compounds containing one or a few hydrogen atoms bonded to a heavier atom. They are fundamental precursor molecules in cosmic chemistry and many hydride ions have become observable in high quality for the first time thanks to the {\it Herschel\ Space\ Observatory}. Ionized hydrides, such as CH$^+$ and OH$^+$, and also HCO$^+$ that affect the chemistry of  molecules such as water,  provide complementary information on irradiation by far UV (FUV) or X-rays and gas temperature.}
{We explore hydrides of the most abundant heavier elements in an observational survey covering young stellar objects (YSO) with different mass and evolutionary state. The focus is on hydrides associated with the dense protostellar envelope and outflows, contrary to previous work that focused on hydrides in diffuse foreground clouds.}
{Twelve YSOs were observed with HIFI on $Herschel$ in 6 spectral settings providing fully velocity-resolved line profiles as part of the `Water in star-forming regions with {\it Herschel}' (WISH) program. The YSOs include objects of low (Class 0 and I), intermediate, and high mass, with luminosities ranging from 4 L$_\odot$ to 2$\times 10^5$ L$_\odot$.}
{The targeted lines of CH$^+$, OH$^+$, H$_2$O$^+$, C$^+$ and CH are detected mostly in blue-shifted absorption. H$_3$O$^+$ and SH$^+$ are detected in emission and only toward some high-mass objects. The observed line parameters and correlations suggest two different origins, related to gas entrained by the outflows and to the circumstellar envelope. The derived column densities correlate with bolometric luminosity and envelope mass for all molecules, best for CH, CH$^+$, and HCO$^+$. The column density ratios of CH$^+/$OH$^+$  are estimated from chemical slab models,  assuming that the H$_2$ density is given by the specific density model of each object at the beam radius. For the low-mass YSOs the observed ratio can be reproduced for an FUV flux of 2 -- 400 times the ISRF at the location of the molecules. In two high-mass objects, the UV flux  is 20 -- 200 times the ISRF derived from absorption lines, and 300 -- 600 ISRF using emission lines.  Upper limits for the X-ray luminosity can be derived from H$_3$O$^+$ observations for some low-mass objects.}
{If the FUV flux required for low-mass objects originates at the central protostar, a substantial FUV luminosity, up to 1.5 L$_\odot$, is required. There is no molecular evidence for X-ray induced chemistry in the  low-mass objects on the observed scales of a few 1000 AU. For high-mass regions, the FUV flux required to produce the observed molecular ratios is smaller than the unattenuated flux expected from the central object(s) at the $Herschel$ beam radius. This is consistent with an FUV flux reduced by circumstellar extinction or by bloating of the protostar.}


   \maketitle
%

\section{Introduction}
The physics and chemistry of the inner few thousand AU of star-forming regions are poorly constrained. {\it Herschel} \citep{2010A&A...518L...1P} has opened up the opportunity for high-quality observations of species that cannot (or only with great difficulties) be observed from the ground. Water is a key molecule, but to fully understand its chemistry related molecules such as hydrides and HCO$^+$ must be characterized as well. Moreover, compared to cold cores, the chemistry in protostellar envelopes is affected by FUV and X-rays from the central source. Some hydrides are particularly good diagnostics of this energetic radiation, which cannot be observed directly in the most deeply embedded phases of star formation.

Specifically, gaseous H$_2$O is destroyed by high levels of FUV (6.2 - 13.6 eV) or X-rays, both through photodissociation and through reactions with ionized species whose abundances are enhanced by irradiation  \citep{2000A&A...358L..79V,2011A&A...527A..69A}.  On the other hand, H$_2$O  in the gas reduces the abundance of species like CH$^+$, OH$^+$, H$_2$O$^+$, and HCO$^+$. Thus their abundances and that of H$_2$O cannot be high at the same location. \citet{2006A&A...453..555S} have analyzed the water abundance under various conditions in star-forming regions. At temperatures $T < 100$ K, irradiation induces the formation of water through ion-molecule reactions. For $100 < T < 250$ K, irradiation decreases the H$_2$O abundance through reactions with H$_3^+$ and HCO$^+$ and local UV emission of excited H$_2$. In the regime $T > 250$ K, endothermic neutral-neutral reactions produce H$_2$O efficiently, but a high X-ray flux at low densities ($n_H = 10^4 - 10^5$ cm$^{-3}$) destroys water and reduces its abundance by orders of magnitude. Constraining the amount of FUV and X-rays across the protostellar envelope is therefore important in tracing the full water chemistry.

Ionizing radiation may have several origins. If the surface temperature of a protostar exceeds 10$^4$ K, the emission of FUV  becomes significant and is proportional to the radiating hot surface area. Accretion and jets produce shocked gas capable of emitting both FUV and X-rays. Finally, protostellar coronae are powerful emitters of both X-rays and high-energy UV radiation. When the envelopes of low-mass YSOs  become transparent in the Class I phase, a median X-ray luminosity of more than $10^{30}$ erg s$^{-1}$ is observed that decreases with age \citep{1999ARA&A..37..363F, 2004A&ARv..12...71G}. Temperatures exceeding 10$^{7}$ K and occasional flares suggest a process in the stellar corona or the star-disk interface releasing magnetic energy \citep{2010ARA&A..48..241B}. The average X-ray luminosity may reach a few times $10^{-3} L_{\rm bol}$ or $10^{32}$ erg s$^{-1}$ \citep{2004A&ARv..12...71G}, limited by the stellar magnetic field strength (few kGauss) and by the finite volume of the reconnection region.

X-rays and FUV radiation are absorbed differently in molecular gas. X-rays beyond 10 keV penetrate a half-power hydrogen column density of some $10^{24}$ cm$^{-2}$ \citep{1996ApJ...466..561M, 2005A&A...440..949S}. The penetration depth of FUV photons is three orders of magnitude less, thus they are absorbed mainly at irradiated surfaces.  For the distant massive objects, the sphere of influence of X-ray emission (if present) is geometrically limited to a small fraction of the {\it Herschel} beam and may be overwhelmed by FUV \citep{2010ApJ...720.1432B}. However, the accreting protostars may bloat and have a low surface temperature \citep{2001A&A...373..190B}. Significant populations of massive stars without H{\tiny{\textsc II}} regions have been reported \citep[e.g. by][]{2013ApJS..208...11L}. On the other hand, massive YSOs known to have ultra-compact H{\tiny{\textsc II}} regions indicate that their UV emission has significantly changed their circumstellar environment. The situation is different for the nearby low-mass objects studied here, where the X-ray dominated region (XDR) may exceed the size of the {\it Herschel} beam. Excess FUV emission has been inferred from observed mid-$J$ CO isotopolog observations interpreted to originate from irradiated cavity walls created by the outflows \citep[][]{2009A&A...501..633V,2012A&A...542A..86Y,2015A&A...576A.109Y}. Here we provide independent tracers of the FUV and X-ray irradiation using observations of hydrides that are particularly sensitive to these radiations.

In addition to ionization, the high-energy photons heat the molecular gas and thus further enhance the abundances of those hydrides whose formation requires extra energy, such as CH$^+$ \citep{1995ApJS...99..565S}. Ionized hydrides are chemically active and can drive substantial chemical evolution. When formed, they generally react fast and without activation energy. Once their chemistry and excitation is understood, they will become tracers of warm and ionized gas in deeply embedded phases of star and planet formation.

\begin{table*}[htb]
\begin{center}
\caption{Observed objects and their parameters}
\begin{tabular}{lcrcrrrrrc}   
\hline \hline
Object &RA&Dec\ \ \ \ &Class&$d$&$V_{\mathrm{LSR}}$&$L_{\rm bol}$ & $T_{\rm bol}$ &$M_{\rm env}$&References \\
 &[h m s]&[$^{\circ\ '\ "}$] & &[pc]&[km s$^{-1}$]&[L$_\odot$]&[K] &[M$_\odot$]& \\
\hline\\
NGC1333 I2A&03:28:55.6&+31:14:37.1&LM 0&235&+7.7&35.7&50&5.1&1,2,10,11\\
NGC1333 I4A&03:29:10.5&+31:13:30.9&LM 0&235&+7.2&9.1&33&5.6&1,2,10,11\\
NGC1333 I4B&03:29:12.0&+31:13:08.1&LM 0&235&+7.4&4.4&28&3.0&1,2,10,11\\
Ser SMM1&18:29:49.8&+01:15:20.5&LM 0&415&+8.5&99.0&39&52.1&1,2,10,11,12\\
L 1489&04:04:43.0&+26:18:57.0&LM I&140&+7.2&3.8&200&0.2&1,2,11,10\\
\\
NGC7129 FIRS2&21:43:01.7&+66:03:24.0&IM&1250&$-$9.8&430&40&50&1,3\\
\\
W3 IRS5&02:25:40.6&+62:05:51.0&mIRb&2000&$-$38.4&1.7$\times10^{5}$&370&424&1,4,7\\
NGC6334 I&17:20:53.3&$-$35:47:00.0&HMC&1700&$-$7.7&2.6$\times10^{5}$&100&500&1,4,8\\
NGC6334 I(N)&17:20:55.2&$-$35:45:04.0&mIRq&1700&$-$3.3&1.9$\times10^{3}$&30&3826&1,4,8\\
AFGL 2591&20:29:24.9&+40:11:19.5&mIRb&3300&$-$5.5&2.2$\times10^{5}$&325&320&1,4\\
S 140 IRS1&22:19:18.2&+63:18:46.9&mIRb&910&$-$7.1&1.0$\times10^{4}$&85&100&1,5,6\\
NGC7538 IRS1&23:13:45.3&+61:28:10.0&UCHII&2800&$-$56.2&1.3$\times10^{5}$&40&433&1,4,9\\
\hline
\end{tabular}
\end{center}
\vskip-0.1cm
{\tiny {\bf Notes.} Reference position chosen in J2000, distance $d$, systemic velocity $V_{\mathrm{LSR}}$ in the Local Standard of Rest, bolometric luminosity $L_{\mathrm bol}$, bolometric temperature $T_{\mathrm bol}$, and envelope mass $M_{\mathrm env}$. The evolutionary stage is abbreviated for low-mass objects as LM 0 and LM I for Class 0 and I. IM stands for intermediate mass objects. High-mass objects are classified according to \citet{2013A&A...554A..83V}, based on the scheme presented by \citet{2008A&A...481..345M}: mIRq for mid-infrared quiet high-mass protostellar object (HMPO), mIRb for mid-infrared bright HMPO, HMC for hot molecular cores, and UCHII for objects reported to contain an ultra-compact H{\tiny{\textsc II}} region. References: (1) \citet{2011PASP..123..138V}, confusion between NGC 6334 I and I(N) corrected; (2) \citet{2012A&A...542A...8K} and references therein ; (3) \citet{2010A&A...516A.102C}; (4) \citet{2013A&A...554A..83V}; (5) \citet{2012ApJ...749L..20H}; (6) \citet{2010A&A...521L..24D}; (7) \citet{2005A&A...431..993V}; (8) \citet{2000A&A...358..242S}; (9) \citet{2004ApJ...600..269S}; (10) \citet{2013A&A...553A.125S}; (11) \citet{2013A&A...556A..89Y}; (12) \citet{2010ApJ...718..610D}.}
\label{table_objects}
\end{table*}

Absorption lines of hydrides have been found in diffuse interstellar clouds both at optical and mm wavelengths \citep[e.g.][]{2002A&A...391..693L,2006ARA&A..44..367S} and were extensively observed by {\it Herschel/HIFI} \citep[][and others]{2010A&A...521L..15F, 2010A&A...521L..16G, 2013ApJ...767...81M,2013ApJ...762...11F}. Relevant for this work are also ground-based SH$^+$ observations by \citet{2011A&A...525A..77M}, HIFI CH$^+$ and  SH$^+$ observations by \citet{2012A&A...540A..87G}. The lines of this diffuse gas are also seen in our data at velocities substantially offset from the YSO.

Here we report on hydrides and their relation to H$_2$O in star-forming regions, where the gas is denser, hotter, and strongly irradiated by the nearby protostar. In the environments of newly forming stars, H$_2$O has been found in various places including disks \citep[e.g.][]{2010ApJ...720..887P,2011Sci...334..338H}, shocked gas \citep{2013A&A...557A..23K, 2014A&A...572A..21M}, outflow lobes \citep{2010A&A...518L.120N,2013A&A...549A..16N, 2010A&A...518L.113L,2012A&A...542A...8K, 2014A&A...561A.120B}, and the envelope \citep[e.g.][]{2012A&A...539A.132C,2012A&A...542A...8K,2013A&A...558A.126M, 2013A&A...554A..83V}. The origin is generally inferred from line widths and Doppler shifts. Water observations were compared to CH$^+$, OH$^+$, and C$^+$ toward low-mass sources by \citet{2013A&A...557A..23K}. The ratio of H$_2$O$^+$ to H$_2$O was measured in high-mass YSOs by \citet{2010A&A...521L..34W}.

We present results of the subprogram `Radiation Diagnostics' \citep{2013JPCA..117.9840B} of the {\it Herschel} guaranteed time key program `Water in Star-forming regions with {\it Herschel}'  \citep[WISH,][]{2011PASP..123..138V}. Here the focus is on possibilities of identifying FUV and X-ray emission through chemistry in deeply embedded objects, where UV and X-rays cannot be observed directly due to a high attenuating column density, but can affect the chemistry of water.

Hydrides and ion molecules in dense star-forming regions are of interest also in red-shifted galaxies of the early universe, where these species can be observed from the ground. The analysis and observational characteristics of these species in nearby objects are important for future observations.

In the following section and in Appendix A, where the details are given, we present the observations. Section 3 describes the method to derive column densities. Its quantitative results are tabulated in Appendix B. In Section 4 the results are analyzed and correlated in various ways; some details are available in Appendix C. Section 5 presents the discussion of the resulting constraints on FUV and X-ray irradiation. Three scenarios on the origin of the ionized molecules are discussed in Section 6. The conclusions can be found in Section 7. The correlation of HCO$^+$ with $L_{bol}$ is interpreted in Appendix D.

\section{Observations}
A selection of 12 star-forming objects was observed with the Heterodyne Instrument for the Far Infrared \citep[HIFI,][]{2010A&A...518L...6D} on {\it Herschel}. The YSOs were selected from the list of the WISH key program \citep{2011PASP..123..138V} and are listed in Table \ref{table_objects}. The low-mass sample contains mostly deeply embedded Class 0 sources and one more evolved Class I source. The high-mass sources were chosen to represent different stages of evolution, but the classification is too uncertain and the number is too small to expect clear evolutionary trends. The observed transitions were selected according to their expected intensity based on model calculations by \citet{2005A&A...440..949S}, \citet{2006MsT..........1B} and \citet{2010ApJ...720.1432B}. They are listed in Table \ref{table_lines} together with molecular and atomic parameters.

The objects are observed in six 4 GHz spectral settings for a total of typically 950 s each (including on and off source plus overhead) in the low frequency bands. In Band 4 (960 -- 1120 GHz) the observing times are typically \mbox{1800 s} for OH$^+$ and H$_3$O$^+$, and 2400 s for H$_2$O$^+$. The data are taken in dual beam switching (DBS) mode. In the DBS mode, the telescope is centered at the object coordinates given in Table \ref{table_objects} within a few arcsec, and the reference positions are offset by 3$'$ on either side of the object. The only exception is S 140, observed in the framework of the WADI key program, where the C$^+$ observation is in on-the-fly (OTF) mode, and HCO$^+$, CH$^+$(1-0) and CH in frequency switching mode \citep{2010A&A...521L..24D}. Table C.1 in Appendix C contains the observing log.

Most of the lines in this exploratory survey are published for the first time. For the two high-mass regions AFGL 2591 and W3 IRS5, first results on CH$^+$, OH$^+$, H$_2$O$^+$, CH, CH$^+$, and H$_3$O$^+$  were published before \citep{2010A&A...521L..44B,2010A&A...521L..35B}, but are reanalyzed and included here. C$^+$ and HCO$^+$ toward S 140 were studied by \citet{2010A&A...521L..24D}. The detection of SH$^+$ toward W3 IRS5 was reported in \citet{2010A&A...521L..35B}. Foreground clouds of some high-mass objects in this selection at velocities offset from the YSO were analyzed by \citet{2015ApJ...800...40I}.

We used primarily the Wide Band Spectrometer, having a spectral resolution of 1.1 MHz, yielding a velocity resolution better than 0.7 km s$^{-1}$ over the entire HIFI range. The {\it Herschel} Interactive Processing Environment (HIPE) 4.4 \citep{2010ASPC..434..139O} and higher was used for the pipeline and version 11.0 for data analysis. We resampled all spectra to 1 km s$^{-1}$, sufficient to resolve most observed lines. Some of the narrow CH and HCO$^+$ lines were analyzed more precisely with the High Resolution Spectrometer yielding a resolution better than 0.36 MHz (0.2 km s$^{-1}$). The accuracy of the velocity calibration is estimated to be better than 0.1 km s$^{-1}$ \citep{2012A&A...537A..17R}.

The antenna temperature was converted to main beam temperature, using the beam efficiencies of \citet{2012A&A...537A..17R} and the forward efficiency of 0.96 for observations after 2011/01/01. A newer calibration of the data came out recently (HIFI memo Oct. 1, 2014), which increases the main beam intensities by around 13\%. The analysis was not repeated since the conclusions would not change significantly as they depend mostly on relative intensities.

\begin{table}[]
\begin{center}
\caption{Frequency, upper level energy, $E_{u}$, and Einstein coefficient $A_{ul}$ of observed lines}
\begin{tabular}{lcrcc}   
\hline \hline
Species &Tran-& Frequency&$E_{u}$&$A_{ul}$\\
 & sition& [GHz]&[K]&[s$^{-1}$]\\
\hline
CH$^+$&$1-0$&835.1375&40.08&6.4(-3)\\
CH$^+$&$2-1$&1669.2813&120.19&6.1(-2)\\
OH$^+$&$1_{1\frac{1}{2}}-0_{1\frac{1}{2}}$&1032.9979&49.58&1.4(-2)\\
&$1_{1\frac{3}{2}}-0_{1\frac{1}{2}}$&1033.0044&49.58&3.5(-3)\\
&$1_{1\frac{1}{2}}-0_{1\frac{3}{2}}$&1033.1118&49.58&7.0(-3)\\
&$1_{1\frac{3}{2}}-0_{1\frac{3}{2}}$&1033.1186$^*$&49.58& 1.8(-2)\\
o-H$_2$O$^+$&1$_{11\frac{3}{2}\frac{3}{2}}$ $-$ 0$_{00\frac{1}{2}\frac{1}{2}}$&1115.1507&53.51&1.7(-2)\\
&1$_{11\frac{3}{2}\frac{1}{2}}$$-$ 0$_{00\frac{1}{2}\frac{1}{2}}$&1115.1862&53.51&2.8(-2)\\
&1$_{11\frac{3}{2}\frac{5}{2}}$$-$ 0$_{00\frac{1}{2}\frac{3}{2}}$&1115.2041$^*$$$&53.51 & 3.1(-2) \\
&1$_{11\frac{3}{2}\frac{3}{2}}$$-$ 0$_{00\frac{1}{2}\frac{3}{2}}$&1115.2629&53.51&1.4(-2)\\
&1$_{11\frac{3}{2}\frac{1}{2}}$$-$ 0$_{00\frac{1}{2}\frac{3}{2}}$&1115.2983&53.51&3.5(-3)\\
o-H$_3$O$^+$&4$_{3}^{+}$ $-$ 3$_{3}^{-}$&1031.2995&232.2 & 5.1(-3) \\
SH$^+$&$1_{2(\frac{3}{2})} - 0_{1(\frac{1}{2})}$&526.0387&25.25&8.0(-4) \\
&$1_{2(\frac{5}{2})} - 0_{1(\frac{3}{2})}$&526.0479$^*$&25.25&9.7(-4) \\
&$1_{2(\frac{3}{2})} - 0_{1(\frac{3}{2})}$&526.1250&25.25&1.6(-4) \\
HCO$^+$&6 $ -$ 5& 535.0616&89.9&1.3(-2) \\
CH&$(\frac{3}{2})_{2-}-(\frac{1}{2})_{1+}$ &536.7611$^*$& 25.76& 6.4(-4)\\
&$(\frac{3}{2})_{1-}-(\frac{1}{2})_{1+}$&536.7820&25.76&2.3(-4)\\
&$(\frac{3}{2})_{2-}-(\frac{1}{2})_{0+}$&536.7956&25.76&4.6(-4)\\
C$^+$&$^2P_{\frac{3}{2}} -\ ^2P_{\frac{1}{2}}$&1900.5369&91.21&2.3(-6) \\
\hline
\end{tabular}
\end{center}
\vskip-0.1cm
{\tiny  {\bf Notes.} The strongest (hyper)fine transition is marked with an asterisk ($^{*}$ symbol) and used to set the velocity scale in Figs. \ref{CHp_obs} -- \ref{CH_obs}. The numbers in parentheses give the decimal power. Molecular and atomic data are taken from CDMS \citep{2001A&A...370L..49M}.}
\label{table_lines}
\end{table}

After visual inspection and defringing, the two polarizations were averaged. The continuum level was corrected for single sideband by halving the measured value. The calibration uncertainty is $\approx 10\%$ and limited to $<15\%$ in all bands \citep{2012A&A...537A..17R}. For some sources in each band, a second observation of equal length was made, using a local oscillator frequency shifted by 10 km s$^{-1}$ to identify interfering lines from the other sideband. All lines of interest could be attributed to a sideband without ambiguity.

Emission at the off-positions can interfere with the background elimination. This is mostly a problem for C$^+$, which is abundant, has a low critical density, and thus is easily excited. All of the C$^+$ detections were checked by comparing the spectra at the reference positions. For the C$^+$ observation of S 140 the reference position was at a distance of 6.57 arcmin from IRS1 in an H{\tiny{\textsc II}} region in the direction of the illuminating star. Some questionable cases of CH$^+$ and OH$^+$ absorptions were also tested with no indications for off-source contamination.

\begin{figure*}[]
\centering
\resizebox{15cm}{!}{\includegraphics{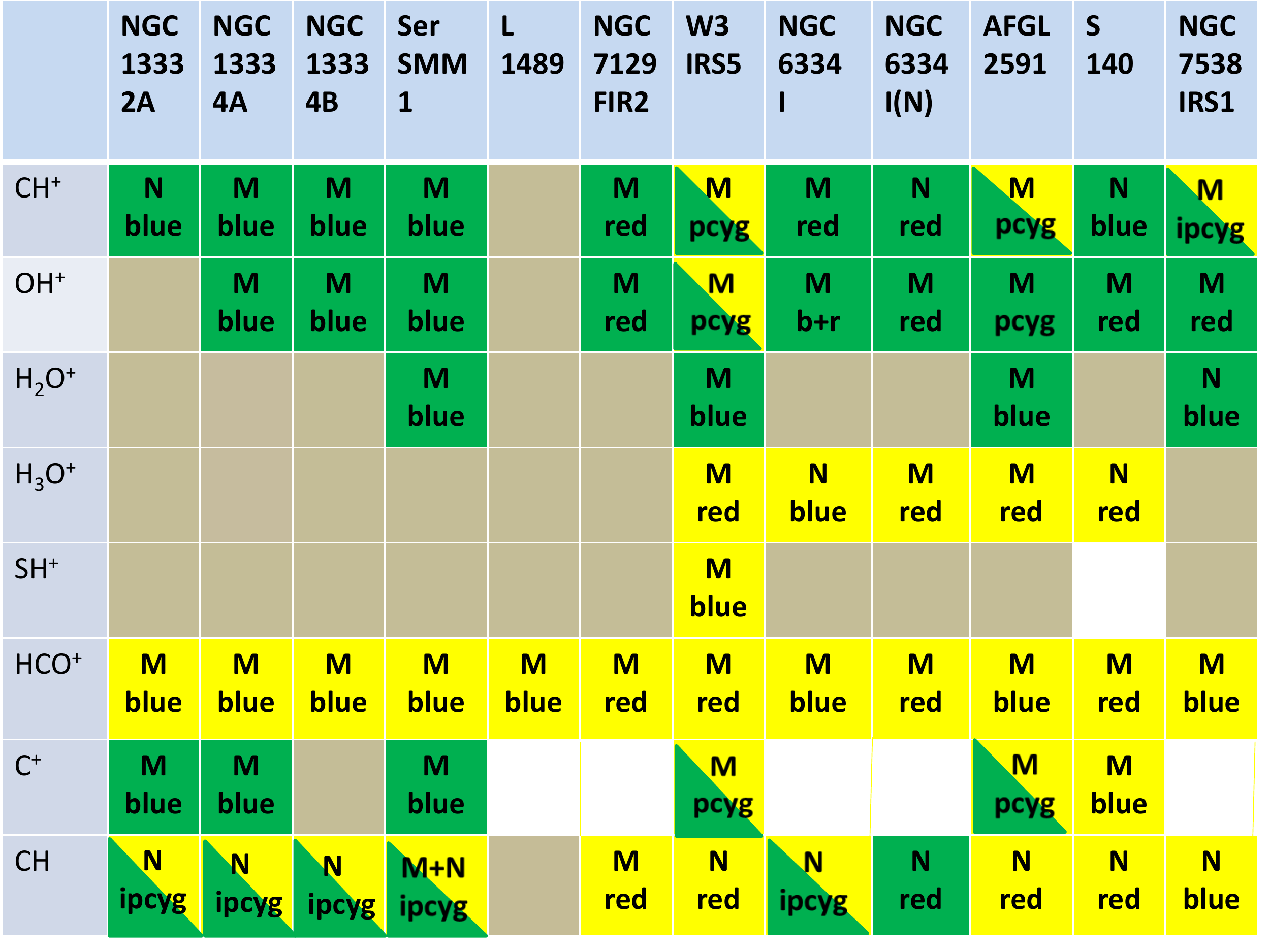}}
\caption{Overview of observed  properties of the main line component. Color code: yellow=emission, green=absorption, beige=non-detection, white=not observed. ``M" and ``N" refer to medium-broad (5 - 20 km s$^{-1}$) and narrow  ($<$5 km s$^{-1}$) line width; ``blue", ``red" to the line shift, ``b+r" to components in both directions, ``pcyg" to P-Cygni, and ``ipcyg" to inverse P-Cygni profiles. The quantitative values are reported in Tables A.1-A.3; line transitions and frequencies are given in Table \ref{table_lines}.}
\label{overview}
\end{figure*}

\section{Derivation of column density}
The column density $N_i$ of molecule $i$ is calculated from the velocity integrated molecular line emission or absorption. For optically thin emission, the column density of the upper level of the transition, $N_i^{u}$, is
\begin{equation}
N_i^{u} = {{8 \pi k \nu^2}\over {h c^3 A_{ul}}}\int{T_{MB} dV}\ \ \ ,
\end{equation}
where $A_{ul}$ is the Einstein-A coefficient, $\nu$ the frequency of the line and $k$ the Boltzmann constant. For absorption, the column density $N_i^{l}$ of the lower level $l$ of the transition ($l\rightarrow u$) is
\begin{equation}
N_i^{l} = {{8 \pi \nu^3 g_{l}}\over {c^3 A_{ul} g_{u}}}\int{\tau dV}\ \ \ ,
\end{equation}
where $g_{u}$ and $g_{l}$ are the statistical weights of the upper and lower level. The optical depth is $\tau = {\rm ln}(T_{cont}/T_{line})$. $T_{cont}$ is the single sideband continuum main-beam temperature. The contributions of different fine or hyperfine transitions are summed up according to their statistical weight. For a ground state line in absorption, the total column density, $N_i$, of a molecule $i$ can be derived from
\begin{equation}
N_i = {{N_i^{0} Q}\over g_0 }\ \ \ .
\label{total_N}
\end{equation}
$N_i^{0}$ is the column density of the ground state, and $g_0$ its degeneracy. $Q$ denotes the partition function at the excitation temperature $T_{\rm ex}$. In the case of line emission from an upper level $u$, the total column density is
\begin{equation}
N_i = {{N_i^{u} Q(T_{\rm ex})}\over g_{u} } \exp({E_{u}\over {k T_{\rm ex}}})\ \ \ ,
\label{total_N}
\end{equation}
where $E_{u}$ is the upper energy level.

In some cases $T_{MB}$ seems to reach values slightly below zero (such as for CH$^+$ in NGC6334 I, AFGL 2591, and NGC6334 I(N)). For C$^+$ toward W3 IRS 5 (Fig. \ref{CII_obs}, third row, middle), the negative value exceeds the 15\% calibration uncertainty of the continuum. Most likely it is an effect of an incomplete sideband correction (in all cases the continuum increases with frequency and the line was observed in the upper sideband). In these cases the integral $\int\tau dV$ cannot be determined. Its column density is given as a lower limit by assuming $\tau = 3$ in the region where the data would suggest $\tau > 3$. The limit on $\tau$ is necessary because the accuracy in the background continuum determines the accuracy of large $\tau$. At $\tau = 3$ the background is reduced to 5\%. Since the bottom of the absorption still has a sharp peak in all cases, we infer that the absorption is not extremely saturated. Note that complete absorption requires that the absorbing region has equal or larger spatial extent than the background emission region behind it.

All column densities given in Tables B.1 - B.3 are beam-averaged values and are not corrected for a beam filling factor. There are two cases of column densities derived from absorption lines: {\it (i)} if the absorbing region is larger than the continuum source, the derived column density refers to the average along the line of sight through which the continuum is observed. {\it (ii)} If the absorbing region is smaller than the continuum emitting region and does not fill the beam, the derived column density underestimates the true value.

There are also two cases of lines in emission: {\it (i)} If the emitting regions are larger than the beam, the derived values refer to the sampled part of that region. {\it (ii)} If the source regions are smaller than the beam, the average column densities are beam diluted and depend on the ratios of the beam size to the source size. The absolute beam size (in AU) is given by the source distance and the frequency. When column densities are compared for species whose emission does not fill the beam, we may correct the column density for the difference in beam size, but not for the usually unknown source size. Equivalently, the line intensities in such comparisons may be normalized to a given distance (we use 1 pc), i.e., multiplied by the square of the distance. Note that this correction if applied to case (i) would spuriously introduce or enhance a correlation due to the relation between YSO luminosity and distance (to be discussed later, see Fig. \ref{D5}, bottom right). In reality, the distribution of the emission over the beam area is heterogeneous and is between the extreme cases of homogeneous and point-like. In the following, we refrain from corrections in general and indicate the exceptions.

\section{Results and analysis}

Figure \ref{overview} presents an overview of the detection of the observed lines. Most of the lines are detected in several sources; SH$^+$ is found only toward one object. Only lines centered within $\pm 12$ km s$^{-1}$ of the systemic velocity were considered. The range excludes gas at high velocity (``bullets") associated with jets and most of the absorption lines caused by the diffuse interstellar clouds in the foreground. The analyzed lines are assumed to be associated with the YSO; exceptions are discussed. The observed line spectra are individually shown in Appendix A (Figs. \ref{CHp_obs} -- \ref{CH_obs}), where details on individual observations, such as line identification, significance of the detection, and line confusion are discussed. The line shapes are fitted with Gaussians. The results are given in Tables B.1 -- B.3.

\begin{table}[]
\begin{center}
\caption{Assumed $T_{\rm ex}$ for reported column density, expected range of  $T_{\rm ex}$, and total error margins (factor) of the reported column density at the lower/upper limits of the expected $T_{\rm ex}$ range}
\begin{tabular}{lrccc}   
\hline \hline
Species &line&assumed& expected &error factors\\
& mode& $T_{\rm ex}$ [K]&$T_{\rm ex}$ [K]&col. density\\
\hline
CH$^+$&emission&38$-$44&19$-$75&1.8 \ 1.2\\
CH$^+$&absorption&3$-$9&5 $-$19&0.8 \ 1.4\\
OH$^+$&emission&38&9$-$75&0.8 \ 2.5\\
OH$^+$&absorption&3$-$9&3$-$19&0.8 \ 1.4\\
H$_2$O$^+$&absorption&3$-$9&3$-$19&0.8 \ 1.5\\
H$_3$O$^+$&emission&225&75$-$225&1.6\ 0.8\\
SH$^+$&emission&38&19$-$75&0.8 \ 1.5\\
HCO$^+$&emission&38&19$-$75&5.7 \ 0.5\\
C$^+$&emission&38&19$-$75&1.3 \ 0.4\\
CH&emission&38&19$-$75&0.8 \ 1.4\\
CH&absorption&6&3$-$9&0.85 \ 1.2\\
\hline
\end{tabular}
\end{center}
\label{table_errors}
\end{table}

The accuracy of the reported column density is limited by the calibration error ($<15\%$) of the line flux  at low frequencies (for CH$^+$ and HCO$^+$) and by the background determination for the other lines. Similar error margins arise from the assumed excitation temperature $T_{\rm ex}$, which is estimated from various sources with multiple transitions detected. The excitation temperature of CH$^+$ in AFGL 2591, W3 IRS5, and NGC6334 I was determined from a 1D slab mode solving the radiative transfer equation along the line of sight and fitting the $J=2-1$ and $J=1-0$ lines for the same $T_{\rm ex}$. The other fitted parameters were column density, velocity and line width \citep{2010A&A...521L..44B, 2013JPCA..117.9840B}. The inferred values, 38 K for emission and 9 K for absorption, were subsequently assumed for the other sources. For OH$^+$, the upper limit of $T_{\rm ex}$ derived from the non-detection of the (2-1) line in absorption is 40 K. We assumed 9 K. Analogously, we assumed 9 K for the H$_2$O$^+$ absorption lines of the ground-state. For the emission lines of SH$^+$, HCO$^+$, C$^+$, and CH we assumed 38 K. The reasons are:  SH$^+$  in analogy to CH$^+$;  HCO$^+$ (6-5)  as reported by \citet{2014A&A...563A.127M} toward the outer regions of NGC6334 I; CH similar to the envelope component of OH as derived by \citet{2011A&A...531L..16W} for W3 IRS5. The adopted $T_{\rm ex}$  is given in Tables B.1 -- B.3 for each case. For the error introduced by the assumption of $T_{\rm ex}$, we explored a range of expected values.  Table \ref{table_errors} summarizes the $T_{\rm ex}$ values, gives the expected $T_{\rm ex}$ range  for each species and reports the resulting uncertainty in the column density at the range limits. In conclusion, the total error margin of the inferred column densities is typically factors of 0.8 - 2 for emission  lines, and factors of 0.8 - 1.5 for lines in absorption.

Figure \ref{overview} gives also a rough overview of line widths and line shifts. The line widths are classified according to \citet{2012A&A...542A...8K} for H$_2$O lines in low-mass objects. No broad line widths ($>$20 km s$^{-1}$) were observed in the lines selected here. More useful are average values for each species and class based on the Tables B.1 - B.3. They are given in Table 3. The indicated range does not represent the uncertainty but rather the spread of the distribution around the average. The line widths of intermediate and high-mass mass objects are on average larger than for low-mass objects (Figs. C.4 -- C.9, top). This trend holds for all lines and was noted by \citet{2013A&A...553A.125S} also for C$^{18}$O ($J=10-9$ and $3-2$)(Table 3). The only exception is C$^+$, for which the line width does not increase with luminosity or envelope mass (see also Fig. \ref{CII}, top row). Furthermore, the scatter in line width increases generally from low- to high-mass objects, again with the exception of C$^+$.

\begin{table}[]
\begin{center}
\caption{Averages of line width, velocity shift, and standard deviation of their distributions. The accuracy of individual line widths and line shifts are better than 1 km s$^{-1}$. First four lines: blue-shifted molecules in absorption; second group: molecules preferentially in emission and unshifted; bottom line: $^{13}$CO (10-9) from \citet{2013A&A...553A.125S} for comparison. }
\begin{tabular}{llll}   
\hline \hline\\
Species &Line  &Velocity &Comment\\
&Width&Shift&\\
 &[km s$^{-1}$]&[km s$^{-1}$]&\\
\hline\\
\multicolumn{4}{c}{Low-mass Objects}\\
\\\hline
CH$^+$&6.0$\pm$0.8&$-$2.0$\pm1.2$&main abs. comp.\\
OH$^+$&4.1$\pm$4.0&$-$3.9$\pm1.0$&main abs. comp.\\
o-H$_2$O$^+$&5.5&$-$3.5&one object, abs.\\
C$^+$&6.0$\pm$0.9&$-$5.8$\pm1.2$&main abs. comp. \\
\hline
H$_3$O$^+$&$-$&$-$&not detected \\
HCO$^+$&2.5$\pm$0.6&$-$0.12$\pm0.12$&narrow em. comp. \\
HCO$^+$&8.5$\pm$3.3&\ \ 0.20$\pm0.86$&broad em. comp. \\
CH&1.0$\pm$0.5&$-$0.35$\pm0.23$&narrow em. comp. \\
&9.4&\ \ 0.5&broad em. comp. \\
\hline
$^{13}$CO(10-9)&3.9$\pm 3.0$&&narrow em. comp.\\
\hline\\
\multicolumn{4}{c}{Intermediate and High-mass Objects}\\
\\\hline
CH$^+$&8.2$\pm$4.3&$-$0.1$\pm4.9$&main abs. comp.\\
OH$^+$&8.7$\pm$2.9&2.8$\pm7.1$&main abs. comp.\\
o-H$_2$O$^+$&9.4$\pm4.1$&$-$7.6$\pm3.3$&blue abs. comp.\\
C$^+$&5.7$\pm$1.0&1.5$\pm2.5$&main em. comp. \\
\hline
H$_3$O$^+$&6.4$\pm$2.7&1.2$\pm1.2$&em. comp.\\
HCO$^+$&4.1$\pm$1.1&$-$0.01$\pm0.46$&narrow em. comp. \\
HCO$^+$&12.9$\pm$8.0&0.59$\pm1.5$&broad em. comp. \\
CH&3.9$\pm$0.8&1.2$\pm2.0$&narrow em. comp. \\
\hline
$^{13}$CO(10-9)&4.3$\pm 0.8$&&narrow em. comp.\\
\hline
\end{tabular}
\end{center}
\vskip-0.1cm
\label{table_line_width}
\end{table}

\subsection{Line classification}

Based on Fig. \ref{overview} and Table 3, two classes of lines can be identified that generally comply with the following trends: \\
$(i)$ The CH$^+$, OH$^+$, H$_2$O$^+$, and C$^+$ lines are mostly in absorption, preferentially blue-shifted and occasionally P-Cygni. Their line widths are generally medium-broad.\\
$(ii)$ The H$_3$O$^+$, SH$^+$, HCO$^+$, and CH lines are mostly in emission. Their shifts relative to the systemic velocity are small ($<$ 1 km s$^{-1}$); positive and negative values are equally frequent. The line widths are generally narrow or have a narrow component, except for some cases of H$_3$O$^+$ and the wing component of HCO$^+$. The line widths of CH are extremely narrow and show inverse P-Cygni profiles in about half of the objects.

\begin{figure}[htp]
\centering
\begin{minipage}{0.7\linewidth}
\centering
\begin{minipage}[t]{1.0\linewidth}
\centering
\includegraphics[width=\linewidth]{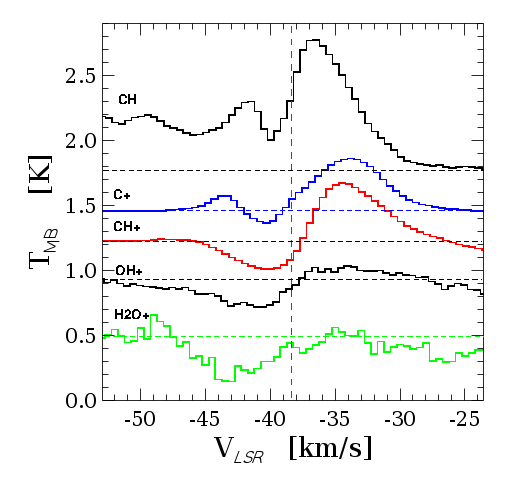}
\end{minipage}
\\[-5pt]
\centering
\begin{minipage}[t]{1.0\linewidth}
\centering
\hspace{-0.2cm}
\includegraphics[width=\linewidth]{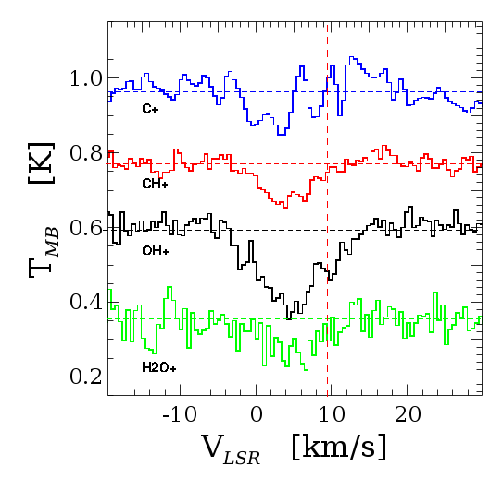}
\end{minipage}
\end{minipage}
\caption{{\it Top:} Superposition of line profiles of various molecules of the first molecular class (see text) toward the high-mass object W3 IRS5. The line intensities are scaled and shifted to allow better comparison (CH: $T_{MB}$+1.5 K; C$^+$: $0.01 T_{MB}$+1.4 K;  CH$^+$: $0.25 T_{MB}$+1.0 K; OH$^+$: $T_{MB}$ - 0.5 K; H$_2$O$^+$: $2 T_{MB}$ - 3.7 K). The systemic velocity of the YSO, $-$38.4 km s$^{-1}$, is indicated with a vertical red dashed line. {\it Bottom:} Same as top for the low-mass object Ser SMM1 (C$^+$: $0.5 T_{MB}$ - 0.2 K;  CH$^+$: $T_{MB}$+0.5 K; OH$^+$: $T_{MB}$+0.5 K; H$_2$O$^+$: $2 T_{MB}$ - 0.45 K).}
\label{super}
\end{figure}

Figure \ref{super} shows two examples of composites of the first class of lines. In the high-mass object (top) both the absorption line width and the line shift decrease from bottom to top, indicating a relation between mean velocity and turbulent velocity. The absorption feature of CH at $-$40 km s$^{-1}$ may also be related to the first group of molecules, but the dip in the CH spectrum at $-$46 km s$^{-1}$ is caused by another CH hyperfine absorption line (see Fig. \ref{CH_obs}, bottom left). Absorption lines toward Ser SMM1 are superposed in Fig. \ref{super} (bottom). We conclude that the absorption lines of our first molecular class are related to each other and originate in the same physical unit, but not at exactly the same place. \citet{2013A&A...557A..23K} report similar properties for the ``offset component" of H$_2$O.

The second molecular class includes emission lines with narrow line width and a velocity shift that is small relative to the systemic velocity and unrelated to each other (as shown later in Fig. \ref{corr2}). Most of the shifts are within the accuracy of the measurements of molecular and systemic motions. The line width of H$_3$O$^+$ is generally larger than that of the narrow component (peak) of HCO$^+$, and CH in emission is even narrower than HCO$^+$. An inverse P-Cygni profile as in the observed CH line toward some objects (see Fig. \ref{overview}) was reported also for the narrow component of the H$_2$O ground-state lines in low-mass objects \citep{2012A&A...542A...8K,2013A&A...558A.126M,2016A&A...585A.103S}.

\begin{figure*}[htb]
\centering
\resizebox{17cm}{!}{\includegraphics{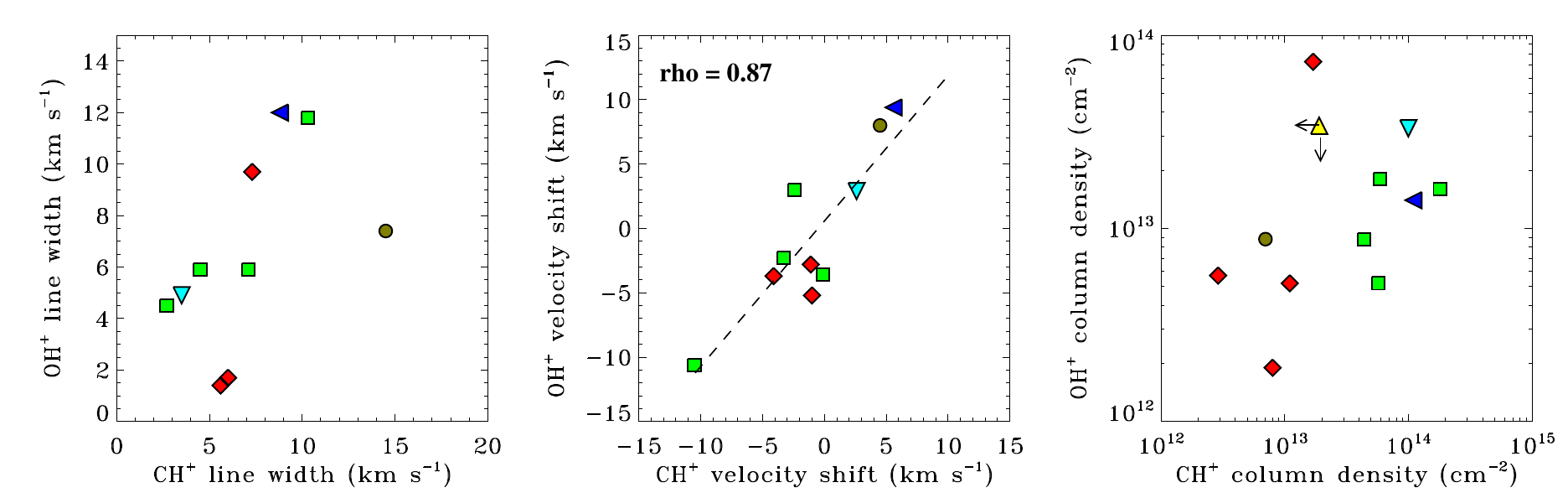}}
\caption{Observed line parameters of OH$^+$ (Table B.1) vs. observed line parameters of CH$^+$. The symbols mark different objects: Red diamonds for Class 0, yellow triangle up for Class I (L 1489), brown circle for intermediate mass, light blue triangle down for high-mass mid-IR quiet (NGC6334 I(N)), green square for high-mass mid-IR bright and hot molecular core, and dark blue triangle left for high-mass ultra-compact H{\tiny{\textsc II}} (NGC7538 IRS1).  In this and subsequent figures uncertainties of individual points are typically the size of the symbols and up to a factor of 2 for column densities. The Pearson correlation coefficient $\rho$ is given only for statistically significant relations.}
\label{corr1}
\end{figure*}

\subsection{Correlations between different molecules}
In the following, correlations are studied between line parameters of different molecules and objects with the aim to test the membership to the two classes defined above. We have used for analysis only the components likely to be associated with the YSO and omitted absorption components marked with ``DC" in Tables B.1 - B.3, for which confusion with diffuse clouds cannot be excluded. Where appropriate (column density, luminosity, mass, etc.), log-log scales are used for statistics to give equal weight to widely different objects. Quantitative studies use the Pearson correlation coefficient $\rho$ to characterize the correlation and test its significance against the null hypothesis (applying the Student's t-test at 99 \% confidence level). A linear fit (regression) quantifies the relation, and chi-square statistics is used to compare the scatter of different correlations. Correlation coefficients are only indicated in the figures where statistically significant.

Figure \ref{corr1} displays the relations between the lines of the first class of molecules (medium width, blue-shifted absorptions). OH$^+$ and CH$^+$ correlate in line shift relative to the systemic velocity. The correlation extends from low-mass to high-mass objects. The scatter is within a few km s$^{-1}$. The apparent correlations between OH$^+$ and CH$^+$ in line width and column density are statistically not significant. The line parameters of C$^+$ correlate less with CH$^+$ than those of OH$^+$, partly due to the P-Cygni line profiles of some sources.

The line shifts of OH$^+$ and H$_2$O$^+$ toward Ser SMM1, AFGL 2591 and W3 IRS5 correlate well (Tables B.1 and B.2). The averaged difference $<\delta V_{\rm{H_2O^+}} - \delta V_{\rm{OH^+}}>$ is $-$0.27 km s$^{-1}$. The difference may result from the inaccuracy of the transition frequency of the H$_2$O$^+$ line; it would amount to a correction of +1.0 MHz, to be added to the H$_2$O$^+$ frequencies listed in Table \ref{table_lines}. A correction was discussed by \citet{2010A&A...521L..10N} and \citet{2015ApJ...800...40I}, who reported a value of +5 MHz.

\begin{figure*}[htb]
\centering
\resizebox{17cm}{!}{\includegraphics{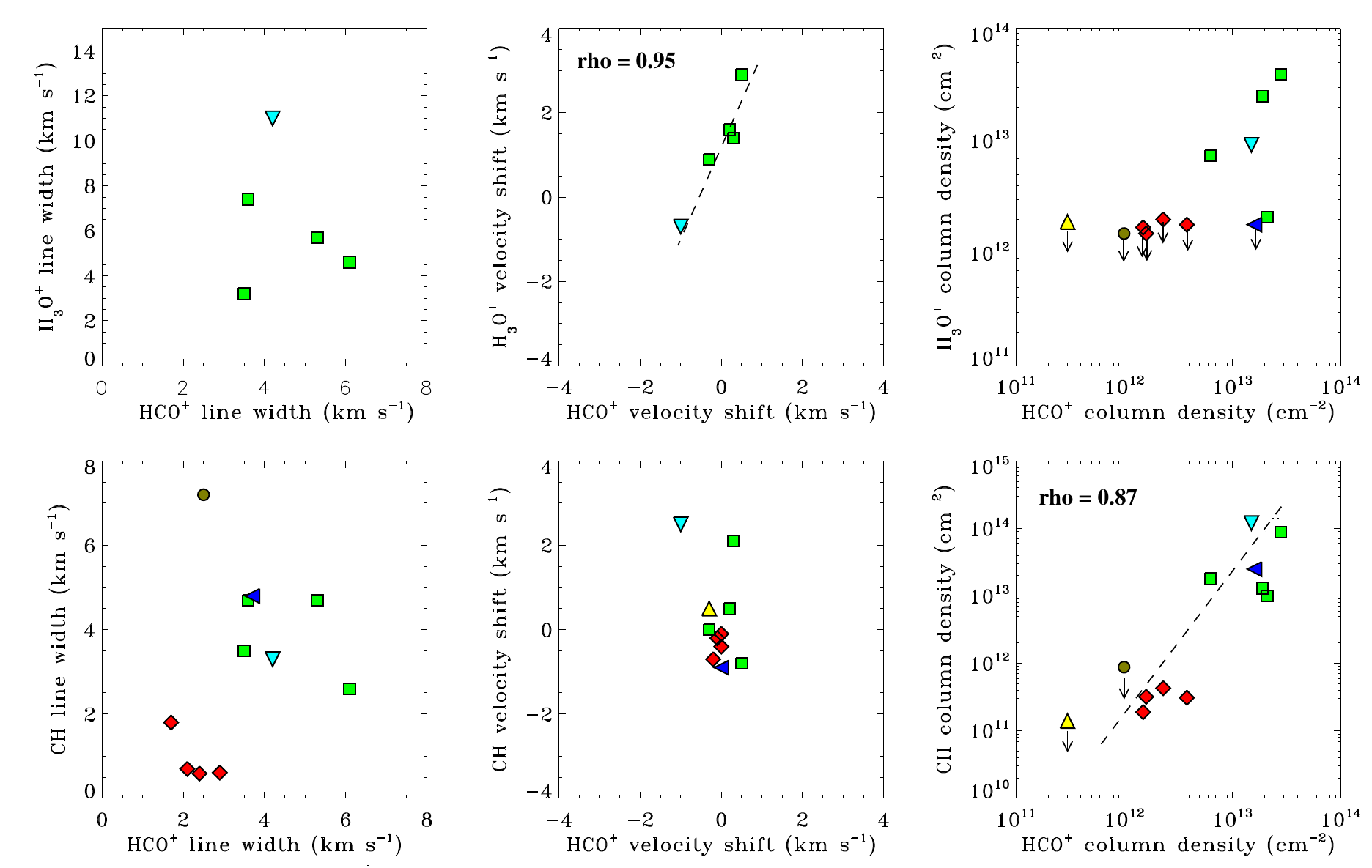}}
\caption{Observed emission line parameters of H$_3$O$^+$ and CH  vs. observed line parameters of HCO$^+$ (narrow component) for line width and shift relative to the systemic velocity, and total for column density. For the notation of the symbols see Fig. \ref{corr1}.}
\label{corr2}
\end{figure*}

The relations between the molecules of the second class are presented in Fig. \ref{corr2}. The line shifts of H$_3$O$^+$ and HCO$^+$ are correlated. Also the  column densities of CH and HCO$^+$ are well correlated. The relation between them is $N({\rm CH}) \propto N({\rm HCO}^+)^{1.7\pm0.3}$.

As expected, there is no correlation in line width and line shift of the second class of molecules with CH$^+$ as representative of the first class (Fig. \ref{H3Op-CHp}). However, the column density of HCO$^+$ (derived from an emission line) and the column density of CH$^+$ (derived from absorption) are well correlated ($\rho = 0.83$). CH and CH$^+$ are also correlated in column density. The relations are not linear, but power laws: $N({\rm HCO^+}) \propto N({\rm CH}^+)^{0.7\pm 0.2}$, and  $N({\rm CH}) \propto N({\rm CH}^+)^{1.5\pm 0.3}$.

The HCO$^+$(6-5) line was fitted by a broad and a narrow Gaussian, representing high and low turbulent velocities. The average line widths are 11.0$\pm$1.0 km s$^{-1}$ and 3.4$\pm$1.3 km s$^{-1}$ for the broad and narrow component, respectively. The distribution of the two components in line width can be fitted with the linear regression $\Delta V_{\rm broad}=3.1(\pm 0.5) \Delta V_{\rm narrow}$ (Fig. \ref{HCOp1-HCOp2}). There is no correlation of the line shifts, which scatter around zero. The column densities fitted to the line peak and the wings correlate well, the narrow one being two times higher on average, indicating that the narrow component amounts to two thirds of the total. Thus the two HCO$^+$ components have many similarities and they seem to be physically related. Therefore the column densities are added in Table B.2 and the two components will not be distinguished in the following.

\cite{2013A&A...553A.125S} have analyzed the $^{13}$CO ($J=10-9$) line in a similar way. The  average line widths towards the objects in comon are 4.1$\pm$1.9 km s$^{-1}$ and 12.6$\pm$7.1 km s$^{-1}$ for the narrow and broad component, respectively. Their average ratio is 3.1$\pm$1.0, the same as found for HCO$^+$. The agreement in velocity dispersion between $^{13}$CO (10-9) and our HCO$^+$(6-5) line (Table 3)  suggests a common region of origin.

\begin{figure*}[htb]
\centering
\sidecaption
\resizebox{14cm}{!}{\includegraphics{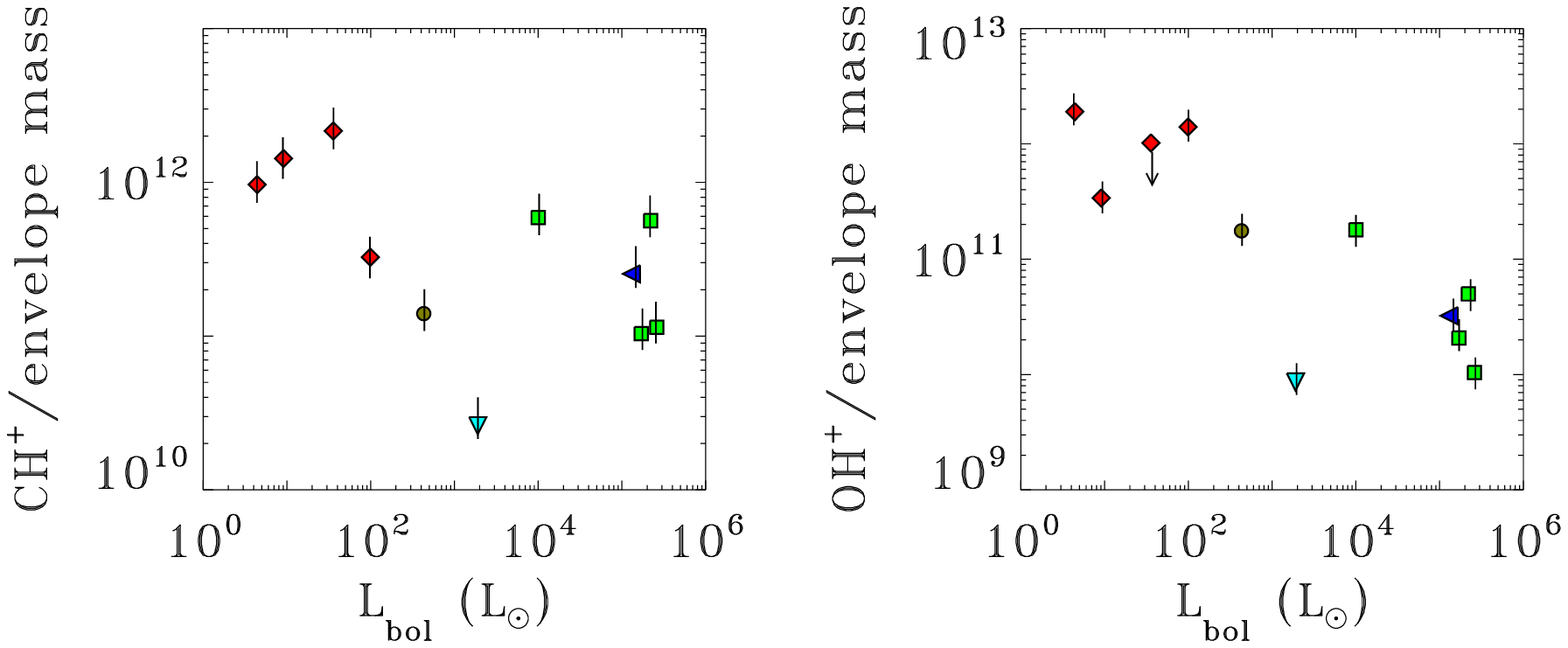}}
\caption{Proxy CH$^+$ abundance. {\it Left:} The ratio of CH$^+$  column density (cm$^{-2}$) to envelope mass (in units of M$_\odot$) versus bolometric luminosity (in units of L$_\odot$). {\it Right:} Same for OH$^+$. For the notation of the symbols see Fig. \ref{corr1}. The error bars show the margins given in Table \ref{table_errors} for the column density and do not include the uncertainty in the envelope mass.}
\label{CHP_abund}
\end{figure*}

\subsection{Relations between molecular lines and object parameters}
Here we summarize the most prominent results on the relations between lines and objects. First, we note that the source parameters are not independent of each other. Already \citet{1996A&A...311..858B} reported an observational relation between bolometric luminosity and envelope mass for Class 0 and Class I low-mass objects. The relation appears clearly in our sample, but now includes also intermediate and high-mass objects and covers 5 orders of magnitude (Fig. \ref{D5} top left); the result is
\begin{equation}
M_{\rm env} \approx 1.1 (L_{\rm bol})^{0.54(\pm 0.12)}
\label{lum_mass-corr}
\end{equation}
using solar units [M$_{\odot}$] and [L$_{\odot}$]. The largest deviations from the regression line are NGC6334 I(N) (a mid-IR quiet object in early evolutionary phase) having the largest envelope mass in the sample, and L 1489, an evolved (Class 1) object with a small envelope mass. Both deviations are consistent with the evolutionary trend reported by \citet{1996A&A...311..858B} and \citet{2008A&A...481..345M} of increasing ($L_{\rm bol})^{0.6}/M_{\rm env}$ with time.

The bolometric temperature is not related to either envelope mass, luminosity or distance (Fig. \ref{D5}). The correlation of luminosity (and thus envelope mass) with distance is conspicuous. It is clearly a selection effect and must be taken into account for the interpretation of  column densities toward unresolved sources (see end of Section 3).

In general, the observed column density of all molecules increases with luminosity. This is the case for column densities based both on absorption and emission. The column density of CH$^+$ has a correlation coefficient of $\rho = 0.87$ with luminosity (Fig. \ref{CHp}, bottom left). The correlation of OH$^+$ with luminosity is not statistically significant. Especially Ser SMM1, which shows exceedingly deep and broad absorptions (Fig. \ref{OHp_obs}, second row, left), is an  outlier. As envelope mass correlates with luminosity (Fig. \ref{D5}, top left), it is not surprising that the column densities of all molecules (except H$_3$O$^+$ and OH$^+$) also correlate with envelope mass with comparable coefficients (Figs. \ref{CHp} - \ref{CH}).

\begin{figure*}[]
\centering
\sidecaption
\resizebox{13cm}{!}{\includegraphics{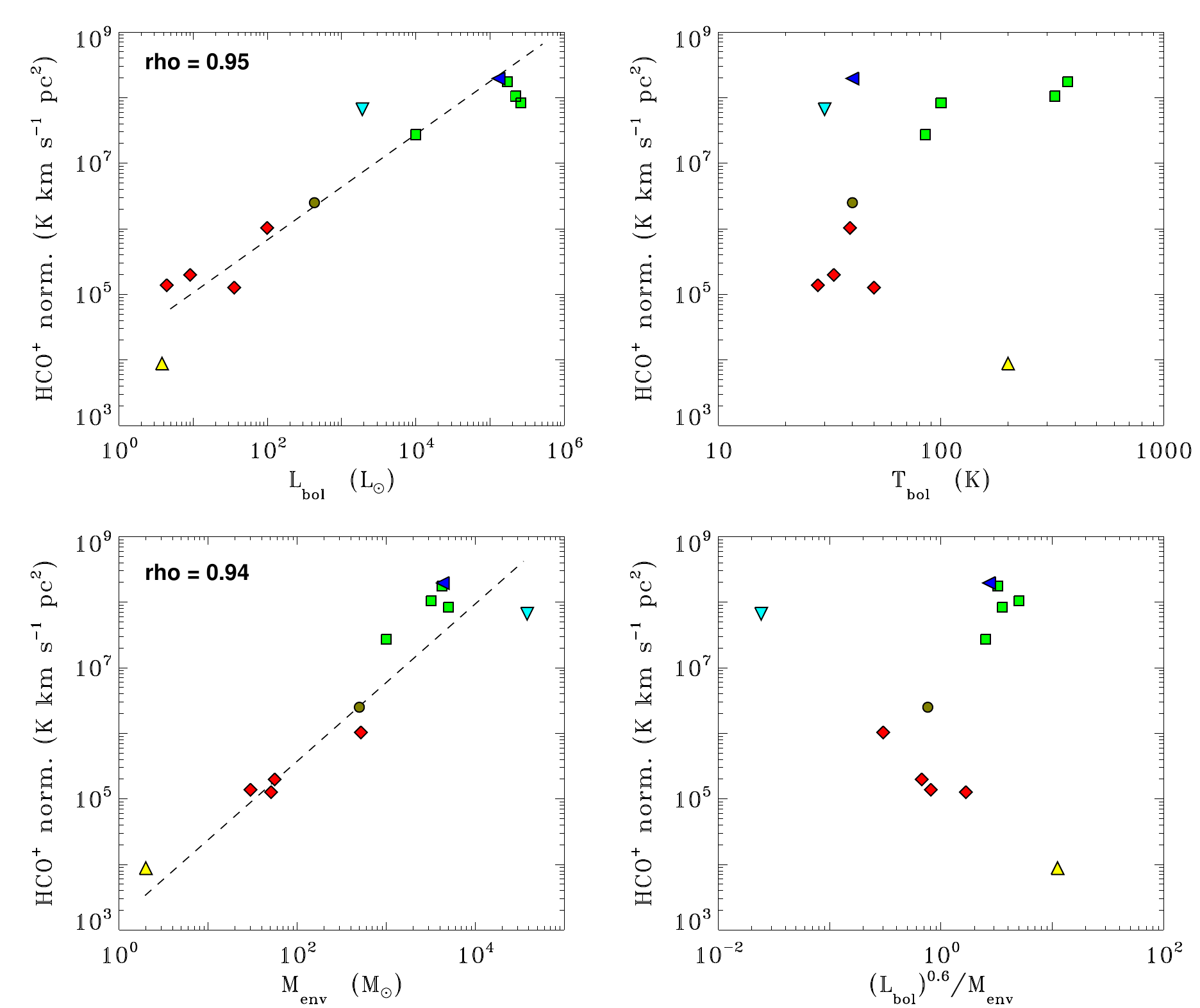}}
\caption{Integrated line intensity, $\int T_{MB} dV$, of HCO$^+$ normalized to 1 pc vs. object parameters given in Table \ref{table_objects}. Notation: Red diamonds for Class 0, yellow triangle up for Class I (L 1489), brown circle for intermediate mass, light blue triangle down for high-mass mid-IR quiet (NGC6334 I(N)), green square for high-mass mid-IR bright and hot molecular core, and dark blue triangle left for high-mass ultra-compact H{\tiny{\textsc II}} (NGC7538 IRS1).\vskip1cm}
\label{HCOp_norm}
\end{figure*}

In Fig. \ref{CHP_abund} the CH$^+$ column density as determined from absorption is divided by the envelope mass (Table \ref{table_objects}). For optically thin emission this ratio is proportional to the ratio of the total number of CH$^+$ ions divided by the total number of hydrogen atoms. With the simplification that the CH$^+$ absorbing region has the same size as that of the H$_2$ region, this ratio is a proxy for the fractional CH$^+$ abundance averaged over the YSO envelope. The ratio is shown vs. luminosity in Fig. \ref{CHP_abund} (left). There is no statistically significant correlation with $L_{\rm bol}$ ($\rho=0.29$), but the low-mass objects ($<$ 100 L$_\odot$) have a factor of 4 larger CH$^+$ abundance on average compared to the intermediate- and high-mass objects. The OH$^+$ abundance (Fig. \ref{CHP_abund} (right) shows a stronger trend ($\rho=0.54$) and a factor of 30 enhancement on average at the low-mass objects. The same inequality between low- and high-mass objects results if instead of the envelope mass the $^{13}$CO ($J$=10-9) and C$^{18}$O ($J$=3-2) line intensities as reported by \citet{2013A&A...553A.125S} are used.

There are no indications for opacity effects in CH$^+$ and OH$^+$ such as rounded or flat peaks in the line shape that could explain Fig. \ref{CHP_abund} Since FUV ionization, heating, and chemistry are surface effects, but the envelope mass is in a volume, the result may be interpreted by a higher surface-to-volume ratio of smaller objects. The very low value for the high-mass mid-IR quiet NGC6334 I(N) (light blue triangle down in both CH$^+$ and OH$^+$ in Fig.  \ref{CHP_abund}) then would indicate that in massive cold envelopes of high-mass YSOs there are large regions where these molecular ions are not enhanced.

\subsection{Correlation of HCO$^+$ with object parameters}
The correlation between the HCO$^+$ (6-5) column density and bolometric luminosity (Fig. \ref{HCOp}) is among the best.  The uncorrected correlation coefficient is 0.83 and the chi-square value is 12.6 relative to the regression line in log-log scale. Radiative transfer modeling of the high-mass AFGL 2591 \citep{2009ApJS..183..179B} and  interferometric observations of HCO$^+$ in the (3-2) transition towards low-mass YSOs indicate a source diameter $\le$30$"$ \citep{1997ApJ...489..293H}. \citet{2014A&A...563A.127M} report optically thick HCO$^+$ (6-5) emission for NGC6334 I and a size of 40$\pm 6"$ for the outer envelope. This suggests that the HCO$^+$ (6-5) emission originates predominantly from an optically thick surface (see also Appendix D). We assume that the emission region is effectively smaller than the {\it Herschel} beam at 535 GHz  (HPBW 44$"$) for all objects.  Thus the HCO$^+$ (6-5) line luminosity must be corrected for the varying beam dilution at different distances. This is done by normalizing the line luminosity to the same distance (see end of Section 3).

The velocity-integrated line intensity of HCO$^+$(6-5), normalized to a distance of 1 pc, is depicted in Fig. \ref{HCOp_norm}. Its correlations with bolometric luminosity and envelope mass are remarkably good. L 1489, the Class I object, and NGC6334 I(N), the high-mass object with highest envelope mass, have the largest deviations from the linear regression line (Fig. \ref{HCOp_norm}, top left). There is no correlation with bolometric temperature (Fig. \ref{HCOp_norm}, top right).

Correcting the integrated line intensity of the HCO$^+$, $I_{\rm HCO+}$ in [K km s$^{-1}$], yields a line luminosity, $L_{\rm HCO+}^{\rm norm}$, normalized to a distance of 1 pc. This increases the correlation coefficient with bolometric luminosity to 0.95; the chi-square value reduces to 4.0, indicating that the data points are close to the regression line. The linear regression in log-log scale amounts to a power-law relation
\begin{equation}
d_{pc}^2\   I_{\rm HCO^{+}}\ \propto\ \ L_{\rm HCO+}^{\rm norm}\ \ \propto\ \ \left(L_{\rm bol}\right)^{0.76\pm 0.08}\ \ .
\label{HCOp_relation}
\end{equation}

This tight relation can be interpreted by noting that the luminosity of an optically thick line depends on the one hand on the radius of the line photosphere, which increases with $L_{\rm bol}$. On the other hand, the temperature of the line photosphere also follows a power law with $L_{\rm bol}$. The properties of the dust radiative transfer become self-similar \citep{1997MNRAS.287..799I} in all models, yielding the observed power-law relations between the relevant parameters. The details are given in Appendix D.

A similar explanation may hold for other optically thick emission lines reported here and elsewhere which correlate between column density and $L_{\rm bol}$. The numerical modeling described in Appendix D indicates that the power-law exponent in the relation between line and bolometric luminosities depends on the location of the line photosphere in relation to the inner and outer envelope radius.

\section{Chemistry constraining FUV and X-rays}
Having characterized the observational properties of the hydrides, we investigated their chemistry in order to explore the origin of the observed absorption or emission within the protostellar envelope and to use them as diagnostics of FUV and X-ray emission. There is an extensive literature on the chemistry of ionized hydrides, especially on CH$^+$ and OH$^+$, in diffuse clouds (see Introduction for references). Although some aspects of the chemistry are similar, star-forming regions differ physically from diffuse clouds. YSOs are expected to be more inhomogeneous in density, temperature and irradiation and may have internal sources of ionizing radiation.

\begin{figure*}[htb]
\centering
\resizebox{18cm}{!}{\includegraphics{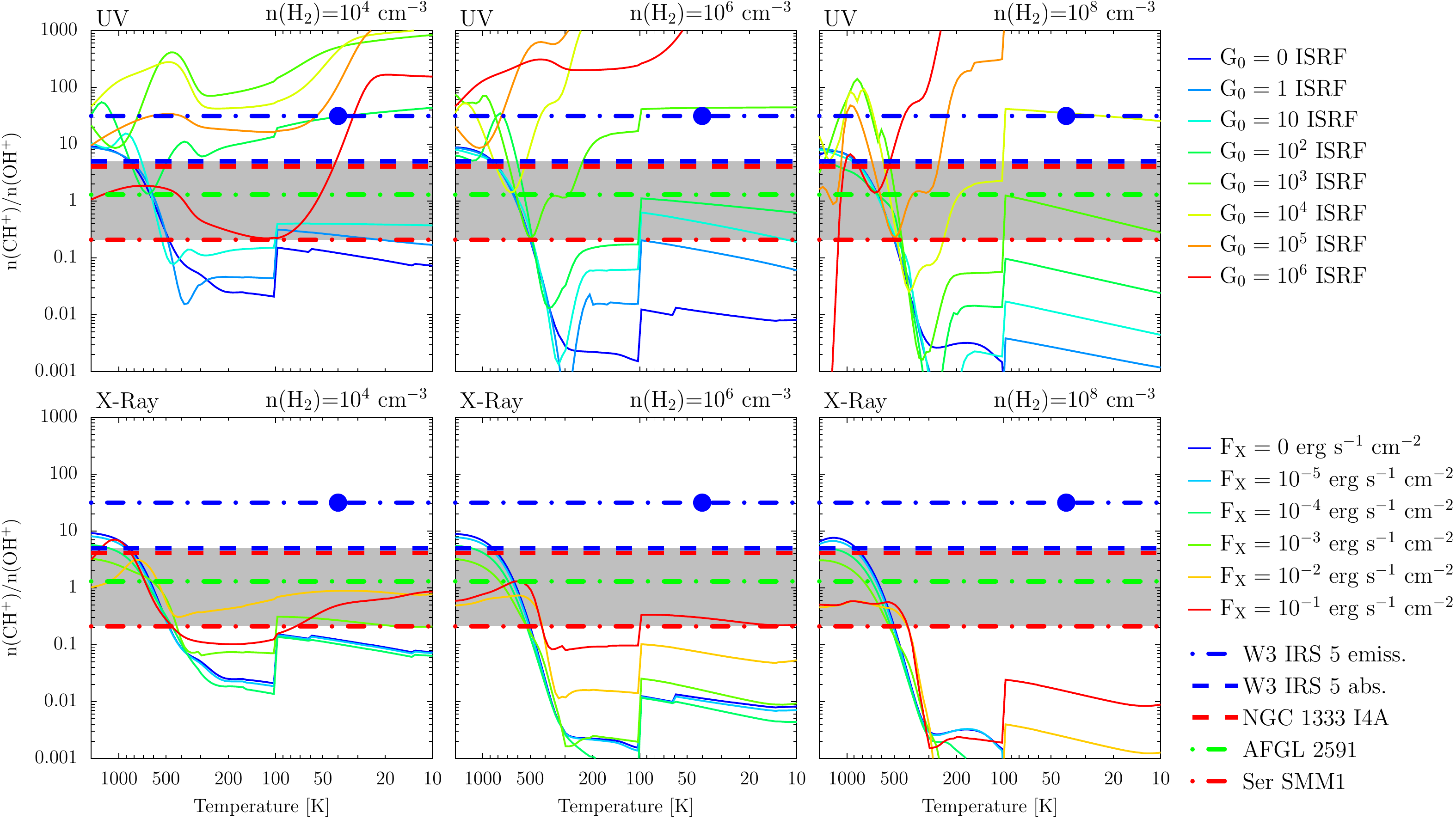}}
\caption{Column density ratio of CH$^+$ to OH$^+$ vs. gas temperature for several chemical toy models  (densities given above the figures) for FUV irradiation without X-rays ({\it top}) and X-ray irradiation without FUV ({\it bottom}). The radiation levels are color coded. The range of observed values for absorption lines is shaded. The excitation temperature for CH$^+$ from 1D slab model fitting of emission lines (Table B.1) is marked with a blue dot.}
\label{D6}
\end{figure*}

\begin{figure*}[htb]
\centering
\resizebox{18cm}{!}{\includegraphics{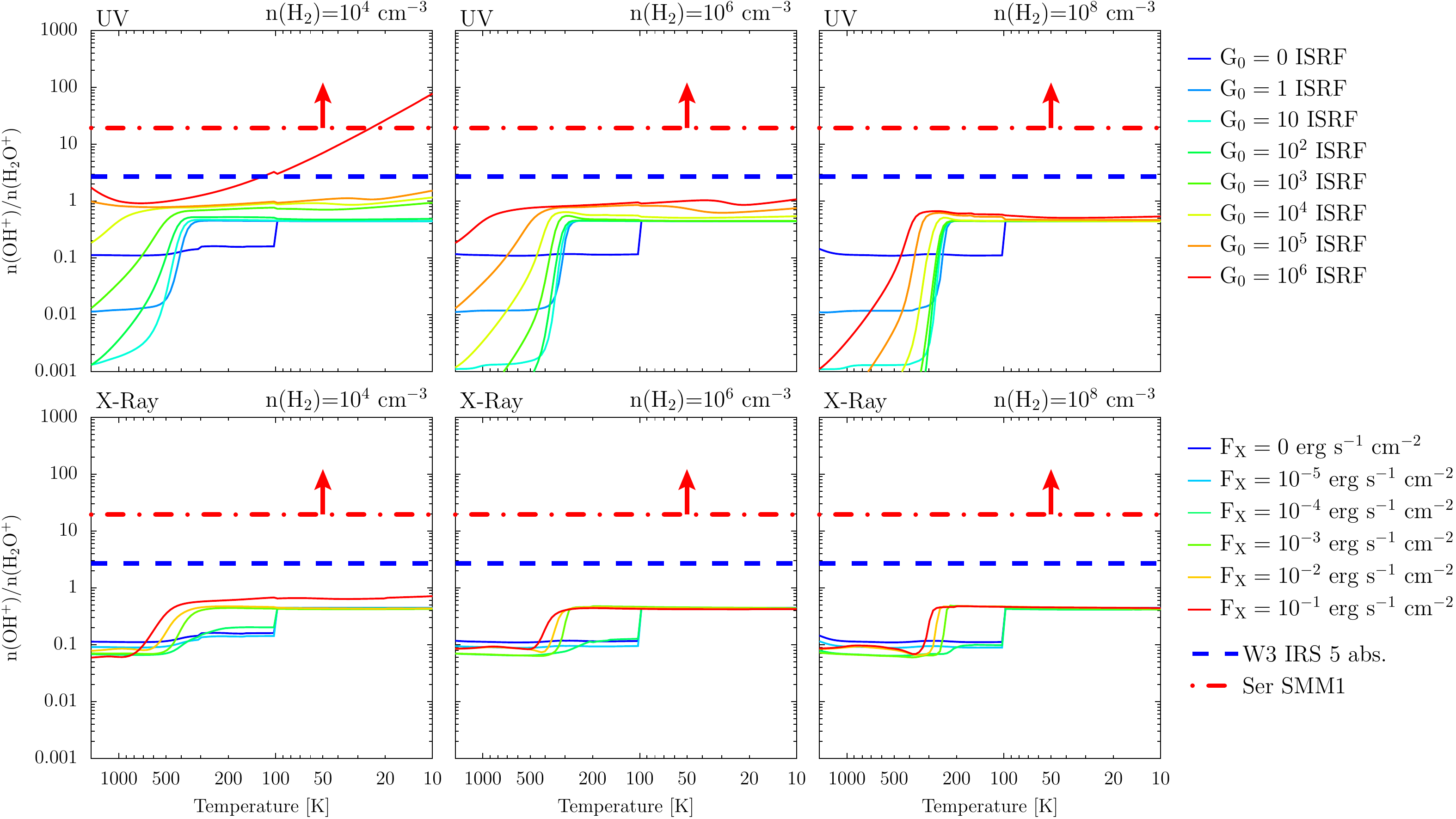}}
\caption{Same as Fig. \ref{D6} for column density ratio of OH$^+$ to H$_2$O$^+$. The ratio given for Ser SMM1 is a lower limit because of the only tentative detection of H$_2$O$^+$.}
\label{D7}
\end{figure*}

\begin{figure*}[htb]
\centering
\resizebox{18cm}{!}{\includegraphics{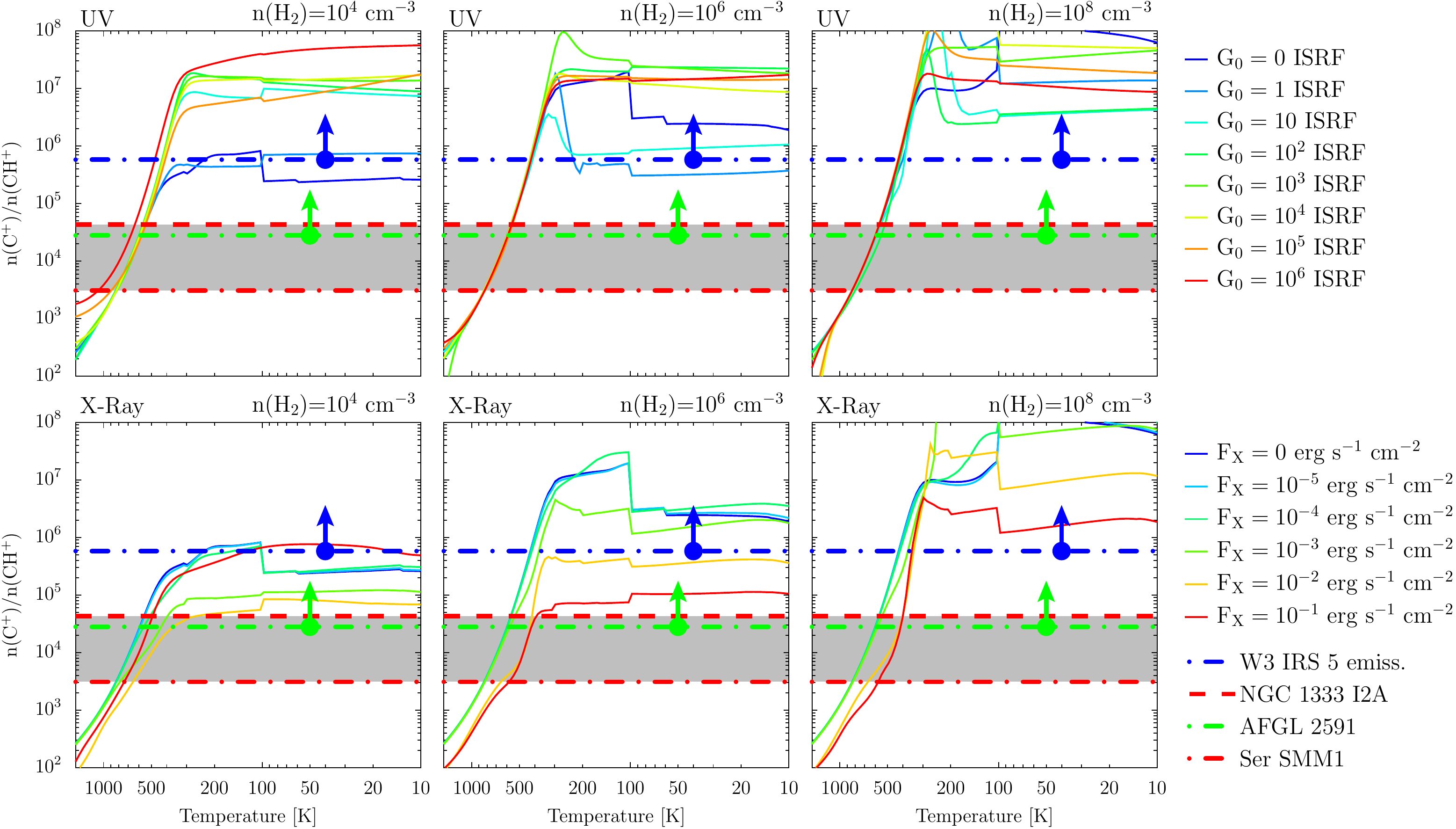}}
\caption{Same as Fig. \ref{D6} for column density ratio of C$^+$ to CH$^+$. }
\label{D8}
\end{figure*}

\subsection{Chemical modeling}
Molecular abundances may be used to probe the internal radiation fields of FUV and X-rays, which add to the interstellar UV and cosmic ray ionization. To explore irradiated hydride chemistry in parameter space, abundances were calculated from chemical models for given values, assuming chemical equilibrium, a given density, temperature, FUV and X-ray irradiation \citep{2005A&A...440..949S}. The temperature is kept as a free parameter. In the bulk of the dense envelope, gas and dust temperatures are well coupled; only along the outflow cavity walls are gas temperatures higher than dust temperatures in a narrow boundary layer \citep[e.g.][]{2009ApJS..183..179B,2012A&A...537A..55V}. The sublimation of ices is controlled by the dust temperature, whereas the rates of gas-phase reactions depend on the gas temperature.

The models use the chemical gas-phase reaction network from the UMIST 06 database \citep{2007A&A...466.1197W}, adopted and updated by \citet{2009ApJS..183..179B}. The sublimation of the most important ices is included (H$_2$O, CO$_2$, and H$_2$S at $T_{\rm dust} >$100 K; H$_2$CO, and CH$_3$OH at $T_{\rm dust} >$60 K). The cosmic ray ionization rate of H$_2$ is assumed to have the value $5\times 10^{-17}$ s$^{-1}$ \citep[e.g.][]{2000A&A...358L..79V}. The elemental abundances are taken from  \citet{2005A&A...440..949S}, Table A1. They practically agree with \citet{1989GeCoA..53..197A} except for the O abundance that is lower by $\sim$40\% and the same as in \citet{2009ARA&A..47..481A}. These models are not intended to quantitatively reproduce the observations, but instead they are single-point toy models to roughly constrain density, gas temperature, and FUV and X-ray irradiation for the parameter ranges that are expected in protostellar envelope models.

The parameter range of toy models (such as that shown in Figs. \ref{D6} -- \ref{D8}) is guided by the assumption that the FUV irradiation originates externally or from the central protostar(s) and that the X-ray fluxes are in the range observed from low-mass Class I and II objects \citep[][]{1999ARA&A..37..363F}. The densities were chosen at three levels to reflect the conditions at the outer edge of the envelope, the average envelope and outflow walls, and the inner envelope and disk atmosphere. Physical models for the envelopes of our sources giving their temperature and density are presented in \citet{2012A&A...542A...8K} and \citet{2013A&A...554A..83V} (see Figs. \ref{fig:hcop_correl1} and \ref{hcop_correl2}). We note that the FUV flux, $G_0$, and gas temperature are connected through FUV absorption. A gas temperature below 300 K is realistic only for $G_0 < 100$ ISRF at $n = 10^4$ cm$^{-3}$; and for $G_0 < 1000$ ISRF at $n = 10^8$ cm$^{-3}$ \citep{2012A&A...537A..55V}. ISRF denotes the FUV flux in terms of the standard interstellar radiation field, $1.6\times 10^{-3}$ erg cm$^{-2}$ s$^{-1}$ \citep{1968BAN....19..421H}.

\begin{table}[htb]
\begin{center}
\caption{Ratios of mean column densities according to Tables B.1 - B.3. The values are derived from lines observed in absorption except where marked ``em." The $-$ sign is set where both lines are undetected.}
\begin{tabular}{lrrr}   
\hline \hline
Object &${N(\rm{CH}^+)}\over {N(\rm{OH}^+)}$&${N(\rm{OH}^+)}\over {N(\rm{H_2O}^+)}$&${N(\rm{C}^+)}\over {N(\rm{CH}^+)}$ \\
\hline\\
NGC1333 I2A&$>$2.1&$-$ &43000\\
NGC1333 I4A& 4.1&$>$0.39&$\le$16000\\
NGC1333 I4B&$>$0.48&$-$&$<$170000\\
Ser SMM1& 0.21&$\ge$15.5& 3100\\
L 1489&$-$&$-$&$-$\\
NGC7129 FIRS2& 0.75&$>$1.5&\\
\\
W3 IRS5&5.0&2.7& $>$130000\\
W3 IRS5 em.&31.5&$-$&$>$580000\\
NGC6334 I&1.8 & $>$124.0&\\
NGC6334 I(N)&1.4& $>$36.1&\\
AFGL 2591&1.3 & 19.4& $>$28000\\
S 140& 1.8& $>$13.6&$\ge$8800\\
NGC7538 IRS1& 0.48& $>$24.5&\\
\hline
\end{tabular}
\end{center}
\label{table_ratios}
\end{table}

\subsection{CH$^+$ to OH$^+$ ratio}
Figure \ref{D6} shows the model abundance ratio of CH$^+$ to OH$^+$ as a function of gas temperature for various types and strengths of irradiation. The range of $G_0$ given in Table 5 for the emission lines is consistent with the assumption of a gas temperature $<$100 K.  Similar studies have been made at lower molecular density to interpret observations of diffuse clouds \citep{2010A&A...521L..44B,2010A&A...518L.110G,2012ApJ...754..105H}.

The formation of CH$^+$ through C$^++ $H$_2 \rightarrow\  $CH$^+$ + H is endothermic by 4640 K, which enhances the ratio above 300 K. At low irradiation and gas temperatures below about 230 K, the abundances of CH$^+$ and OH$^+$ are below $10^{-12}$ relative to H$_2$, and the ratio decreases with higher density. The CH$^+$ to OH$^+$ ratio is an excellent tracer of FUV at a gas temperature $T < 300$ K (top row of Fig. \ref{D6}), as indicated by the systematic increase in CH$^+$/OH$^+$ up to $G_0 = 10^3$ ISRF. The evaporation of H$_2$O at dust temperatures above 100 K reduces the abundance of CH$^+$ more than OH$^+$ and causes a step in CH$^+$/OH$^+$ at that temperature \citep{2005A&A...440..949S}.

The ratios of molecules calculated from observed column densities (Tables B.1 - B.3) of the first class are listed in Table 4. Some observed values taken from Table 4 are indicated in Fig. \ref{D6} with horizontal lines. They are in the range from 0.2 to 5.0 for lines observed in absorption. The only ratio available from emission lines (toward W3 IRS5) yields a value well above that range. In contrast to the FUV case, the CH$^+$/OH$^+$ ratio is less sensitive to the value of the X-ray flux (bottom row of Fig. \ref{D6}).

The CH$^+$/OH$^+$ ratio may be underestimated in our simplified modeling because of two effects. Vibrationally excited H$_2$ reacts exothermically with C$^+$, favoring CH$^+$ formation \citep{2010ApJ...713..662A} and enhancing the CH$^+$ column density by a factor of two \citep{2013ApJ...766...80Z}. Another factor of two may result from formation pumping, which is more important for CH$^+$ \citep[][Table 9]{2010ApJ...720.1432B} than for OH$^+$ \citep{2014ApJ...794...33G}. However, temperature and irradiation affect the CH$^+$/OH$^+$ ratio
more, thus justifying the qualitative conclusions drawn above.

Figure \ref{D6} suggests that there are two temperature ranges that can match the observed CH$^+$ to OH$^+$ ratios. At temperatures above about 500 K, where O+H$_2 \ \rightarrow$ OH $\rightarrow$ H$_2$O, no irradiation is required to achieve the observed values. The second possible range is $T<300$ K, which is consistent with the rather low excitation temperatures of even the emission lines derived from full chemical and radiation transfer modeling \citep[e.g.][]{2010A&A...521L..44B}. In case of low density  ($n({\rm H_2}) \approx 10^4$ cm$^{-3}$), the required  $G_0$ value is between 0 and a few times 10 ISRF for all objects seen in absorption. In the case of high density  ($n({\rm H_2}) \approx 10^6$ cm$^{-3}$ and higher), the FUV range is from $G_0 = 1$ to a few 10$^5$ ISRF.  For both low- and high-mass YSOs, the envelope densities at the half-power beam radius (HPBR) of {\it Herschel} at 1 THz are typically $\sim 10^5$ cm$^{-3}$.  Densities at the cavity wall can be lower, but are unlikely to be so by more than an order of magnitude. On the other hand, shocks can compress the walls and actually lead to higher densities. Thus, the entire range of $10^4-10^6$ cm$^{-3}$ is plausible. The lack of emission suggests that these hydrides are not present in the innermost part of the envelope where densities can be as high as $10^8$ cm$^{-3}$.

\begin{table}[htb]
\begin{center}
\caption{{\it Herschel} HPBR(10.7$''$ at 1 THz)  at the source distance, density of envelope model at HPBR, line mode, and derived FUV flux range at the site of the molecules inferred from the observed ratio.}
\begin{tabular}{lrccc}   
\hline \hline
Object & radius&density&line&$G_0$ \\
& [AU]&[cm$^{-3}]$&mode&ISRF\\
\hline\\
NGC1333 I4A&2500&1.3$\times 10^6$& abs.&200 - 400\\
Ser SMM1&4400& 6.0$\times 10^5$& abs. & 2 - 8\\
\\
AFGL 2591&35000&7.0$\times 10^4$& abs. &20 - 80\\
W3 IRS5 &21000&1.1$\times 10^5$ & abs. & 80 - 200\\
W3 IRS5  &21000& 1.1$\times 10^5$& em. & 300 - 600\\
\hline
\end{tabular}
\end{center}
\label{G0}
\end{table}

In the following we discuss five well-observed cases. The inferred $G_0$ ranges for these sources, in the temperature range of 10 -- 100 K, are given in Table 5, assuming the density at the HPBR. The uncertainty of the density is assumed to be smaller than a factor of 2 and is included in the $G_0$  range.

1. In the low-mass object {\it NGC1333 I4A} a protostellar X-ray luminosity $L_x = 10^{32}$ erg s$^{-1}$, only geometrically attenuated, would yield a flux of only $5.3\times10^{-3}$ erg s$^{-1}$ cm$^{-2}$ at the HPBR. According to the toy models (Fig.  \ref{D6}), this is not sufficient to reproduce the observed CH$^+$/OH$^+$ ratio at the HPBR density according to the envelope model \citep{2012A&A...542A...8K}. The interpretation by FUV irradiation, on the other hand, needs an FUV flux $G_0\approx 300$  ISRF at  the HPBR density and at a gas temperature of 30 K.

2. Similarly, to reproduce the observed CH$^+$/OH$^+$ ratio toward  {\it Ser SMM1}, X-rays emitted by the protostar could provide a flux of  $< 10^{-2}$ erg s$^{-1}$ cm$^{-2}$ at the HPBR and are unlikely according to Fig.\ref{D6}. In contrast, an FUV flux of only a few ISRF is required due to the low ratio observed.

3. Based on interferometric observations of sulfur-containing molecules, \citet{2007A&A...475..549B} have proposed that the high-mass object {\it AFGL 2591} is an X-ray emitter. Corrected for the updated distance (Table \ref{table_objects}), the estimated X-ray luminosity amounts to $L_x = 9\times 10^{32}$ erg s$^{-1}$. At {\it Herschel's} HPBR, the X-ray luminosity, geometrically attenuated, would yield a flux of $2.6\times10^{-4}$ erg s$^{-1}$ cm$^{-2}$, which  is not sufficient to reproduce the observed ratio (density model of \citet{2013A&A...554A..83V}). On the other hand, \citet{1999ApJ...522..991V}  suggest the central object to be a B star with an effective temperature of $3\times 10^4$ K having a luminosity of $2.2\times 10^{5}$ L$_\odot$ (corrected for new distance), emitted mostly in FUV. If only geometrically attenuated, it yields $G_0 = 1.5\times 10^5$ ISRF at the {\it Herschel} HPBR, three orders of magnitude higher than required. Thus an attenuated FUV flux can readily reproduce the observed ratio.

4. The conclusion that FUV dominates over X-rays is similar for the high-mass object {\it W3 IRS5}, for which the reported X-ray luminosity is $5\times 10^{30}$ erg s$^{-1}$  \citep{2002ApJ...579L..95H}. At {\it Herschel's} HPBR, this X-ray flux is geometrically attenuated to a level of  $4\times10^{-6}$ erg s$^{-1}$ cm$^{-2}$, which is not sufficient to produce the observed CH$^+$/OH$^+$  ratio.  The FUV flux required to match observations at the HPBR density and 30 K amounts to $G_0 \approx 90$ ISRF.

5. For {\it W3 IRS5 in emission}, the excitation temperature of CH$^+$ was determined from 1D slab modeling, which indicates a value of 38 K (Table B.1). It is represented in all plots of Fig. \ref{D6} by a blue dot.

\subsection{OH$^+$ to H$_2$O$^+$ ratio}

Figure \ref{D7} shows the modeling results for the OH$^+$/H$_2$O$^+$ ratio. It generally decreases with temperature. Contrary to the CH$^+$/OH$^+$ ratio, the observations clearly exclude high temperatures. Thus combining the two observed ratios, the hot temperature range is ruled out.

In the absence of irradiation, the model ratio of OH$^+$ to H$_2$O$^+$ (Fig. \ref{D7}) is around 0.4 for temperatures $<$100 K. In the parameter range presented in Fig. \ref{D7}, the OH$^+$/H$_2$O$^+$ ratio is near unity or below for all irradiation fluxes except very high $G_0$ at low density. The value of $\ge$15.5 observed for the low-mass object Ser SMM1 can only be reproduced at a density of $10^4$ cm$^{-3}$ and an FUV flux $G_0$ around 10$^6$, or an X-ray flux of 10 erg s$^{-1}$ cm$^{-2}$ (not shown in Fig. \ref{D7}). Both fluxes exceed the expected upper limits for low-mass objects at the {\it Herschel} half-power beam radius.

The requirement for the high-mass object W3 IRS5 OH$^+$/H$_2$O$^+$ ratio can only be met with $G_0 > 10^6$ at a density of $8\times 10^4$ cm$^{-3}$. The required X-ray flux of 10 erg s$^{-1}$ cm$^{-2}$ is far above the observed value from \citet{2002ApJ...579L..95H} if the emitting X-ray source is near the center of the YSO. Thus, we cannot explain the observed ratios with irradiation of the central object(s) and the density expected from envelope models.

As OH$^+$ is mainly destroyed through reactions with H$_2$ to form H$_2$O$^+$, a high OH$^+$ to H$_2$O$^+$  abundance ratio points at a low density. Indeed, in diffuse clouds  OH$^+$/H$_2$O$^+$  ratios of 1 - 15 have been observed, suggesting densities $<100$ cm$^{-3}$ with a molecular fraction of only a few percent  \citep{2010A&A...521L..16G,2015ApJ...800...40I}.  Such low densities are unlikely to be present in protostellar systems on the scales covered by the Herschel beam.  We therefore  consider also the alternative scenario of non-equilibrium chemistry in irradiated dissociative shocks which may enhance both OH and OH$^+$ in star-forming regions  \citep{1989ApJ...340..869N,2013A&A...557A..23K}  (see Sect. 6).

\subsection{C$^+$ to CH$^+$ ratio}

The modeled C$^+$/CH$^+$ ratio (Fig. \ref{D8}) is low at high temperature due to the rapid formation of CH$^+$ above 300 K (see above) and strongly increases from 2000 K to 300 K. For temperatures below 300 K, the ratio traces well the irradiating FUV and X-ray fluxes. X-rays reduce the ratio. For low irradiation, the ratio strongly increases with increasing density.

The observed ratio of C$^+$ to CH$^+$ is generally lower for low-mass objects (Table 4). For Ser SMM1 and NGC1333 I2A, it is below the values predicted by the models except for temperatures around 900 and 500 K, respectively (Fig. \ref{D8}). Several interpretations are possible: {\it (i)} A gas temperature higher than 500 K may be expected from FUV irradiation of gas along the outflow walls \citep[e.g.][]{2012A&A...537A..55V}. {\it (ii)} The two molecular regions may not be co-spatial. If the C$^+$ absorbing or emitting region is more extended than the CH$^+$ absorbing region, but smaller than the beam and the continuum source, the true C$^+$/CH$^+$ ratio of the column densities in the co-spatial regions would be larger. {\it (iii)}  The observed C$^+$ line is self-absorbed in high-mass objects, thus column densities are underestimated (see Appendix A). {\it (iv)} The H$_2$ density may be lower than 10$^4$ cm$^{-3}$ in the C$^+$ emitting region and be comparable to intercloud conditions. {\it (v)} Alternatively, chemistry in a shock may again play a role.

In conclusion, CH$^+$/OH$^+$  are the most reliable ratios and their comparison with models suggests that substantial protostellar FUV fluxes are needed at the site of the molecules to fit the observations. This statement is based on the assumption that the molecular hydrogen density is approximately given by the source models at the {\it Herschel} HPBR.  X-ray emission cannot play a role unless it originates closer to the line absorption/emission region than the protostar.

\subsection{The case of X-ray irradiation}
X-rays are effective destroyers of water in the inner envelope \citep{2006A&A...453..555S}, so it is important to put limits on their flux in cases where no direct X-ray observations are possible.  In general, we find no evidence for strong X-ray emission based on the observations of the hydrides presented here. This could be viewed to be in contradiction with \citet{2007A&A...466..977S}, who find three indicators for X- rays: CO$^+$, CN, and SO$^+$. The abundances of the first two molecules can also be explained by FUV irradiation of outflow cavity walls \citep{2010ApJ...720.1432B}, whereas the sulfur chemistry is in general poorly understood and may not be co-spatial.

The non-detection of H$_3$O$^+$ toward the low-mass objects imposes limits on their X-ray luminosity. According to \citet{2006PhDT..........1S} (Table 6.5), who derived the H$_3$O$^+$ line intensities for the prototypical Class 0 object IRAS16293-2422, our observed upper limits ($<0.14$ to $<0.19$ K km s$^{-1}$) suggest that the X-ray luminosities are $<10^{30}$ erg s$^{-1}$.  For the prototypical Class I object TMC1 however, an X-ray luminosity of $10^{31}$ erg s$^{-1}$ is predicted to produce a line intensity  below the sensitivity of these H$_3$O$^+$ observations \citep[][Table 6.6]{2006PhDT..........1S}. Thus the upper limit for TMC1 is more than $10^{31}$ erg s$^{-1}$ and does not constrain the X-ray luminosity further.

\section{Origin of ionized hydrides}
The first group of molecules, CH$^+$, OH$^+$, H$_2$O$^+$, and C$^+$, is observed preferentially in blue-shifted absorption. 1D slab modeling indicates that the excitation temperature of the ions is low ($\le$40 K, Section 4), below the radiation temperature of the background. However, a gas temperature of several hundred K would be necessary for efficient formation of CH$^+$ through C$^+$ + H$_2$. Thus the lines must be subthermally excited and the molecules are mostly in the ground state. Since the ions are destroyed by collisions with molecular hydrogen, a low density and high irradiation are required for enhanced abundances. This is consistent with the abundance ratios reported in Section 5. The collisional destruction may also explain why the ion hydride lines are rarely in emission and seem to originate from further out, where they absorb the continuum.

Several line characteristics indicate that the observed molecules are located not in diffuse clouds in the foreground but are related to star-forming regions:  {\it (i)}  The column densities correlate with source luminosity and envelope mass (Figs. \ref{CHp} - \ref{CH}). {\it (ii)} The line velocities correlate with the source velocity, yielding consistently small line shifts (Figs. \ref{CHp} - \ref{CH}). {\it (iii)} All lines (except H$_2$O$^+$) are occasionally detected in emission, which requires high density in cases like CH$^+$ and SH$^+$.

\citet{2014A&A...563A.127M} reported blue-shifted components of several HCO$^+$,  CO and N$_2$H$^+$ emission lines toward NGC6334 I. They interpret them by a shell of the envelope irradiated by external FUV, having a density of $10^5$ cm$^{-3}$, and expanding at a velocity of $-1.5$ km s$^{-1}$. We consider such an interpretation for the first group of molecules unlikely, which have line shifts in the range of $-10$ to 10 km s$^{-1}$, exceeding the values expected for the outer edge of the envelope. Such a scenario may apply to CH, but not to the molecular  ions which are clearly different in line shape, width and velocity from CH and to the other molecules in the second group.

Blue-shifted emission and absorption of H$_2$O has been discussed by \citet{2013A&A...557A..23K} as connected with the outflow directed toward Earth, where the continuum emitted by a disk or the dense inner envelope is absorbed. In the low-mass objects studied, this ``offset component" is shifted by $-$12.7 to 0.9 km s$^{-1}$ relative to the YSO velocity and has a line width between 4 $-$ 40 km s$^{-1}$. Several of the ionized molecules have similar broad and shifted profiles as those of H$_2$O.

\begin{figure}[]
\centering
\resizebox{10cm}{!}{\includegraphics{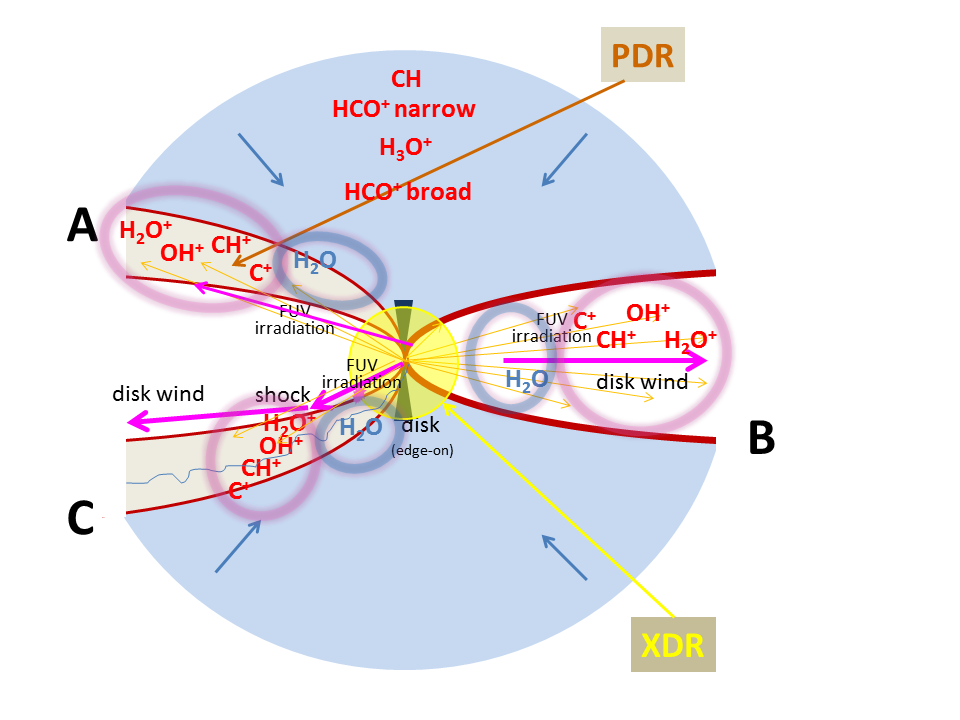}}
\caption{Cartoons of three scenarios described in the text for the origin of the observed CH$^+$, OH$^+$, H$_2$O$^+$, and C$^+$ absorptions.  {\it A:} Irradiated outflow walls with slow shocks entraining the outer layers of the walls;  {\it B:} disk wind irradiated by protostellar FUV; {\it C:} fast dissociative shocks irradiated by protostellar FUV.}
\label{cartoons}
\end{figure}

The observed abundance ratios suggest either an H$_2$ density lower than $10^4$ cm$^{-3}$ or an enhanced FUV flux by an internal source. Such low H$_2$ densities were proposed for diffuse clouds (Section 5) but not for star-forming regions (see Fig. \ref{fig:hcop_correl1}). For the well-studied high-mass object AFGL 2591 a detailed physical-chemical model has been developed including non-LTE line radiative transfer by Bruderer et al. (2010a), which indicates that the enhanced FUV from the central B star is key in explaining the abundances of species like CO$^+$ and CH$^+$. As a result of the high critical densities of diatomic hydrides, they found CH$^+$ to be subthermally excited using densities of the order of 10$^6$ cm$^{-3}$ in the molecular region. Depending on the continuum radiation field, the lines can thus be in absorption even though they form in dense regions. The low density option cannot be excluded but the line profiles favor the enhanced FUV flux interpretation. This FUV radiation may be emitted either from the surface of the protostar or from fast shocks in the jets. In the following we discuss three origins for these ions: {\it (A)} the irradiated cavity walls, {\it (B)} the disk wind, or {\it (C)} the slow shock of the wind impacting the cavity wall. Cartoons are presented in Fig. \ref{cartoons} for illustration.

\subsection{Scenario A: Irradiated cavity walls}
The walls of the outflow cavities may be illuminated by freely propagating or only partially absorbed FUV radiation. In concave cavities, a protostellar FUV flux irradiates the walls, heats and ionizes the gas at the large surface to the envelope \citep{1995ApJ...455L.167S,2007A&A...466..977S,2009ApJS..183..179B}. Detailed modeling yields good agreement with observed column densities in two high-mass objects \citep{2010ApJ...720.1432B,2013JPCA..117.9840B}. The scenario is possible for lines that are broadened, but not shifted. The observed blue-shift is however not compatible with stationary outflow walls. Another element is needed: the ionized molecules must be coupled to the outflow and be entrained.

\subsection{Scenario B: Disk wind}
 Another possible location of the CH$^+$, OH$^+$, H$_2$O$^+$, and C$^+$ ions is protostellar wind. \citet{2012A&A...538A...2P} have studied the survival of molecules in disk winds of low-mass objects. In MHD driven winds irradiated by FUV from accretion shocks, the wind density at 100 AU in Class 0 objects is $10^6$ cm$^{-3}$, about one order of magnitude below the envelope density \citep[][]{2012A&A...542A...8K}. The temperature at this distance is around 1000 K, and the velocity reaches 40 km s$^{-1}$. None of our observed lines show such high velocities. According to the \citet{2012A&A...538A...2P} models, the wind densities are high enough so that dust absorption shields the molecules sufficiently from being photo-dissociated. H$_2$O forms via the endothermic neutral-neutral reactions. At slightly larger distances (10 - 1000 AU) ionized molecules form. Figure 8 in \citet{2012A&A...538A...2P} suggests a column density of about $5\times 10^{14}$ cm$^{-2}$ for CH$^+$ in Class I objects, which exceeds the observed upper limit for L 1489 by more than an order of magnitude (Table B.1). A more serious discrepancy is the copious amount of SH$^+$ predicted by the wind model. A column density of $2\times 10^{12}$ cm$^{-2}$ was predicted for Class 0 and $10^{14}$ cm$^{-2}$ for Class I, exceeding some of the observed upper limits by more than two orders of magnitude (Table B.2). \citet{2012A&A...538A...2P} do not make predictions for the other species observed here and not for high-mass objects.

\subsection{Scenario C: Shocked cavity wall}
The disk wind interacts with the cavity wall; such interaction is most likely a shock. Having a velocity component perpendicular to the wall of about 10 km s$^{-1}$, the shock produces gas temperatures of a few thousand K, which is not high enough for FUV emission. \citet{2013A&A...557A..23K} and \citet{2014A&A...572A..21M} considered this scenario as the origin of the blue-shifted component of H$_2$O toward low-mass YSOs. The shock front and internal layers are uneven, turbulent, and driven by the outward motion of the wind. Thus emission and absorption lines are medium-broad and blue shifted, the closer they originate from the cavity wall, the broader and more shifted the lines.

\citet{2013A&A...557A..23K} note that the chemical abundances predicted by the fast shock model of \citet{1989ApJ...340..869N} for CO, H$_2$O, and OH agree with observed values within an order of magnitude. However, the abundances of CH$^+$, OH$^+$, and HCO$^+$ are under-predicted by more than an order of magnitude. The low velocities of the blue-shifted components ($\ll$ 100 km/s) also suggest that the UV irradiation is not created within the shock itself. Molecular ions like CH$^+$ and OH$^+$ thus are proposed to form already in the pre-shock material by FUV irradiation from some source external to the shock.

The FUV radiation with luminosity $L^{FUV}$ may originate from the accretion shock onto the protostar covering a fraction of its surface or (for high-mass objects) from the hot surface of the protostar. At a distance $r$ and without absorption, the FUV flux $G_0$ is given by
\begin{equation}
G_0\ =\ {{L^{FUV}L_\odot}\over 4 \pi r^2 G_0^{ISRF}}\ =\ 850 {L^{FUV}\over r_{1000}^2}\ \ {\rm [ISRF],}
\label{equ_G}
\end{equation}
where $L^{FUV}$ is the FUV luminosity in units of L$_\odot$, $r_{1000}$ is the distance of the radiation source in units of 1000 AU, and $G_0^{ISRF} = 1.6\times 10^{-3}$ erg cm$^{-2}$ s$^{-1}$. If the source of this radiation is located at the NGC1333 I4A protostar for example, the luminosity required by the observed CH$^+$/OH$^+$ ratio (given in Section 6.2) would be $L_{FUV}=1.5 L_\odot$. This is compatible with the reported bolometric luminosity of $L_{\rm bol} = 9.1 L_\odot$. A hot spot of $3\times 10^4$ K would have to have a diameter of 0.13 R$_\odot$ to emit this luminosity. Similarly, one derives $L_{FUV}=0.023 L_\odot$ for Ser SMM1. \citet{2012A&A...537A..55V} independently report $L_{\rm FUV} < 0.1 L_\odot$ for their three low-mass sources based on highly excited CO and H$_2$O lines measured with {\it Herschel}/PACS. Such a source of FUV originating from accretion at the star-disk interface is not implausible, and thus we favor Scenario C.

\section{Conclusions}
This paper presents an exploratory high-resolution survey of hydrides in star-forming regions in far infrared and submillimeter wavelengths. Objects range from low-mass to high-mass. The observed lines reported here include CH$^+$, OH$^+$, H$_2$O$^+$, H$_3$O$^+$, SH$^+$, HCO$^+$(6-5), C$^+$, and CH. The selected lines bring the effects of ionization by internal sources into focus. The observed line properties suggest molecules in protostellar envelopes, entrainments of outflows, and diffuse interstellar foreground clouds. Only the components likely to be associated with the YSO are studied here.

\begin{enumerate}
\item The detected lines can be grouped in two sets (see summary in Fig. \ref{overview}): the first group includes CH$^+$, OH$^+$, and possibly H$_2$O$^+$ and C$^+$. These ion lines are predominantly in absorption, have line widths $5-15$ km s$^{-1}$, and are blue-shifted relative to the systemic velocity. All these species (except H$_2$O$^+$) have also an emission component in one or more sources.

\item Our preferred scenario for the first group of molecules is a slow shock heating and entraining envelope material combined with protostellar FUV irradiation (Scenario C) predicted for high-mass stars. It is compatible with our observations and previous modeling of cavity walls. An alternative interpretation where the ions exist in hypothetical low H$_2$ density regions ($< 10^4$ cm$^ {-3}$) cannot be excluded but is less likely based on the line profiles.

\item A second group including CH and HCO$^+$(6-5) is observed predominantly in emission. The line widths are $\le$5 km s$^{-1}$, and the line shifts w.r.t. the systemic velocities are small. H$_3$O$^+$ and SH$^+$ may also be part of this group, but have larger average line widths (Table \ref{table_line_width}). This group of lines is suggested to originate in the bulk of the envelope.

\item The correlations of line characteristics confirm the grouping. CH$^+$, OH$^+$ (and to a lesser extent C$^+$) correlate in their velocity shifts, and tentatively in line widths and column densities (Fig. \ref{corr1}). The relationships are further illustrated in the superposition of spectral profiles, showing similar emission and absorption features (Fig. \ref{super}). H$_3$O$^+$ and CH correlate with HCO$^+$ in column density and scatter around zero shift in velocity relative to the systemic velocity (Fig. \ref{corr2}).

\item The beam-averaged column density of all molecules correlates with bolometric luminosity (Figs. C.4 - C.9) and thus increases from low-mass to high-mass objects. The best correlations are found for CH, CH$^+$, and HCO$^+$. As the envelope mass correlates with luminosity \citep{1996A&A...311..858B}, it is not surprising that the column density of all molecules also correlates  with envelope mass (except H$_3$O$^+$).

\item The correlation between the distance-corrected line intensity and bolometric luminosity reaches a correlation coefficient of 0.95 for HCO$^+$(6-5) (Fig. \ref{HCOp_norm}). Appendix D offers an interpretation based on optically thick emission and the self-similarity of the dust radiative transfer.

\item Comparing observational and modeled abundance ratios provides a rough indication of FUV and X-ray irradiation. For absorption lines, the observed CH$^+$/OH$^+$ ratios can be reproduced in simple chemical models for reasonable FUV fluxes. For the emission lines of W3 IRS5, the required FUV flux is $G_0$=300--600 ISRF at a density of $1.1\times 10^5$ cm$^{-3}$. This value is far below that expected from unabsorbed FUV emission by an unbloated central high-mass object.

\item Our {\it Herschel} data yield no evidence for X-ray irradiation from any of the studied line ratios, even for sources where X-rays have been observed. Also in low-mass objects, the spatial resolution at scales of a few 1000 AU is not sufficient to detect molecular traces of X-rays. The effect of X-rays (if any) is limited to the innermost part of the envelope and disk, requiring higher spatial resolution. Upper limits of the X-ray luminosity can be derived from H$_3$O$^+$ observations for some objects.
\end{enumerate}

{\it Herschel}/HIFI observations enabled a big step forward in our knowledge on hydrides. Some of them suggest strong irradiation of low molecular density regions, thus regions where H$_2$O is destroyed by photodissociation (see Introduction). It is consistent with low H$_2$O abundances in outflows as inferred by \citet{2012A&A...537A..55V}, \citet{2013A&A...557A..23K},  \citet{2014A&A...572A..21M}, and \citet{2014A&A...572A...9K}. Surprisingly, the first group of ion molecules shows similarity with the ``offset" component of H$_2$O, indicating proximity of origin. Inhomogeneity of the inner region of star-forming regions may explain the contradiction. A major next step can be expected from imaging observations of CH$^+$ and OH$^+$ in lines that are observable from the ground. Such observations will soon be possible with ALMA and may confirm the suggested scenario. High spatial resolution HCO$^+$ observations at various $J$ transitions will help to find evidence for position dependent ionization rates in the innermost region including the disk, where X-rays and other ionization sources \citep[e.g.][]{2014ApJ...790L...1C} may be important.

\begin{acknowledgements}
We thank the WISH team for inspiring discussions and support. We acknowledge  the  referee for helpful criticism and G\"oran Sandell for the confirmation of the NGC6334 I(N) object parameters. This program is made possible thanks to the Swiss {\it Herschel} guaranteed time program. HIFI has been designed and built by a consortium of institutes and university departments from across Europe, Canada and the United States under the leadership of SRON Netherlands Institute for Space Research Groningen, The Netherlands and with major contributions from Germany, France, and the US. Consortium members are: Canada: CSA, U.Waterloo; France: CESR, LAB, LERMA, IRAM; Germany: KOSMA,MPIfR, MPS; Ireland: NUI Maynooth; Italy: ASI, IFSI-INAF, Osservatorio Astrofisico di Arcetri-INAF; Netherlands: SRON, TUD; Poland: CAMK, CBK; Spain: Observatorio Astronomico Nacional (IGN), Centro de Astrobiologia (CSIC-INT); Sweden: Chalmers University of Technology, Onsala Space Observatory, Stockholm University; Switzerland: ETH Zurich, FHNW Windisch; USA: Caltech, JPL, NHSC. The work on star formation at ETH Zurich was partially funded by the Swiss National Science Foundation (grant nr. 200020-113556). Astrochemistry in Leiden is supported by the Netherlands Research School for Astronomy (NOVA), by a Royal Netherlands Academy of Arts and Sciences (KNAW) professor prize, and by the European Union A-ERC grant 291141 CHEMPLAN. Support for this work was also provided by NASA (Herschel OT funding) through an award issued by JPL/ Caltech.
\end{acknowledgements}

\bibliographystyle{aa}
\bibliography{survey}

\Online

\begin{appendix}

\section{Analysis of individual observations}

\begin{figure*}[htb]
\centering
\sidecaption
\begin{minipage}{0.7\linewidth}
\begin{minipage}[t]{0.33\linewidth}
\centering
\includegraphics[width=\linewidth]{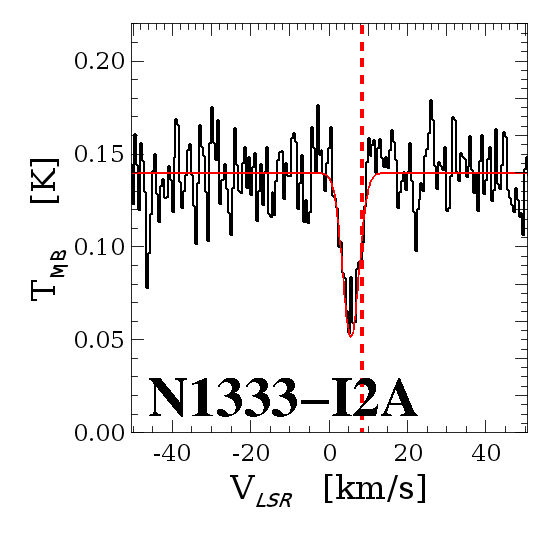}
\end{minipage}
\hspace{-0.2cm}
\centering
\begin{minipage}[t]{0.33\linewidth}
\centering
\includegraphics[width=\linewidth]{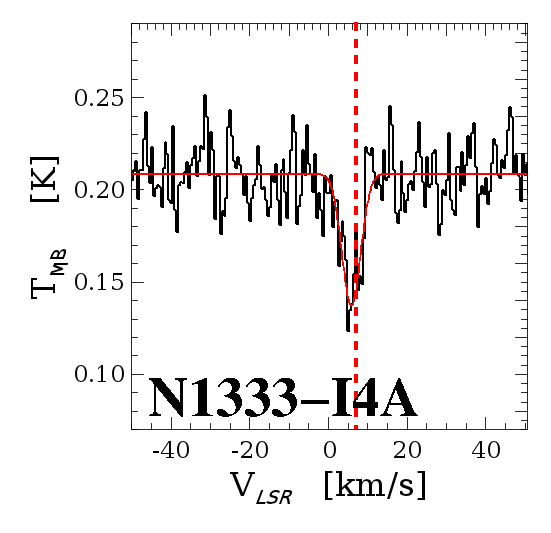}
\end{minipage}
\hspace{-0.2cm}
\begin{minipage}[t]{0.33\linewidth}
\centering
\includegraphics[width=\linewidth]{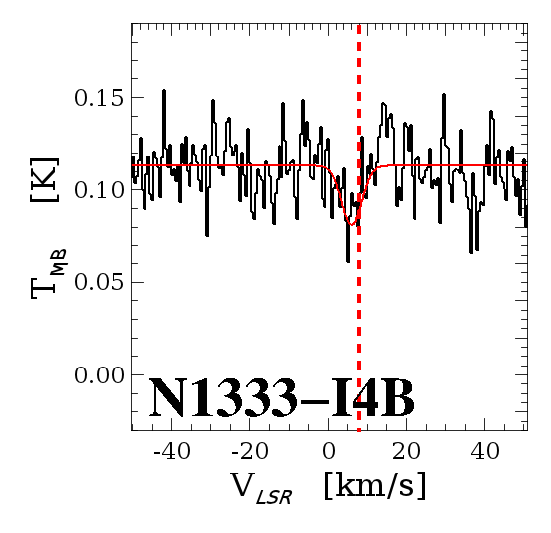}
\end{minipage}\\[-5pt]
\centering
\begin{minipage}[t]{0.33\linewidth}
\centering
\includegraphics[width=\linewidth]{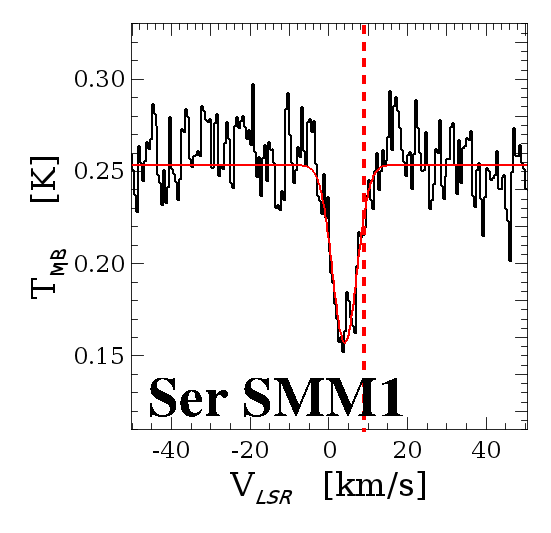}
\end{minipage}
\hspace{-0.2cm}
\begin{minipage}[t]{0.33\linewidth}
\centering
\includegraphics[width=\linewidth]{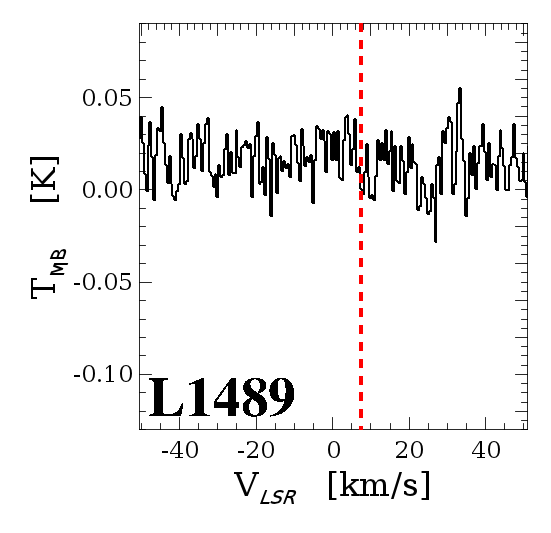}
\end{minipage}
\hspace{-0.2cm}
\begin{minipage}[t]{0.33\linewidth}
\centering
\includegraphics[width=\linewidth]{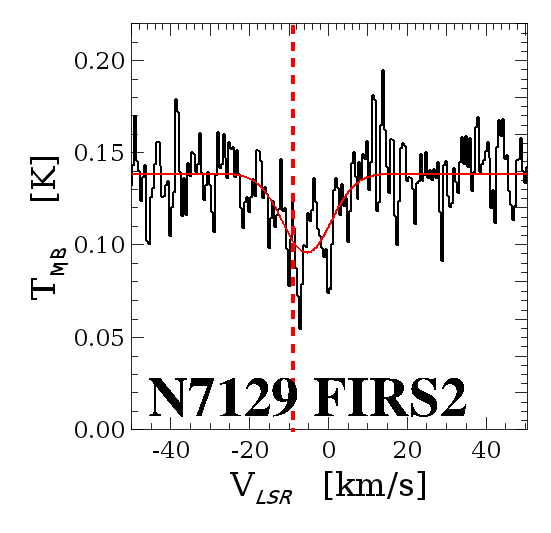}
\end{minipage}\\[-5pt]
\begin{minipage}[t]{0.33\linewidth}
\centering
\includegraphics[width=\linewidth]{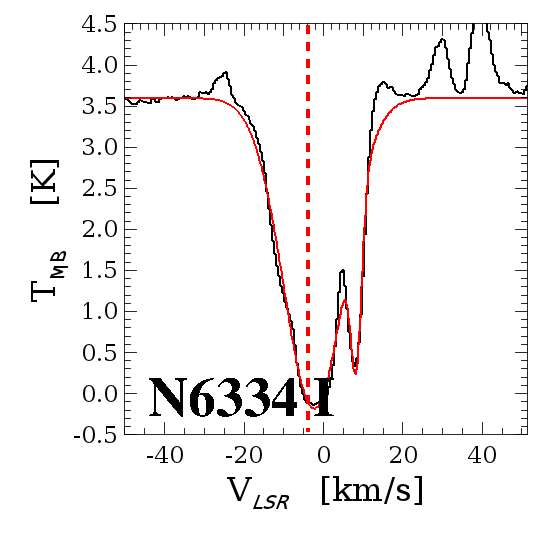}
\end{minipage}
\hspace{-0.2cm}
\centering
\begin{minipage}[t]{0.33\linewidth}
\centering
\includegraphics[width=\linewidth]{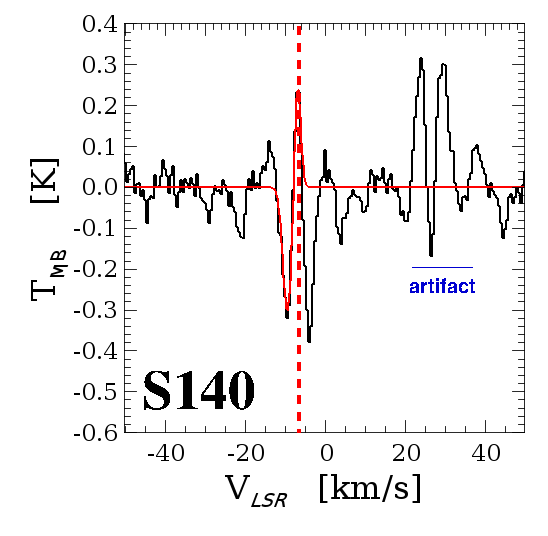}
\end{minipage}
\hspace{-0.2cm}
\begin{minipage}[t]{0.33\linewidth}
\centering
\includegraphics[width=\linewidth]{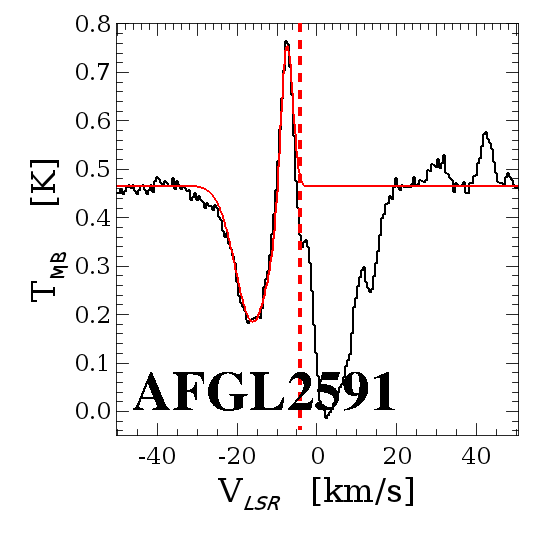}
\end{minipage}\\[-5pt]
\centering
\begin{minipage}[t]{0.33\linewidth}
\centering
\includegraphics[width=\linewidth]{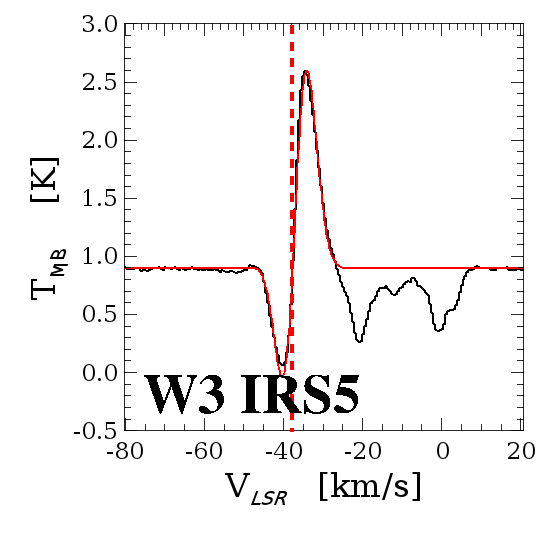}
\end{minipage}
\hspace{-0.2cm}
\begin{minipage}[t]{0.33\linewidth}
\centering
\includegraphics[width=\linewidth]{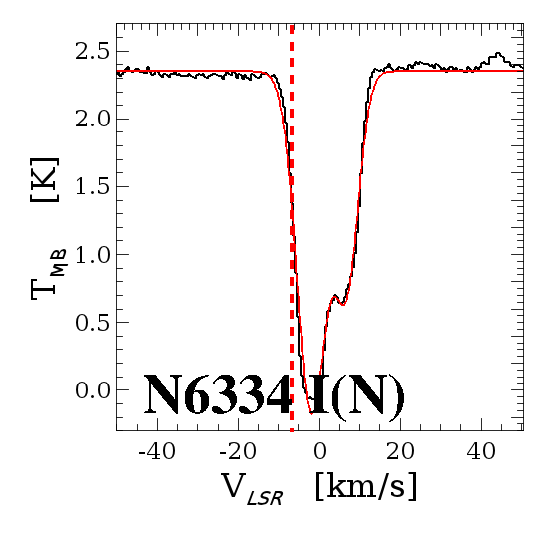}
\end{minipage}
\hspace{-0.2cm}
\begin{minipage}[t]{0.33\linewidth}
\centering
\includegraphics[width=\linewidth]{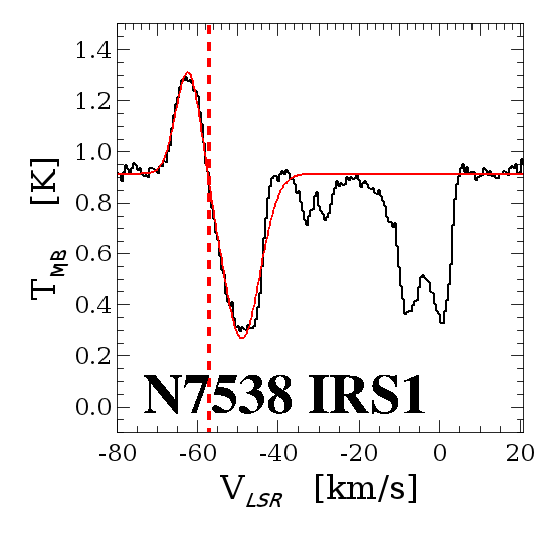}
\end{minipage}
\end{minipage}
\caption{CH$^+$(1-0) line at 835 GHz. The position of the line shifted by the systemic velocity of the YSO is indicated with a vertical red dashed line. Where the line is detected, a Gaussian in red is fitted and its parameters in width and shift are given in Table B.1. The feature at 25 km s$^{-1}$ toward S 140 is an artifact of the frequency switching mode. For the same reason the background continuum is not available for this object.}
\label{CHp_obs}
\end{figure*}

\begin{figure*}[t]
\centering
\sidecaption
\begin{minipage}{0.7\linewidth}
\begin{minipage}[t]{0.33\linewidth}
\centering
\includegraphics[width=\linewidth]{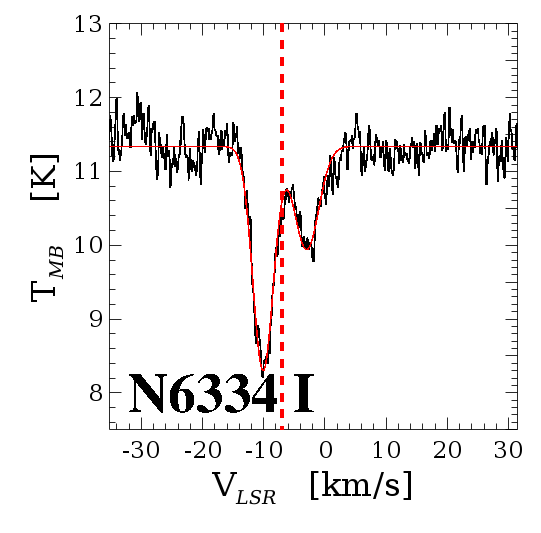}
\end{minipage}
\hspace{-0.2cm}
\centering
\begin{minipage}[t]{0.33\linewidth}
\centering
\includegraphics[width=\linewidth]{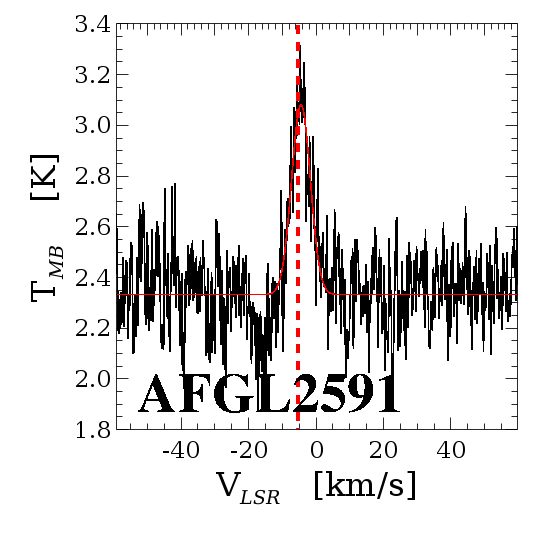}
\end{minipage}
\hspace{-0.2cm}
\begin{minipage}[t]{0.33\linewidth}
\centering
\includegraphics[width=\linewidth]{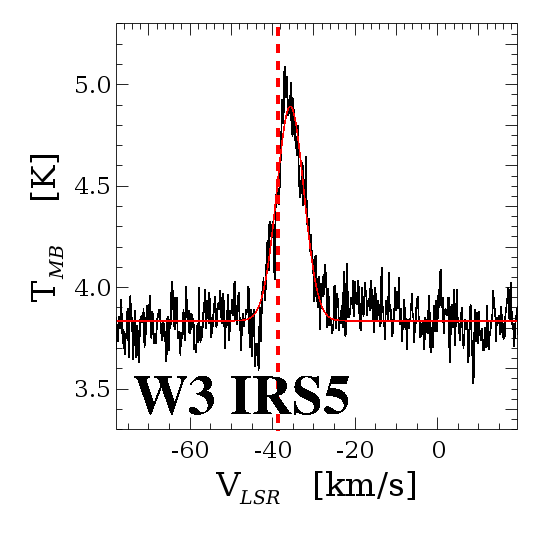}
\end{minipage}\\[-5pt]
\centering
\begin{minipage}[t]{0.33\linewidth}
\centering
\includegraphics[width=\linewidth]{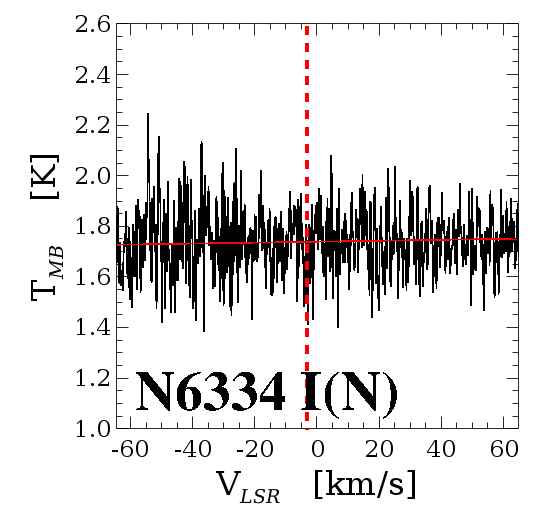}
\end{minipage}
\centering
\begin{minipage}[t]{0.33\linewidth}
\centering
\includegraphics[width=\linewidth]{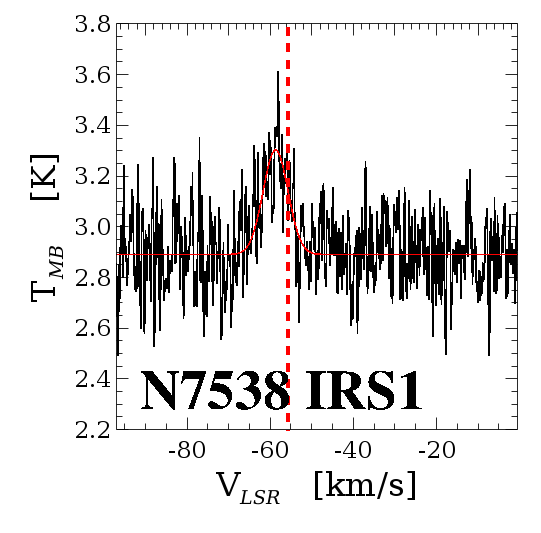}
\end{minipage}
\end{minipage}
\caption{CH$^+$(2-1) line at 1669 GHz observed toward the high-mass objects. Where detected, a Gaussian fit is shown in red and the parameters are given in Table B.1.}
\label{CHp2}
\end{figure*}

\begin{figure*}[t]
\centering
\sidecaption
\begin{minipage}{0.7\linewidth}
\begin{minipage}[t]{0.32\linewidth}
\centering
\includegraphics[width=\linewidth]{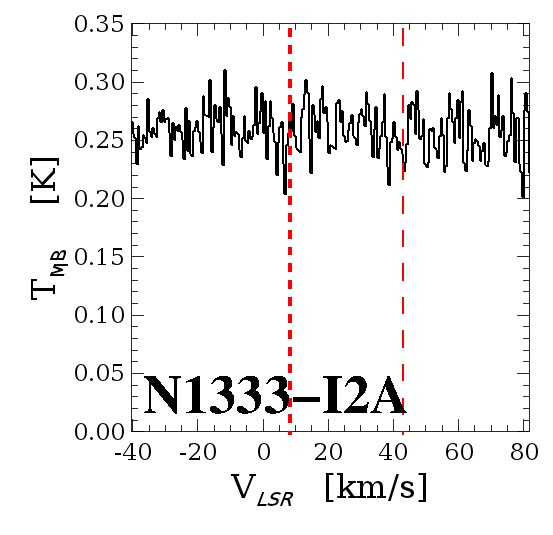}
\end{minipage}
\hspace{-0.2cm}
\centering
\begin{minipage}[t]{0.34\linewidth}
\centering
\includegraphics[width=\linewidth]{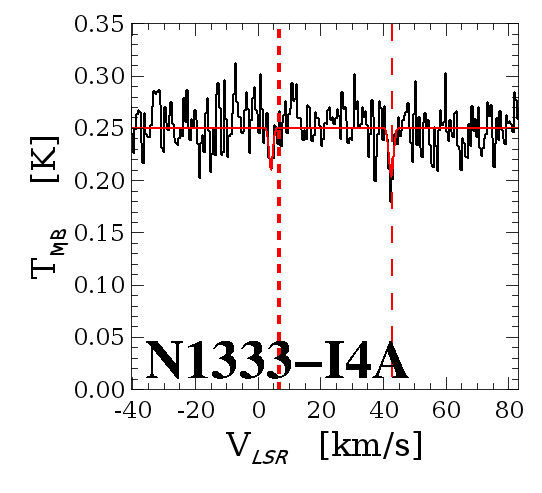}
\end{minipage}
\hspace{-0.2cm}
\begin{minipage}[t]{0.32\linewidth}
\centering
\includegraphics[width=\linewidth]{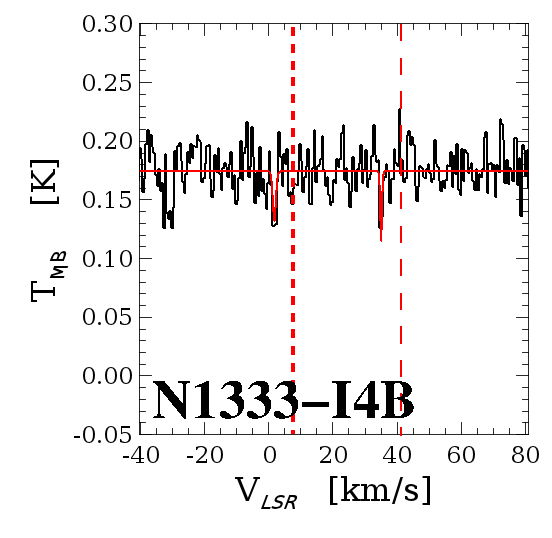}
\end{minipage}\\[-5pt]
\centering
\begin{minipage}[t]{0.33\linewidth}
\centering
\includegraphics[width=\linewidth]{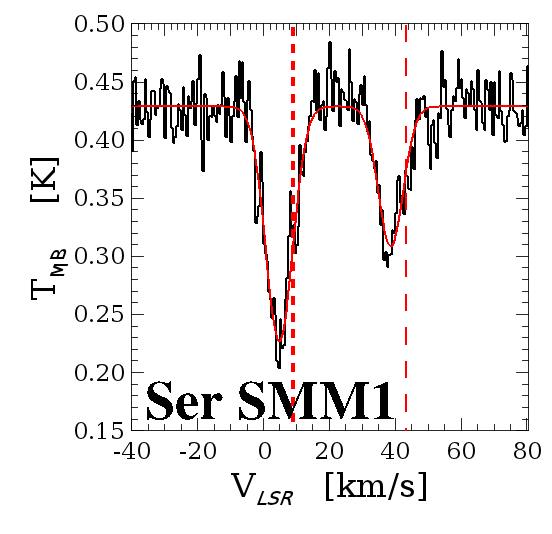}
\end{minipage}
\hspace{-0.2cm}
\begin{minipage}[t]{0.33\linewidth}
\centering
\includegraphics[width=\linewidth]{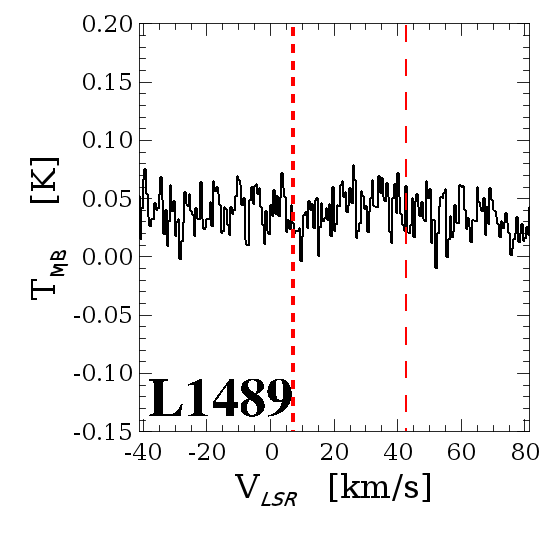}
\end{minipage}
\hspace{-0.2cm}
\begin{minipage}[t]{0.33\linewidth}
\centering
\includegraphics[width=\linewidth]{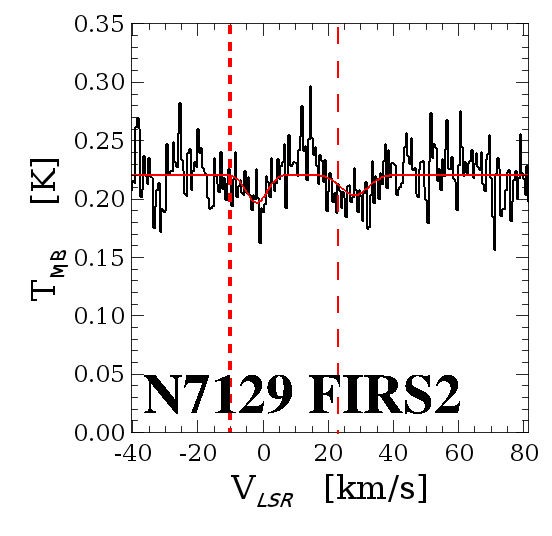}
\end{minipage}\\[-5pt]
\begin{minipage}[t]{0.33\linewidth}
\centering
\includegraphics[width=\linewidth]{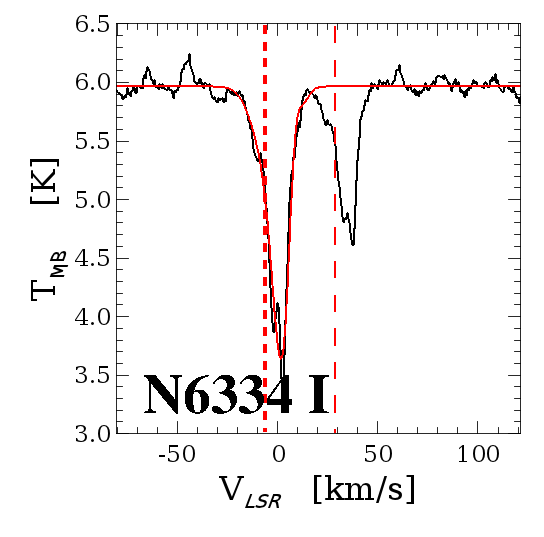}
\end{minipage}
\hspace{-0.2cm}
\centering
\begin{minipage}[t]{0.32\linewidth}
\centering
\includegraphics[width=\linewidth]{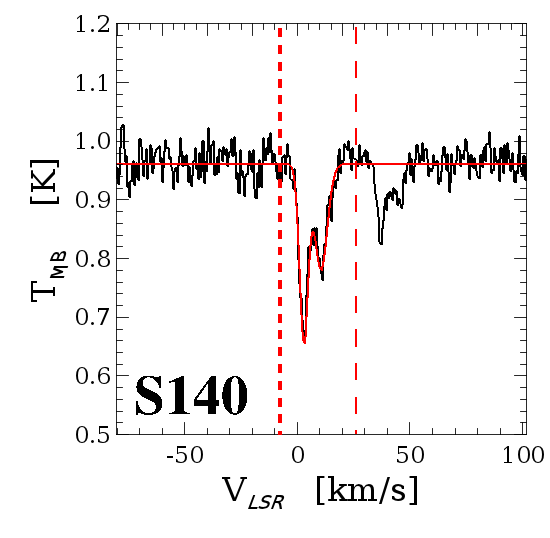}
\end{minipage}
\hspace{-0.2cm}
\begin{minipage}[t]{0.35\linewidth}
\centering
\includegraphics[width=\linewidth]{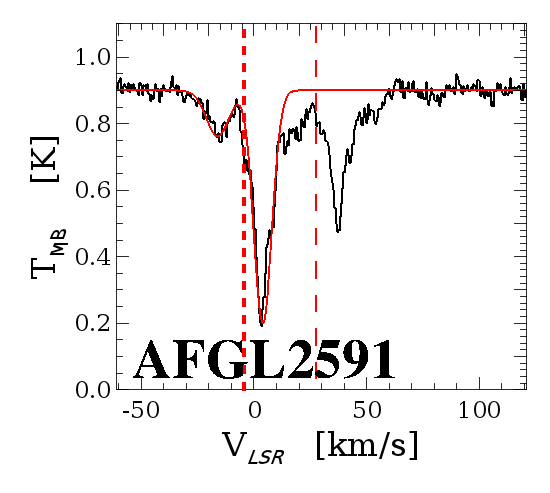}
\end{minipage}\\[-5pt]
\centering
\begin{minipage}[t]{0.32\linewidth}
\centering
\includegraphics[width=\linewidth]{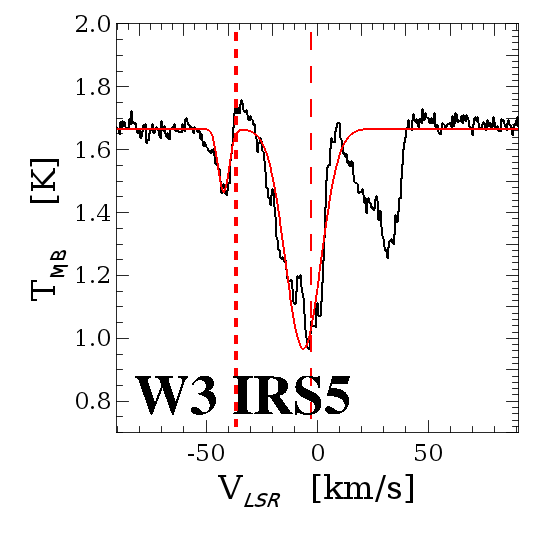}
\end{minipage}
\hspace{-0.2cm}
\begin{minipage}[t]{0.33\linewidth}
\centering
\includegraphics[width=\linewidth]{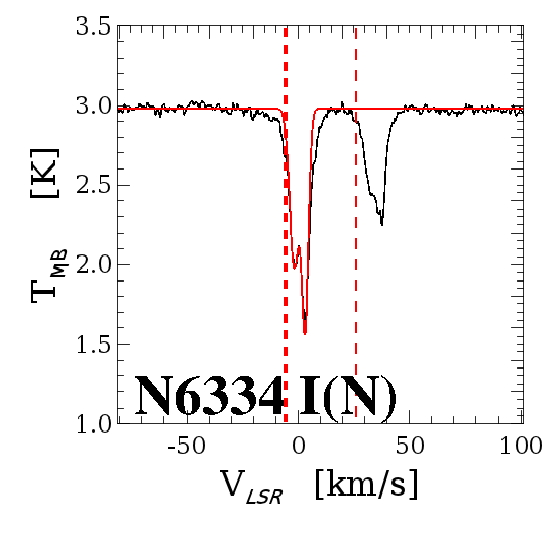}
\end{minipage}
\hspace{-0.2cm}
\begin{minipage}[t]{0.33\linewidth}
\centering
\includegraphics[width=\linewidth]{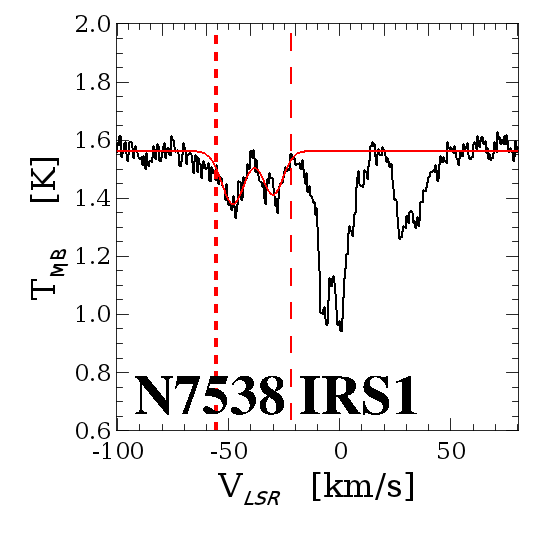}
\end{minipage}
\end{minipage}
\caption{OH$^+$($N_J=1_1-0_1$) lines at 1033 GHz. The positions of the two strongest fine structure lines, ($N_{JF}=1_{1\frac{1}{2}}-0_{1\frac{1}{2}}$) and ($1_{1\frac{3}{2}}-0_{1\frac{3}{2}}$), are indicated, shifted by the systemic velocity of the YSOs. The thickness of the two lines indicates the relative theoretical intensities. Where the lines are detected, two Gaussians in red are fitted.}
\label{OHp_obs}
\end{figure*}

\begin{figure*}[t]
\sidecaption
\begin{minipage}{0.7\linewidth}
\centering
\begin{minipage}[t]{0.33\linewidth}
\centering
\includegraphics[width=\linewidth]{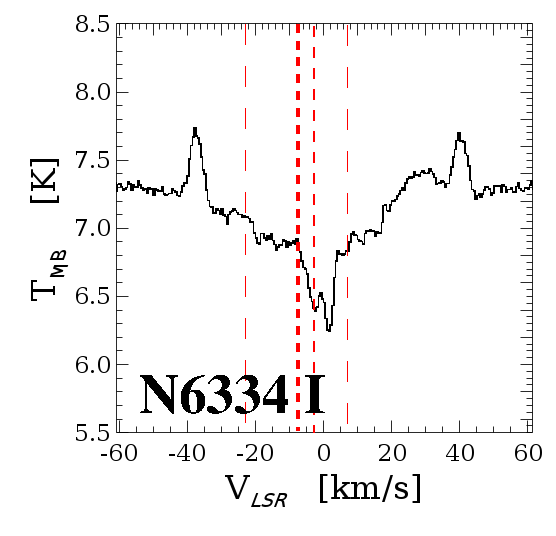}
\end{minipage}
\hspace{-0.2cm}
\centering
\begin{minipage}[t]{0.33\linewidth}
\centering
\includegraphics[width=\linewidth]{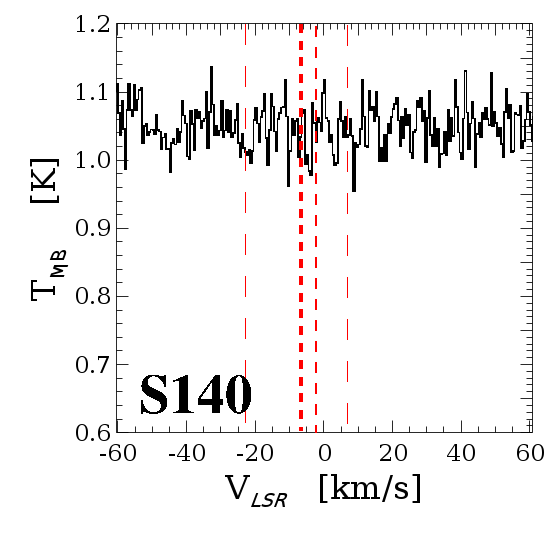}
\end{minipage}
\hspace{-0.2cm}
\begin{minipage}[t]{0.33\linewidth}
\centering
\includegraphics[width=\linewidth]{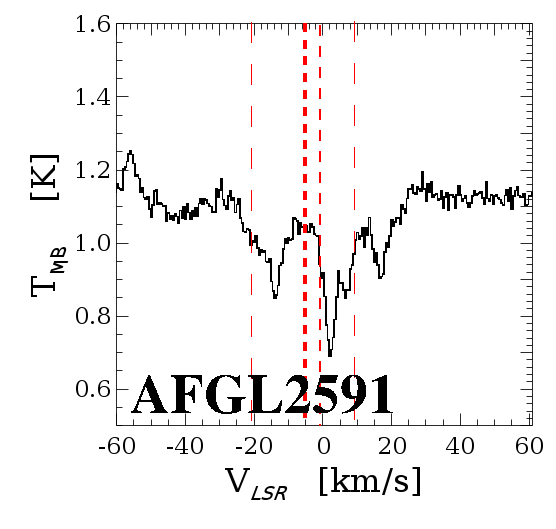}
\end{minipage}\\[-5pt]
\centering
\begin{minipage}[t]{0.33\linewidth}
\centering
\includegraphics[width=\linewidth]{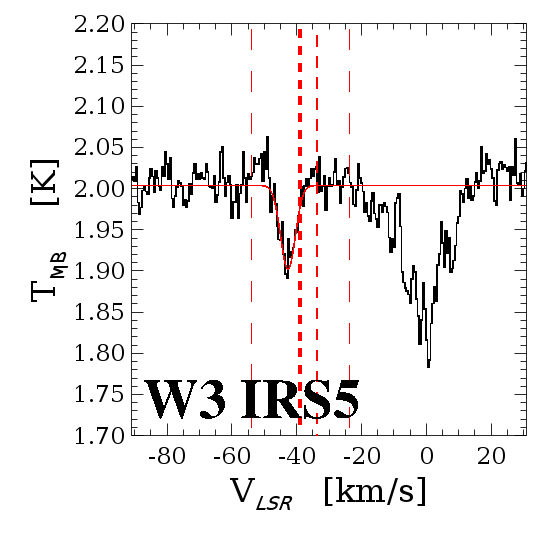}
\end{minipage}
\hspace{-0.2cm}
\begin{minipage}[t]{0.33\linewidth}
\centering
\includegraphics[width=\linewidth]{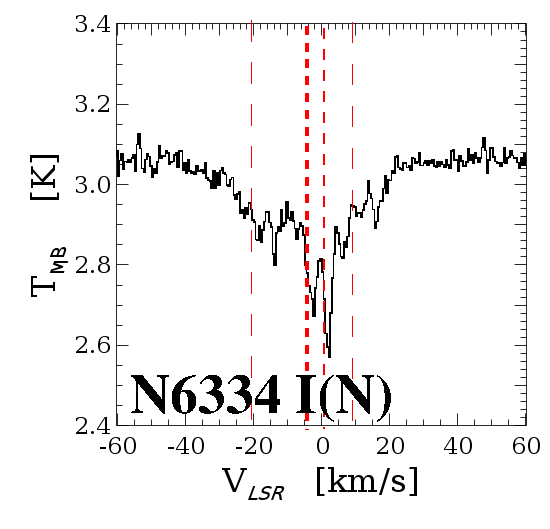}
\end{minipage}
\hspace{-0.2cm}
\begin{minipage}[t]{0.33\linewidth}
\centering
\includegraphics[width=\linewidth]{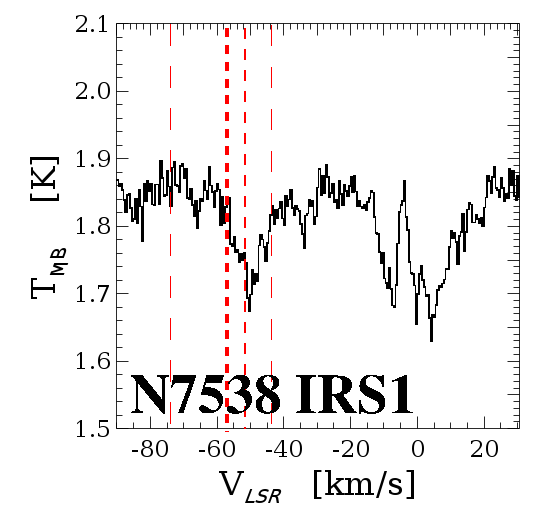}
\end{minipage}
\end{minipage}
\caption{Ortho-H$_2$O$^+$ ($N_{K{_a}K{_b}J}=1_{11\frac{3}{2}}-0_{00\frac{1}{2}}$) lines at 1115 GHz observed toward the high-mass objects. The position of the strongest lines shifted by the systemic velocity of the YSO is indicated with vertical red dashed lines. Their thickness indicates the theoretical intensities. Only the blue-shifted line of W3 IRS5 is fitted by a Gaussian (red, see text). Its parameters are given in Table B.1.}
\label{H2Op}
\end{figure*}

\begin{figure*}[t]
\centering
\sidecaption
\begin{minipage}{0.7\linewidth}
\begin{minipage}[t]{0.33\linewidth}
\centering
\includegraphics[width=\linewidth]{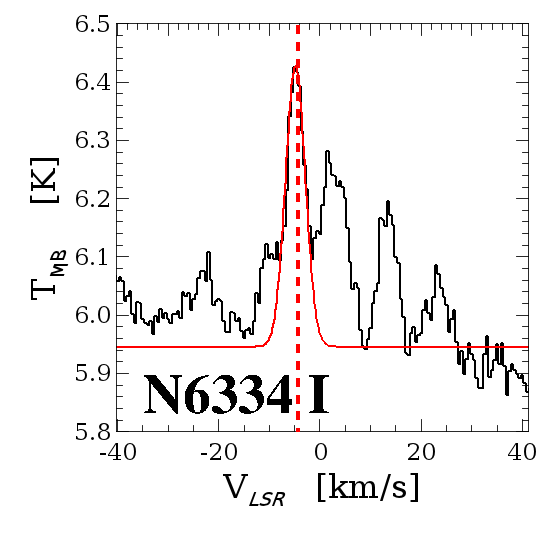}
\end{minipage}
\hspace{-0.2cm}
\centering
\begin{minipage}[t]{0.33\linewidth}
\centering
\includegraphics[width=\linewidth]{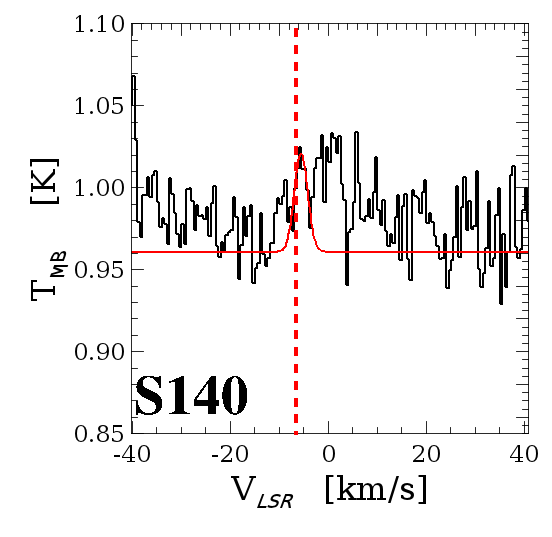}
\end{minipage}
\hspace{-0.2cm}
\begin{minipage}[t]{0.33\linewidth}
\centering
\includegraphics[width=\linewidth]{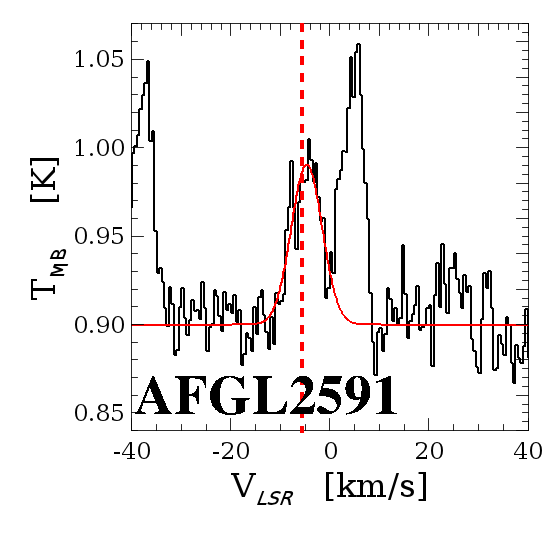}
\end{minipage}\\[-5pt]
\centering
\begin{minipage}[t]{0.33\linewidth}
\centering
\includegraphics[width=\linewidth]{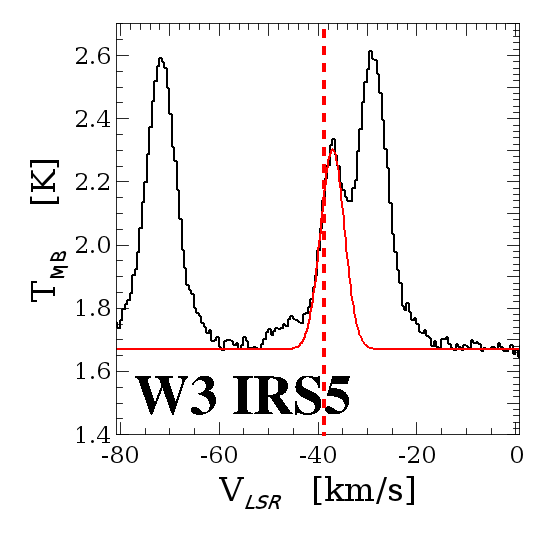}
\end{minipage}
\hspace{-0.2cm}
\begin{minipage}[t]{0.33\linewidth}
\centering
\includegraphics[width=\linewidth]{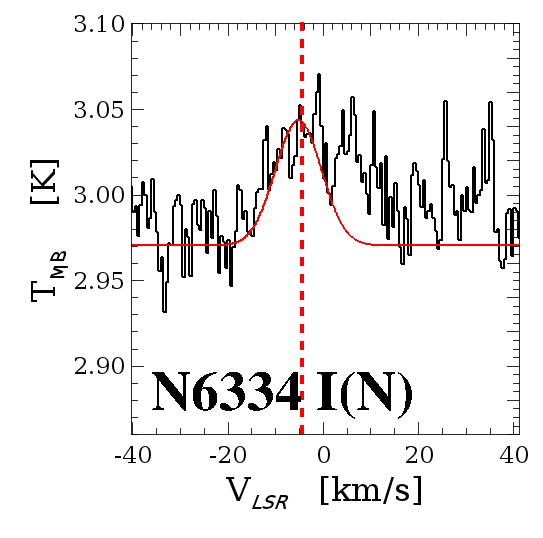}
\end{minipage}
\hspace{-0.2cm}
\begin{minipage}[t]{0.33\linewidth}
\centering
\includegraphics[width=\linewidth]{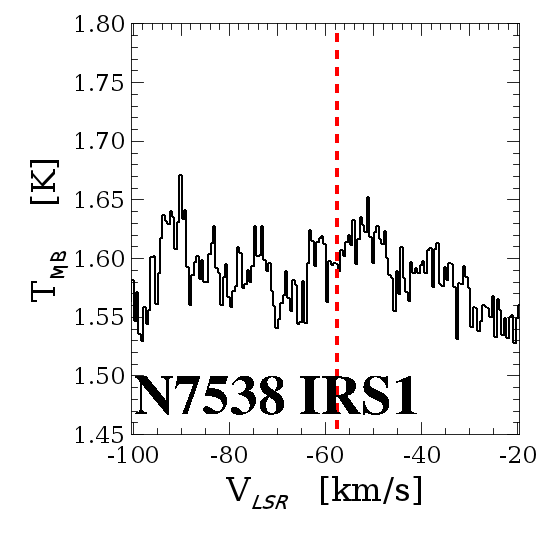}
\end{minipage}
\end{minipage}
\caption{Ortho-H$_3$O$^+$(4$_{3}^{+} - 3_{3}^{-}$) line at 1031 GHz observed toward the high-mass objects. The observed line is fitted by a Gaussian (red) and its parameters are given in Table B.2. An SO line at +10.1 km s$^{-1}$ relative to the systemic velocity and lines from the other sideband appearing at 6 and 8 km s$^{-1}$ are blended (see text).}
\label{H3Op}
\end{figure*}

\begin{figure*}[t]
\sidecaption
\begin{minipage}{0.7\linewidth}
\centering
\begin{minipage}[t]{0.33\linewidth}
\centering
\includegraphics[width=\linewidth]{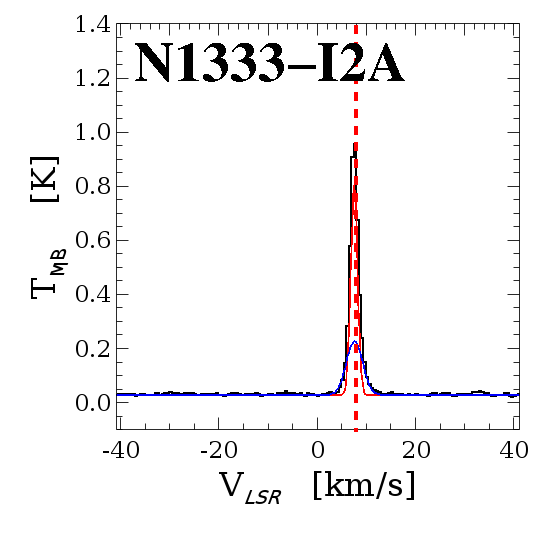}
\end{minipage}
\hspace{-0.2cm}
\centering
\begin{minipage}[t]{0.33\linewidth}
\centering
\includegraphics[width=\linewidth]{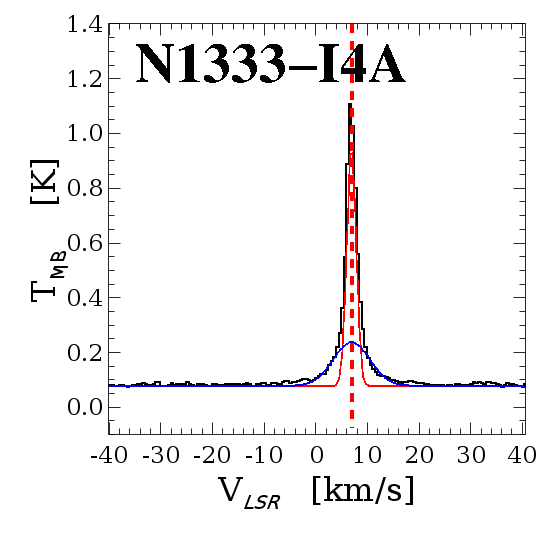}
\end{minipage}
\hspace{-0.2cm}
\begin{minipage}[t]{0.33\linewidth}
\centering
\includegraphics[width=\linewidth]{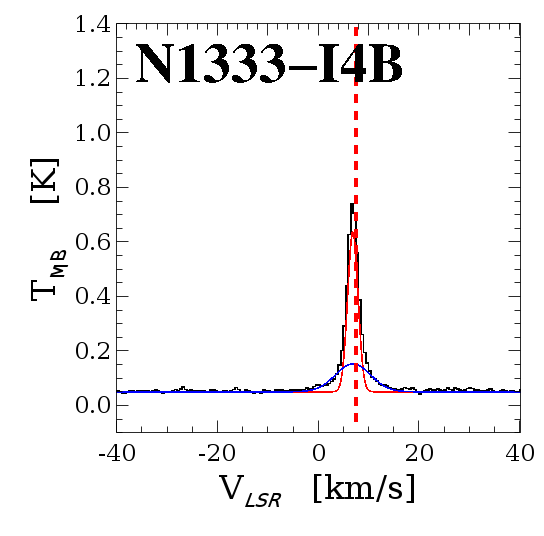}
\end{minipage}\\[-5pt]
\centering
\begin{minipage}[t]{0.33\linewidth}
\centering
\includegraphics[width=\linewidth]{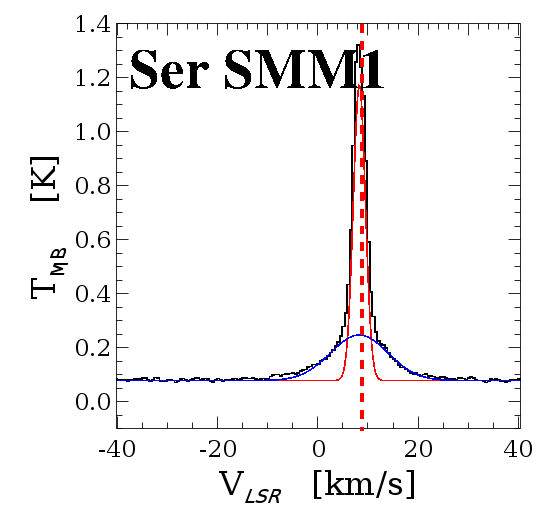}
\end{minipage}
\hspace{-0.2cm}
\begin{minipage}[t]{0.33\linewidth}
\centering
\includegraphics[width=\linewidth]{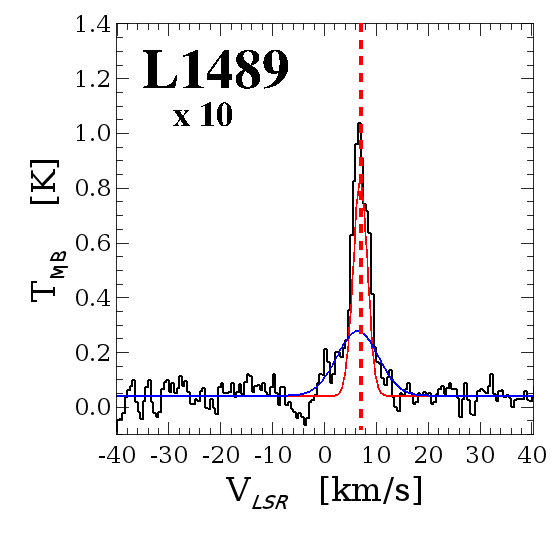}
\end{minipage}
\hspace{-0.2cm}
\begin{minipage}[t]{0.33\linewidth}
\centering
\includegraphics[width=\linewidth]{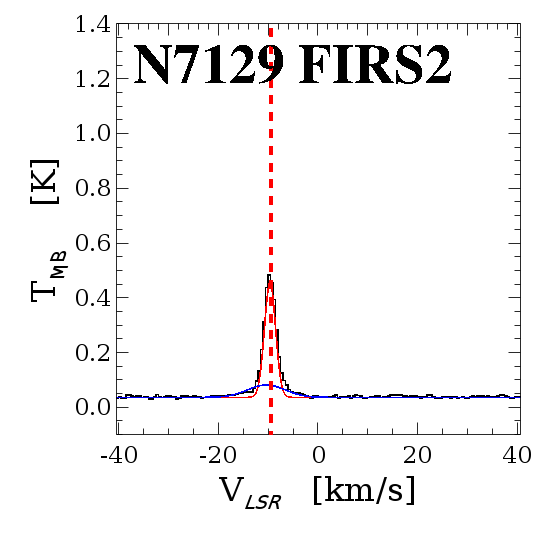}
\end{minipage}\\[-5pt]
\centering
\begin{minipage}[t]{0.33\linewidth}
\centering
\includegraphics[width=\linewidth]{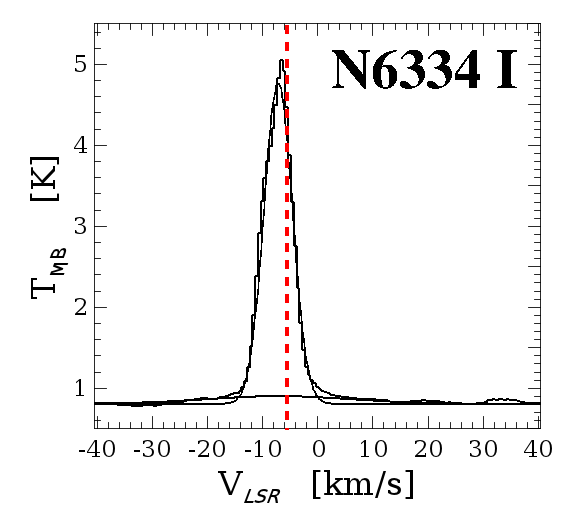}
\end{minipage}
\hspace{-0.2cm}
\centering
\begin{minipage}[t]{0.33\linewidth}
\centering
\includegraphics[width=\linewidth]{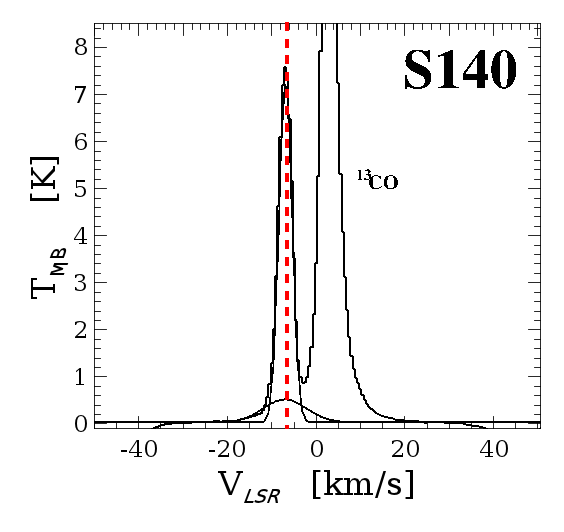}
\end{minipage}
\hspace{-0.2cm}
\begin{minipage}[t]{0.33\linewidth}
\centering
\includegraphics[width=\linewidth]{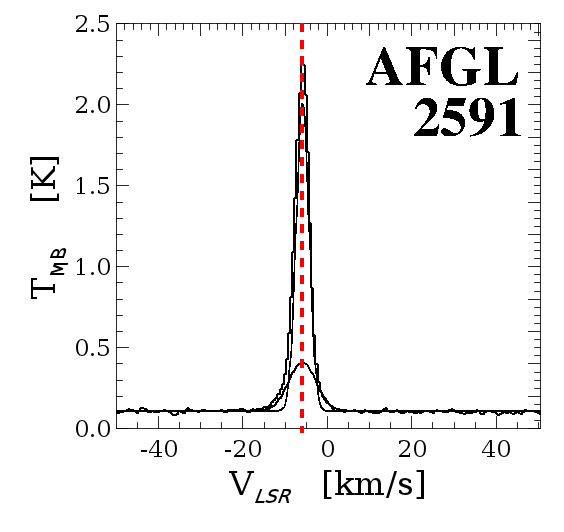}
\end{minipage}\\[-5pt]
\centering
\begin{minipage}[t]{0.33\linewidth}
\centering
\includegraphics[width=\linewidth]{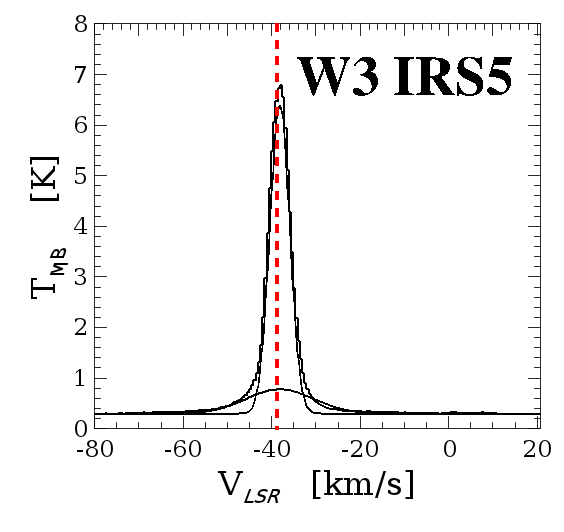}
\end{minipage}
\hspace{-0.2cm}
\begin{minipage}[t]{0.33\linewidth}
\centering
\includegraphics[width=\linewidth]{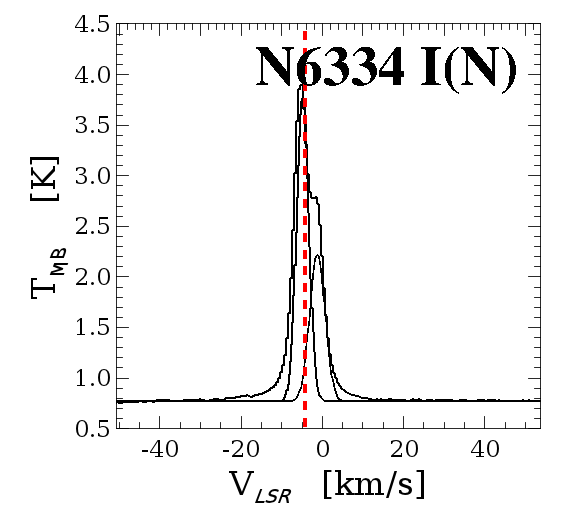}
\end{minipage}
\hspace{-0.2cm}
\begin{minipage}[t]{0.33\linewidth}
\centering
\includegraphics[width=\linewidth]{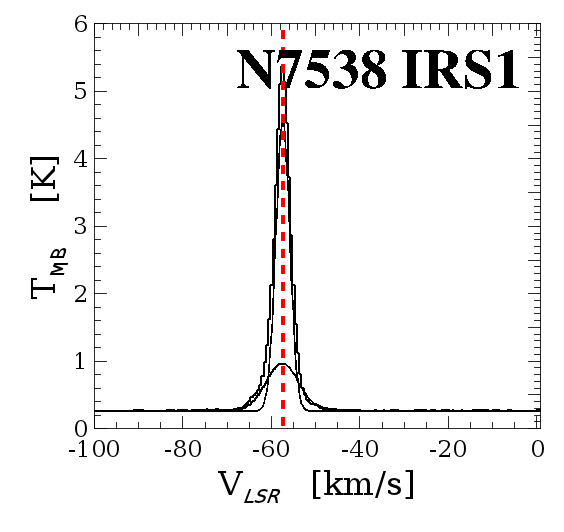}
\end{minipage}
\end{minipage}
\caption{HCO$^+$(6-5) line at 535 GHz. A two-Gaussian fit is shown, emphasizing the center and the wings of the line. Both sets of parameters are given in Table B.2. The strong line at 3 km s$^{-1}$ in S140 is $^{13}$CO ($J$=5-4) from the other sideband.}
\label{HCOp_obs}
\end{figure*}

\begin{figure*}[t]
\centering
\sidecaption
\begin{minipage}{0.7\linewidth}
\begin{minipage}[t]{0.33\linewidth}
\centering
\includegraphics[width=\linewidth]{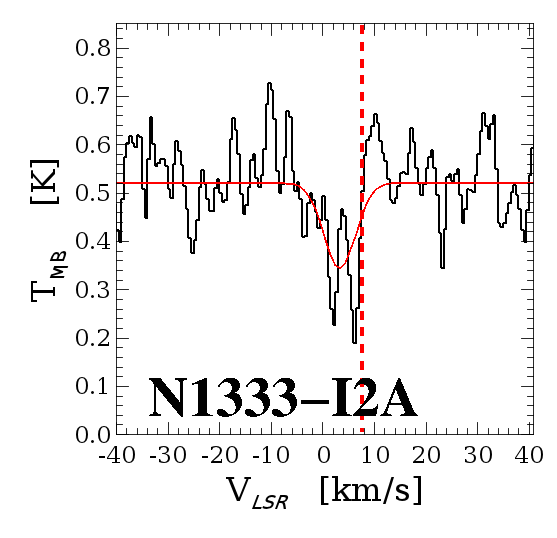}
\end{minipage}
\hspace{-0.2cm}
\centering
\begin{minipage}[t]{0.33\linewidth}
\centering
\includegraphics[width=\linewidth]{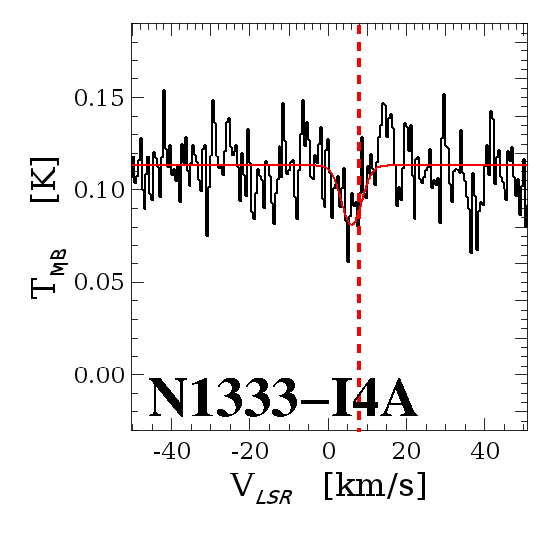}
\end{minipage}
\hspace{-0.2cm}
\begin{minipage}[t]{0.33\linewidth}
\centering
\includegraphics[width=\linewidth]{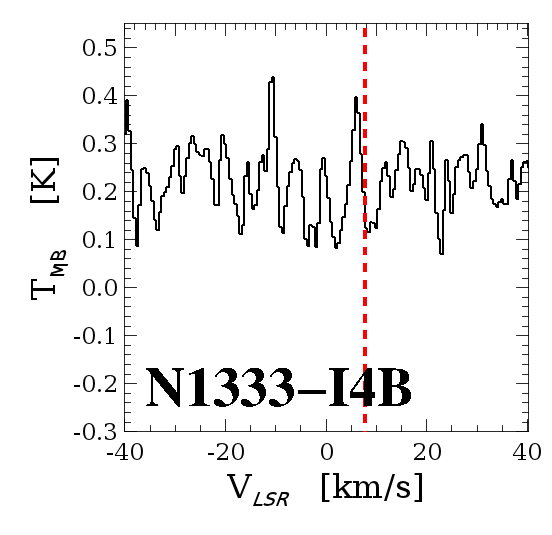}
\end{minipage}\\[-5pt]
\centering
\begin{minipage}[t]{0.33\linewidth}
\centering
\includegraphics[width=\linewidth]{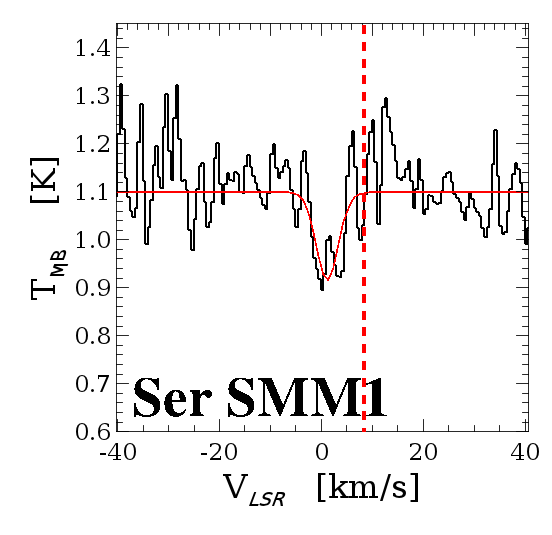}
\end{minipage}\\[-5pt]
\centering
\begin{minipage}[t]{0.33\linewidth}
\centering
\includegraphics[width=\linewidth]{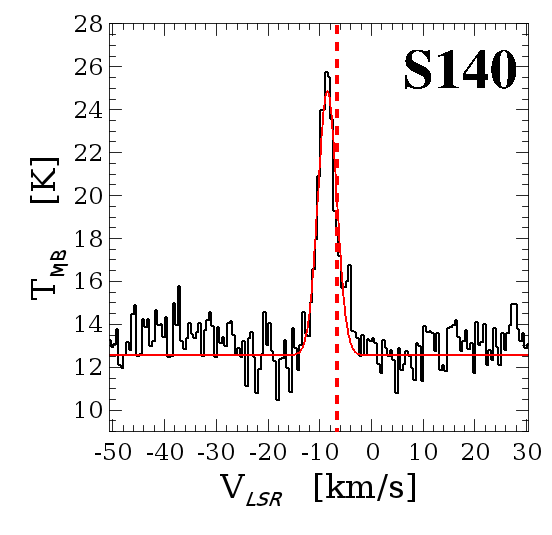}
\end{minipage}
\hspace{-0.2cm}
\begin{minipage}[t]{0.33\linewidth}
\centering
\includegraphics[width=\linewidth]{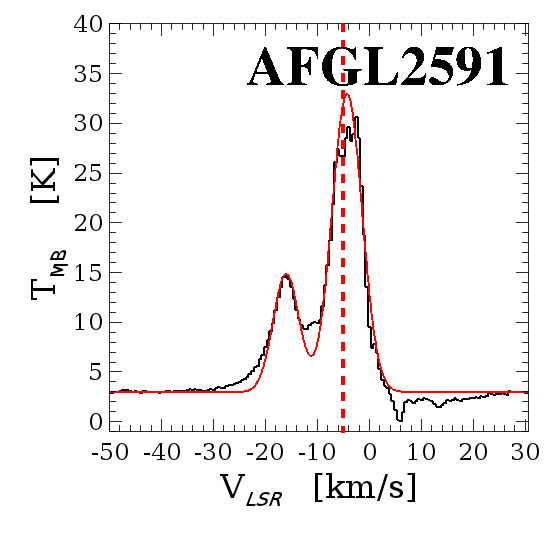}
\end{minipage}
\centering
\begin{minipage}[t]{0.33\linewidth}
\centering
\includegraphics[width=\linewidth]{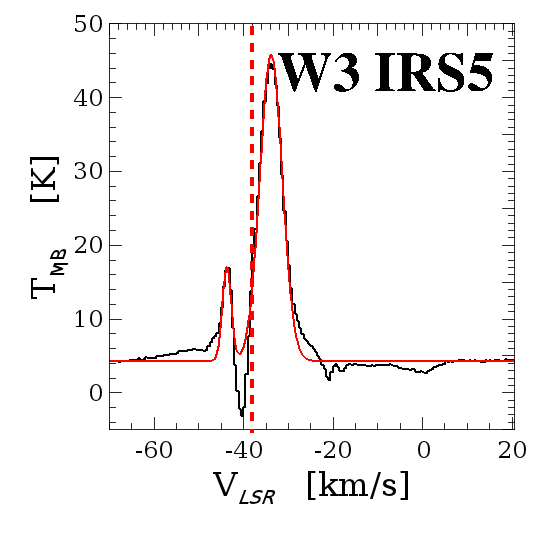}
\end{minipage}
\end{minipage}
\caption{C$^+$ fine structure line at 1901 GHz. Where detected, a Gaussian fit is shown in red and the parameters are given in Table B.3.}
\vspace{0.5cm}
\label{CII_obs}
\end{figure*}

The observed line spectra are individually shown in Figs. \ref{CHp_obs} -- \ref{CH_obs}. The velocity scale refers to the line frequency in the LSR. If several line components exist, the scale is relative to the most intense fine or hyperfine structure line marked by an asterisk in Table \ref{table_lines}. The systemic velocity of the YSO as given in Table \ref{table_objects} is marked with a vertical dashed red line. For lines with several fine or hyperfine structure components, the line thickness and dash length indicate the theoretical intensities in decreasing order (arbitrary scale). The line shapes are modeled with Gaussian fits. If more than one component exists, several Gaussians are fitted to estimate the line parameters, which are given in Tables B.1 -- B.3.

\subsection{CH$^+$}
CH$^+$ ($J = 1-0$) is detected in absorption toward the low-mass objects of Class 0  (Fig. \ref{CHp_obs}).  The detection is only tentative toward NGC1333 I4B, and there is no hint of the line toward L 1489 (Class I), where the continuum is weak and the upper limit is higher than for the absorptions detected in the Class 0 objects (Table B.1). Toward the intermediate-mass object NGC7129 FIRS2, the absorbed line width is relatively large and more similar to some high-mass sources. Contrary to the low-mass objects, where the line is shifted to the blue of the systemic velocity, the line toward NGC7129 FIRS2 is shifted to the red.

CH$^+$ is predominantly observed in absorption, but toward  high-mass some objects also in emission (Fig. \ref{CHp_obs}). AFGL 2591 and W3 IRS5 show pronounced P-Cygni profiles, with absorption to the blue of an emission peak.  In S 140 the continuum is removed by frequency switching. The peaks between 20 and 30 km s$^{-1}$ are the negative ghosts of the lines observed near the systemic velocity. The spectrum can be interpreted by an emission and an absorption on each side or as an emission in the middle of a broad (about 9 km s$^{-1}$) and probably saturated absorption. We assume the first option to derive lower limits for the line intensity and absorption, and the second option for upper limits. The column densities in Table B.1 refer to these limits. NGC7538 IRS1, on the other hand, shows an inverse P-Cygni profile. Emission and absorption are listed separately, including the column depths of each transition. NGC6334 I and I(N) show a double absorption dip, separated by about 10 km s$^{-1}$. The deeper one is wider, the other one is more red-shifted and narrower. The similarity of the two dips toward the two adjacent objects suggests absorption by diffuse clouds in the foreground \citep{2010A&A...521L..28E}. However, the spectra of NGC6334 I were fitted by a 1D slab model \citep[][Appendix B]{2010ApJ...720.1432B} including the CH$^+$(2-1) transition (see below), which resolved the deeper dip into a blue-shifted and a red-shifted component. The blue-shifted component of the deeper dip is assumed to originate from the YSO, but the red-shifted component is questionable. We interpret the other dip as due to diffuse clouds in the foreground.

The CH$^+$ lines presented in Fig. \ref{CHp_obs} and attributed to the YSO consist of one or two Gaussian components. A significant deviation from symmetry, however, is noted toward AFGL 2591, which has extended absorption between $-$25 and $-$35 km s$^{-1}$, thus from $-$19 to $-$29 km s$^{-1}$ relative to the systemic velocity. This tail has been noted before by \citet{2010A&A...521L..44B}, and as well in $^{12}$CO and $^{13}$CO ($\nu = 1-0$) infrared absorption by \citet{1990ApJ...363..554M} and \citet{1999ApJ...522..991V}.

Methanol emission overlaps in the presented frequency range toward the massive objects NGC6334 I and AFGL 2591 at 36.1 and 47.5 km s$^{-1}$ relative to the systemic velocity. The respective transitions are at 824.7197 GHz ($20_{-1,0} - 19_{0,0}$) in the lower sideband, and at 835.0039 GHz ($4_{5,5} - 5_{4,0}$) in the upper sideband. The line at 835.1851 GHz ($17_{-3,0}$ $-$ $17_{2,0}$) is the only methanol line that is directly in the range of CH$^+$ emission. It is prominent toward NGC6334 I at $-$24.4 km s$^{-1}$ and minuscule toward AFGL 2591 at $-$22.5 km s$^{-1}$. CH$^+$ absorption by foreground clouds is noticeable in all high-mass sources. It lines up with absorptions of the ground-state H$_2^{18}$O line in the second dip of W3 IRS5, AFGL 2591, and NGC6334 I(N) \citep{2013A&A...554A..83V}.

CH$^+$ ($J=2-1$) was observed in some of the sources (Fig. \ref{CHp2}). None of the low-mass (Class 0) observations detected the line. The high-mass mid-IR quiet object NGC6334 I(N) was also not detected, but AFGL 2591, W3 IRS5, and NGC7538 IRS1 show emission. The spectra of the first two objects also contain a weak blue-shifted absorption, indicating a P-Cygni profile \citep[as noted already in][]{2010ApJ...720.1432B,2011EAS....52..239B}. The non-detections of the low-mass objects in CH$^+$(2-1) are consistent with the assumed excitation temperature (9 K) and total derived CH$^+$ column densities given in Table B.1.

\subsection{OH$^+$}
OH$^+$ ($N_J=1_1-0_1$) absorption is clearly detected in the low-mass object Ser SMM1 (Fig. \ref{OHp_obs}, second row, left). The two lines (each a doublet, see Table \ref{table_lines}) have similar shape and the flux ratio is consistent with optically thin. They are possibly detected toward NGC1333 I4A in absorption, but only at a signal-to-noise of 4 $\sigma$ (sigma). The detection is doubtful since the hyperfine components are 2 km s$^{-1}$ less apart than expected. We use the fitted line intensity as an upper limit in Table B.1. Even weaker are the lines in NGC1333 I4B, but the similarity in blue shift with the CH$^+$ absorption lines (Fig. \ref{CHp_obs}, top right) corroborates the detection. The line widths toward the NGC1333 I4A and I4B  (if real) are extremely narrow with a FWHM of 1.4 and 1.7 km s$^{-1}$. The fitted lines in NGC7129 FIRS2 have a width of $>$7.4 and 10 km s$^{-1}$ (Fig. \ref{OHp_obs}, second row, right). OH$^+$ is not detected toward NGC1333 I2A and L 1489. The low continuum of L 1489 makes a detection difficult and yields a high upper limit on the column density (Table B.1).

Toward massive objects, OH$^+$ is mostly in absorption. W3 IRS5 (Fig. \ref{OHp_obs}, bottom left) has an emission peak at a 12$\sigma$ level \citep{2010A&A...521L..35B}. It confirms the protostellar origin of OH$^+$ in this source. The emission is slightly red-shifted but close to a blue-shifted absorption, thus suggesting a P-Cygni profile similar to but less pronounced than that in CH$^+$. We note also a tail in absorption blue-shifted by 10 -- 17 km s$^{-1}$ w.r.t. to W3 IRS5, similar to CH$^+$ toward AFGL 2591. As the two main OH$^+$ fine structure lines each have substructures, only the strongest feature is fitted with a Gaussian. The absorption integrated over all fine structure lines and the total column density integrated over all levels are given in Table B.1. NGC6334 I and AFGL 2591 show also an absorption component that is blue-shifted by 4 -- 10 km s$^{-1}$ relative to the systemic velocity and two components red-shifted by typically 4 -- 15 km s$^{-1}$. The blue-shifted component is missing in S 140, NGC6334 I(N), and NGC7538 IRS1. W3 IRS5 and NGC7538 IRS1 have a second, deeper pair of absorption lines shifted to the red by 27.5 and 55 km s$^{-1}$, respectively. They partially coincide with absorptions in CH$^+$(1-0), suggesting diffuse interstellar foreground \citep{2015ApJ...800...40I}.

\begin{figure*}[t]
\centering
\sidecaption
\begin{minipage}{0.7\linewidth}
\begin{minipage}[t]{0.34\linewidth}
\centering
\includegraphics[width=\linewidth]{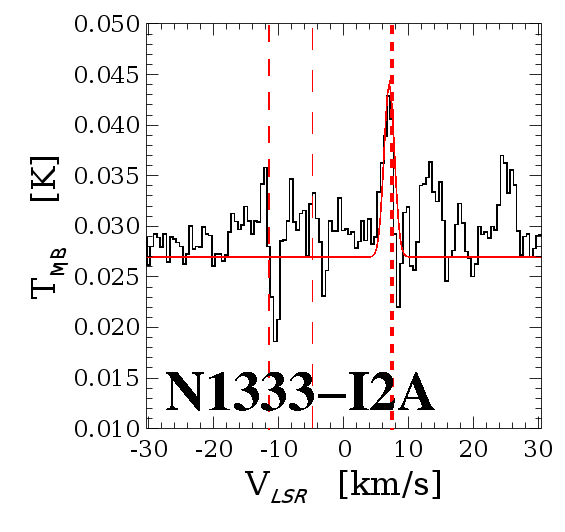}
\end{minipage}
\hspace{-0.2cm}
\centering
\begin{minipage}[t]{0.33\linewidth}
\centering
\includegraphics[width=\linewidth]{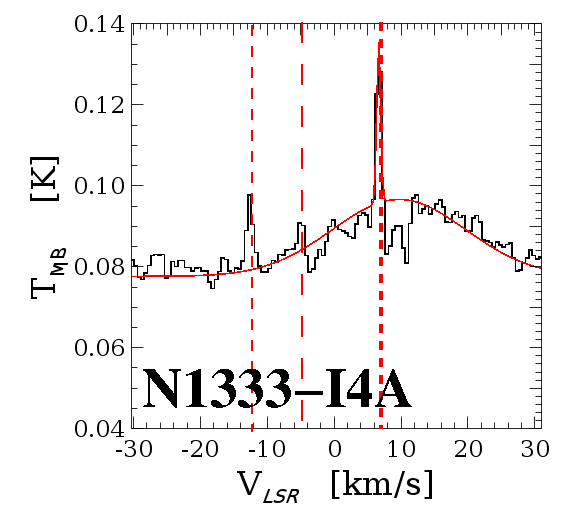}
\end{minipage}
\hspace{-0.2cm}
\begin{minipage}[t]{0.33\linewidth}
\centering
\includegraphics[width=\linewidth]{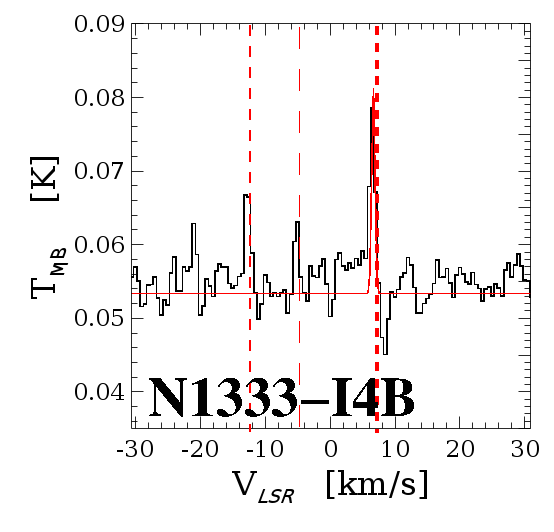}
\end{minipage}\\[-5pt]
\centering
\begin{minipage}[t]{0.32\linewidth}
\centering
\includegraphics[width=\linewidth]{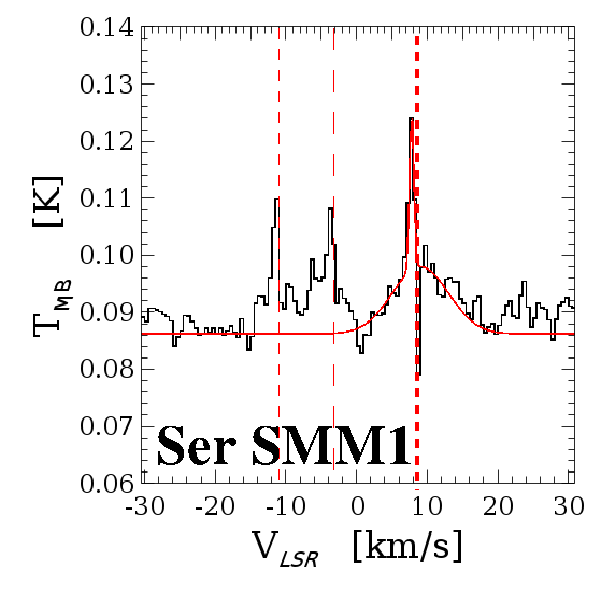}
\end{minipage}
\hspace{-0.2cm}
\begin{minipage}[t]{0.335\linewidth}
\centering
\includegraphics[width=\linewidth]{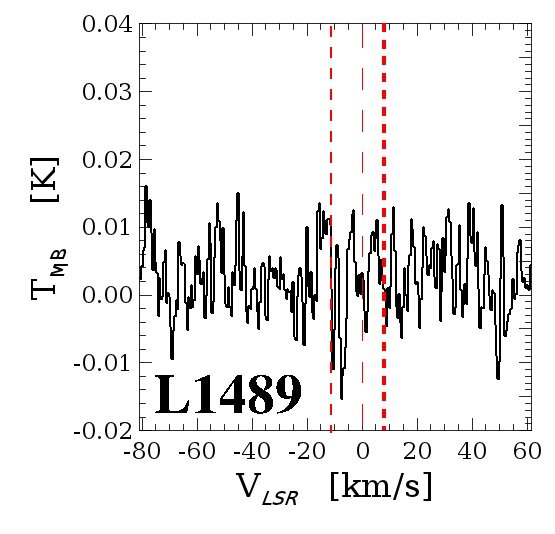}
\end{minipage}
\hspace{-0.2cm}
\begin{minipage}[t]{0.335\linewidth}
\centering
\includegraphics[width=\linewidth]{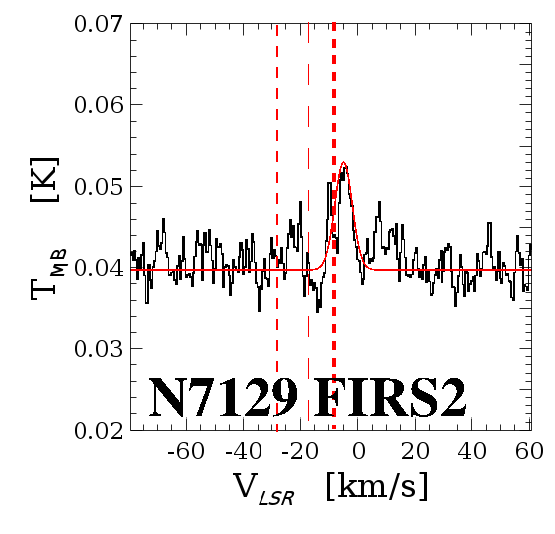}
\end{minipage}\\[-5pt]
\centering
\begin{minipage}[t]{0.32\linewidth}
\centering
\includegraphics[width=\linewidth]{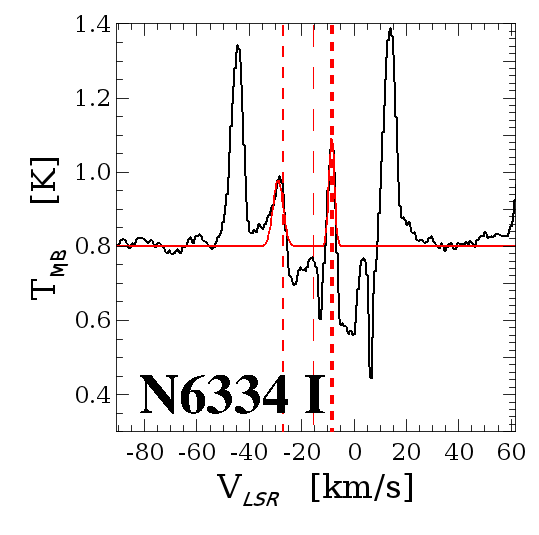}
\end{minipage}
\hspace{-0.2cm}
\centering
\begin{minipage}[t]{0.335\linewidth}
\centering
\includegraphics[width=\linewidth]{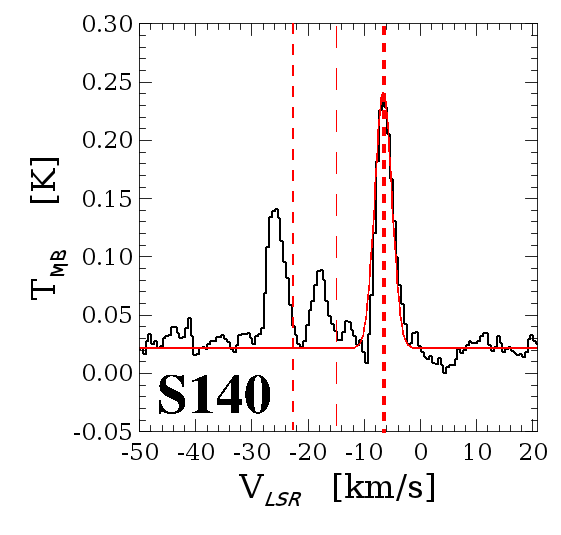}
\end{minipage}
\hspace{-0.2cm}
\begin{minipage}[t]{0.32\linewidth}
\centering
\includegraphics[width=\linewidth]{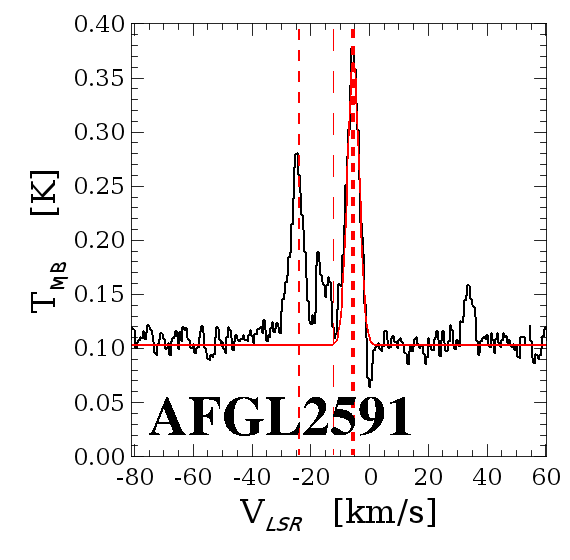}
\end{minipage}\\[-5pt]
\centering
\begin{minipage}[t]{0.33\linewidth}
\centering
\includegraphics[width=\linewidth]{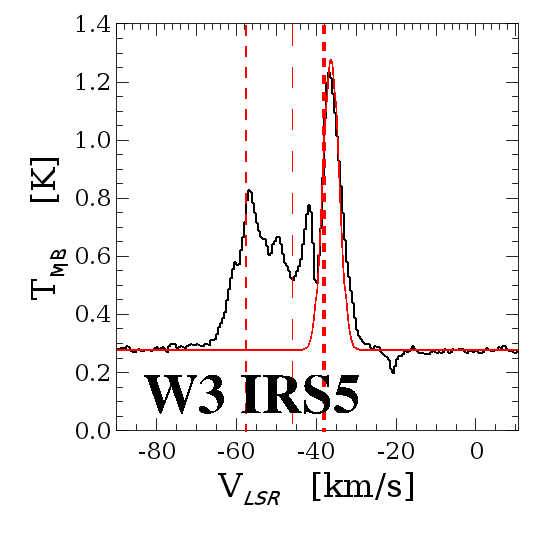}
\end{minipage}
\hspace{-0.2cm}
\begin{minipage}[t]{0.33\linewidth}
\centering
\includegraphics[width=\linewidth]{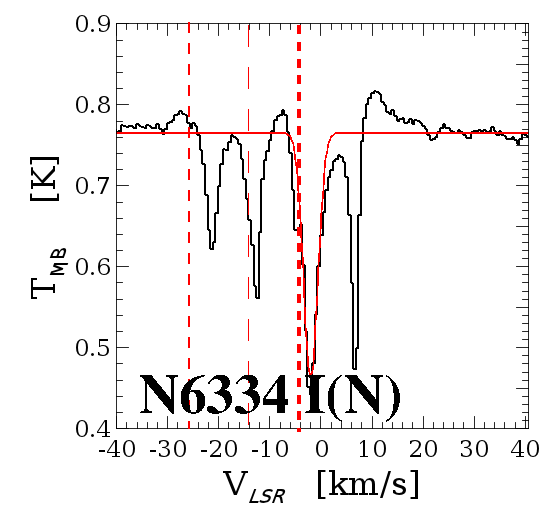}
\end{minipage}
\hspace{-0.2cm}
\begin{minipage}[t]{0.33\linewidth}
\centering
\includegraphics[width=\linewidth]{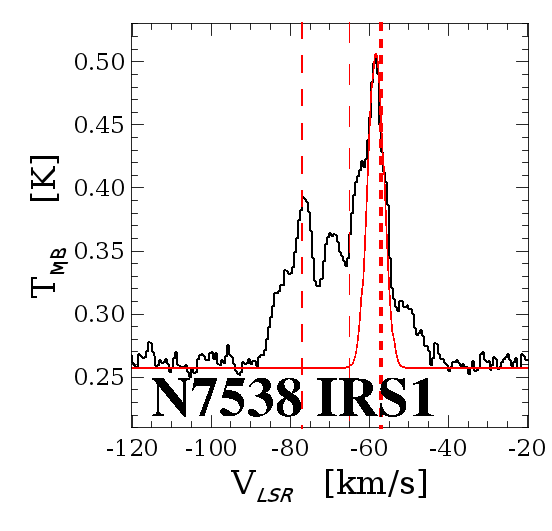}
\end{minipage}
\end{minipage}
\caption{CH lines at 537 GHz. The positions of the fine structure lines are shown shifted by the systemic velocity of the YSOs. The thickness of the three lines indicates the relative theoretical intensities. Where the lines are detected, a Gaussian curve in red is fitted to the strongest transition and its parameters are given in Table B.3. The broad feature in N1333 I4A is an interfering H$_2$$^{18}$O line from the other sideband.}
\label{CH_obs}
\end{figure*}

\subsection{H$_2$O$^+$}
The ortho-H$_2$O$^+$ ($N_{K{_a}K{_b}}=1_{11}-0_{00}, J=\frac{3}{2}-\frac{1}{2}$) hyperfine transitions were not detected toward the low- and intermediate-mass objects. The rms noise is between 20 and 24 mK, yielding upper limit column densities around $5\times 10^{12}$ cm$^{-2}$ for Class 0 objects (Table B.2). The only notable exception is Ser SMM1, for which there is a tentative absorption at the 5$\sigma$ level, blue-shifted by $-$3.5 km s$^{-1}$ (see Fig. \ref{super}, bottom). We use it as an upper limit in Table B.2.

Observations of the high-mass sources are presented in Fig. \ref{H2Op}. Where detected, H$_2$O$^+$ is in absorption. Deep absorptions are found near zero LSR velocity; they are probably caused by diffuse interstellar foreground clouds. The component intrinsic to the object is not always clearly detected. Exceptions are blue-shifted absorptions in AFGL 2591 and W3 IRS5 that have been attributed to the YSO by \citet{2010A&A...521L..44B} and \citet{2010A&A...521L..35B}, respectively.  Absorptions at $-$54, $-$65, and around $-$81 km s$^{-1}$ toward NGC7538 IRS1 may be associated with the YSO, but blue-shifted by $-$9.9 km s$^{-1}$. We use this tentative detection as an upper limit of the intensity. H$_2$O$^+$ toward NGC7538 IRS1 shows a second complex set of absorption lines red-sifted from the systemic velocity. It is attributed to diffuse clouds in the foreground \citep{2015ApJ...800...40I}. Relatively broad and blue-shifted absorption lines of H$_2$O$^+$ have been reported in other sources by \citet{2010A&A...521L..34W} and attributed to the outflows. Consistent with this suggestion, we have not detected any narrow absorptions of H$_2$O$^+$, excluding the envelope as the source. A tentative emission at 0 - 6 km s$^{-1}$ \citep{2010A&A...521L..35B} is not confirmed by the new analysis. For NGC6334 I and I(N) the absorption profiles are nearly identical in $V_{\rm LSR}$, strongly suggesting that the absorption is not intrinsic to the source but caused by foreground diffuse interstellar clouds. There are nearby methanol lines at $-$37.2 and +40.1 km s$^{-1}$ toward NGC6334 I from the lower sideband (1100.469236 GHz, (13$_{8,0}$ $\rightarrow$ 14$_{7,0}$) and 1100.757207 GHz (13$_{5,1}$ $\rightarrow$ 13$_{6,1}$), but no strong methanol line interferes with H$_2$O$^+$.

The column densities listed in Table B.2 were multiplied by 1.33 to account for the non-observed para-H$_2$O$^+$ lines, assuming an ortho-to-para ratio of 3 \citep{2015ApJ...800...40I}.

\subsection{H$_3$O$^+$}
Para-H$_3$O$^+$ ($J^p_K=3^+_2 - 2^-_2$) emission has been previously detected in high-mass YSOs from the ground by \citet{1991ApJ...380L..79W} and \citet{1992ApJ...399..533P}. Recently \cite{2014ApJ...785..135L} observed H$_3$O$^+$ toward W31C in absorption in the inversion transitions and derived $T_{\rm rot} \approx 380$ K.

There are no detections of ortho-H$_3$O$^+$(4$_{3}^{+}$ - 3$_{3}^{-}$) in low- and intermediate-mass objects. The rms noise is between 17 and 20 mK, yielding upper limit column densities $<2.0\times 10^{12}$ cm$^{-2}$ for an assumed excitation temperature of 225 K (Table B.2). The  line is detected in massive objects (Fig. \ref{H3Op}), but blended with the SO ($N_J=24_{24}$ - $23_{23}$) line at 10.1 km s$^{-1}$ and by lines from the lower sideband. All lines are fitted, and the fit of the H$_3$O$^+$ line is reported in Table B.2. The identifications of H$_3$O$^+$ in S 140 and NGC6334 I(N) are tentative because of limited sensitivity. The fitted values are considered to be upper limits. The line was not detected in NGC7538 IRS1.

\subsection{SH$^+$}
The detection of SH$^+$ $(N_{JF}=1_{2\frac{5}{2}}-0_{1\frac{3}{2}}$) and ($1_{2\frac{5}{2}}-0_{1\frac{3}{2}}$) in emission toward W3 IRS5 was previously reported \citep{2010A&A...521L..35B}. We confirm here the detection, but cannot find it either in emission or absorption toward the other objects of low, intermediate and high mass. The upper limit on the SH$^+$ column density toward AFGL 2591 is $<1.5\times 10^{11}$ cm$^{-2}$, a factor of 30 lower than the upper limit derived by \citet{2007A&A...466..977S} from the transition  $N_{JF}=1_{0\frac{1}{2}} - 0_{1\frac{3}{2}}$ at 345.944 GHz, but 6 times smaller than detected toward W3 IRS5. The reason for the difference between the two YSOs may be the high gas-phase sulphur abundance in W3 IRS5, noted before  e.g. by \citet{1997A&AS..124..205H} and \citet{2003A&A...412..133V}. \citet{2007A&A...466..977S} report a 9 times stronger SO$^+$ line intensity toward W3 IRS5 than for AFGL 2591 and attribute it to different sulphur abundances in the gas. SH$^+$ is also not detected toward the similar object NGC6334 I, where the background is high and a line, probably methanol (14$_{2,13}-14_{1,14}$) at 526.02589 GHz, overlaps.

We have not seen, nor especially searched for, SH$^+$ absorption lines that are strongly shifted from the systemic velocity suggesting absorption by diffuse interstellar clouds.

\subsection{HCO$^+$}
HCO$^+$ ($J=6-5, v=0$) is clearly detected in emission toward all objects. It is fitted here with two Gaussians, one for the line peak and one for the wings (Figs. \ref{HCOp_obs}). The properties of the two components are discussed in Section 5.2. Table B.2 gives the values of the fits in peak, width, and shift; the line intensity and column density are integrated over both components. A red-shifted shoulder toward L 1489 is noticeable in Fig. \ref{HCOp_obs}, second row, middle; and a double peak is clearly visible in HRS resolution. A similar feature was reported for this object in HCO$^+$ (3-2) and (4-3)  by \citet{2007A&A...475..915B} and interpreted as a disk contribution.

\subsection{C$^+$}
If C$^+$ ($^2$P $J=\frac{3}{2}-\frac{1}{2}$) is detected toward low-mass objects, it is in absorption (Fig. \ref{CII_obs}). The detection level is 5 $\sigma$ toward Ser SMM1. The red-shifted emission peak is statistically not significant. We note, however, that PACS observations of this source report C$^+$ in emission  preferentially from the outflows \citep{2012A&A...548A..77G}. The Gaussian curve fitted for NGC1333 I2A indicates a statistical significance of 7 $\sigma$ for an absorption feature. The C$^+$ line is tentatively detected toward NGC1333 I4A. A 4.1 $\sigma$ absorption is found at the expected velocity and yields an upper limit for the column density. C$^+$ is not detected toward NGC1333 I4B. Where detected in low-mass objects, the line is shifted on average by $-$5.9$\pm$0.8 km s$^{-1}$ and has an average line width of 6.0$\pm$1.0 km s$^{-1}$. The differences between the two off-source reference positions for the low-mass objects are small and do not indicate significant variations in large-scale C$^+$ emission. Thus the absorption lines seem to be inherent to the source.

The line profiles of the three observed high-mass objects (Fig. \ref{CII_obs}) can be interpreted by two emission peaks or by a broad peak, reduced by a narrow absorption or emission at the reference position. Toward W3 IRS5 the spectrum falls below the continuum level, suggesting that the line is self-absorbed. This interpretation is also assumed for the other two objects. An absorption feature is seen near the systemic velocity in all three objects. The integrated emission line intensities in Table B.3 thus yield lower limits in column density. For line width and line shift, we have fitted two Gaussian emission peaks. Both peaks are integrated to measure the column density.

A difference between the off-position emissions was detected for AFGL 2591 at $-$11.1, $-$5.2, 4.7, and 5.7 km s$^{-1}$ in $V_{LSR}$. The latter is most significant and has an integrated intensity of 8.6 K km s$^{-1}$, thus about 3$\%$ of the total emission. These differences are interpreted as slightly different large-scale emissions, which cause spurious dips in the spectrum when subtracted. In addition, AFGL 2591 seems to have several intrinsic red-shifted absorption features. In particular, the feature at 13 km s$^{-1}$ is not detectable in the difference of the off-positions. It is also visible in the CH$^+$ (1-0) line (Fig. \ref{CHp_obs}, fourth row, right) and was reported by \citet{2010A&A...521L..44B}.

The C$^+$ spectrum of W3 IRS5 is similar and also has many spectral components. Most prominent are the peak at $-$43.9 and the main peak at $-$33.9 km s$^{-1}$ interrupted by a minimum at $-$40.3 km s$^{-1}$ amounting to $-3.14\pm0.06$ K. The difference between the emissions of two reference positions amounts to 2.1 K km s$^{-1}$, but peaks at $-$38.4$\pm0.2$ km s$^{-1}$, the systemic velocity. Nevertheless, the minimum observed at $-$40.3 km s$^{-1}$  seems to be caused mostly by off-source emission. For S 140, observed in OTF mode, the reference position does not emit C$^+$ emission above 1.2 K km s$^{-1}$ \citep{1996A&A...315L.285E}; thus the off-beam contamination is small.  We conclude that the integrated line emission of the C$^+$ line in the observed high-mass objects is not severely altered by emission at the reference position.

\subsection{CH}
The CH ($^2\Pi_{\frac{3}{2}} J = \frac{3}{2}-\frac{1}{2}$) hyperfine transitions were detected in emission toward all the low-mass and intermediate-mass objects except toward the Class I object L 1489 (Fig. \ref{CH_obs}, second row, middle). The line widths of the narrow component toward NGC1333 I4A, NGC1333 I4B and Ser SMM1 are remarkably small (average 0.63$\pm$0.06 km s$^{-1}$) and were analyzed using the High Resolution Spectrometer (HRS/HIFI) data. In these cases, the line width appears to be reduced by a narrow red-shifted absorption, forming an inverse P-Cygni profile. Toward NGC1333 I2A the line is not absorbed and its width is 1.9 km s$^{-1}$.  This value is comparable to the line widths found previously from the outer part of quiescent envelopes in  C$^{17}$O and C$^{18}$O ($J$=2-1) lines \citep[e.g.][: 0.7 - 2.1 km s$^{-1}$]{2002A&A...389..908J}. The same objects as in this study were observed in C$^{18}$O ($J$=3-2) by \citet{2013A&A...553A.125S}, who report line widths of 1.47$\pm$0.53 km s$^{-1}$. Values between 2 - 3 km s$^{-1}$ were reported by \citet{2012A&A...542A...8K} for the narrow component of the 557 GHz H$_2$O (1$_{10}$ - 1$_{01}$) line mostly seen in absorption. The similarities suggests that the narrow component of the CH lines originate in the same region of the outer envelope as the low-$J$ CO lines.

In addition, Ser SMM1 (Fig. \ref{CH_obs}, second row, left) has also a broad emission component (FWHM 9.4 km s$^{-1}$). Its column density is an order of magnitude larger than for the narrow component.

In massive objects (Fig. \ref{CH_obs}), the CH lines at 537 GHz are observed in emission as well as in absorption. Emission and absorption are mixed toward NGC6334 I, as previously reported by \citet{2010A&A...521L..43V}. CH absorption dominates in NGC6334 I(N) having a more massive envelope. In both objects at least one other system of absorption lines at positive velocity ($\approx$ 7 km s$^{-1}$) is observed, originating probably in diffuse interstellar clouds in the foreground. For \mbox{S 140}, AFGL 2591, and NGC7538 IRS 1 emission dominates. In W3 IRS5 emission also dominates, but combines with a blue-shifted absorption to an apparent P-Cygni profile. Methanol line emission in the lower sideband interferes toward NGC6334 I. In the velocity range of Fig. \ref{CH_obs}, bottom left, they appear at $-$40.0 and 18.4 km s$^{-1}$. There is no other obvious interfering line in-between. Thus we assume that all minima and maxima between $-$34 and 14 km s$^{-1}$ are due to CH emission and absorption.

\newpage
\pagebreak[4]
\clearpage

\section{Observed line parameters vs. objects, quantitative results}

The line profiles were fitted by one or two Gaussians and their parameters are given in Tables B.1 - B.3. Only the line components moving with the systemic velocity of the YSO within $\pm$ 12 km s$^{-1}$ are listed. Some components even within that range may originate from diffuse clouds in the foreground. Suspected cases are marked ``DC" in the following tables. The YSO origin of the lines is discussed in Appendix A. The {\it line peak}, $T_{max}$, is given above (or below) the continuum background, $T_b$, in main beam temperature and refers to the strongest fine structure transition. For non-detections the upper limit is given by 3 times the measured rms noise. The {\it line width}, $\Delta V$, is the full width at half maximum (FWHM) above (or below) background of the Gaussian fit for the strongest component. The {\it line shift}, $\delta V$, relative to the LSR also refers to the Gaussian fit of the strongest component. If the line fit was greatly improved using two Gaussians, both are given on separate lines in the tables. Errors are a few units of the last decimal. The accuracies of the line width and line shift are better than 1 km s$^{-1}$. Line peak intensities have uncertainties of 10 - 20\%, which is mostly caused by calibration errors, but above 1 THz due to noise or baseline fitting errors.

{\it Intensity}, $\int T_{MB} dV$, refers to the integrated line profile in the observed main beam temperature above the background of lines in emission. Integration usually goes from$-$40 km s$^{-1}$ to +40 km s$^{-1}$ relative to the line peak. In cases of line blending, the interfering line is also characterized by a Gaussian; the target line fitted parameters are used for line integration and marked with an asterisk (*). In cases of several fine or hyperfine lines, they were all integrated. If not detected, the upper limit for lines expected in emission refers to 1.2 $\sqrt{\delta v\Delta V}\times 3\sigma$, where $\delta v$ is the spectral resolution (1.1 km s$^{-1}$), $\sigma$ is the rms noise of the background, and $\Delta V$ is the expected line width (put to 5 km s$^{-1}$).

For lines in absorption, the integral in optical depth $\tau$ over velocity is given, where $\tau(V)=\ln [T_b/T_{MB}(V)]$. $T_b$ is the background main beam temperature and $T_{MB}(V)$ the observed main beam temperature at velocity $V$. For $\tau\ll 1$, the upper limit of an undetected absorption line amounts to the integrated line intensity below background divided by the background. Thus the upper limit for an undetected absorption line is $\int \tau(V) dV$ $\approx$ 1.2 $\sqrt{\delta V\Delta v}\times 3\sigma/T_b$. It assumes 20 $\%$ calibration uncertainty.

Level column densities are extracted by integrating the observed lines or absorption profiles, neglecting re-emission or re-absorption of the final state (thus optical depth $\tau \ll 1$) and applying either Eq.(3) or (4). The column density of a particular level was extended to the {\it total column density} of a species $i$,  $N_i$, summed over all levels using Eq.(6) and assuming an appropriate excitation temperature (given also in Tables B.1 - B.3). The reasons for the assumed value of $T_{\rm ex}$  are discussed in the text (beginning of Section 4).  The errors introduced based on an expected range of $T_{\rm ex}$, as well as by noise and background subtraction are given in Table \ref{table_errors} by factors that define the error range of column densities around the value given in Tables B.1 -- B.3. The column densities are not corrected for beam filling or optical thickness.

\newpage

\vskip-0.1cm
\begin{table*}[]
\begin{center}
\caption{Observed line parameters of objects}
\begin{tabular}{lrrrrrrrrr}   
\hline \hline
\\
&$T_{max}$&rms&$T_b$&$\Delta V$&$V_{line}$&$\delta V$&$\int T_{MB} dV$&$T_{\rm ex}$&$N_i$\\
&[K]&[K]&[K]&[km s$^{-1}$]&[km s$^{-1}$]&[km s$^{-1}$]&$@=\int \tau dV$&[K]&[cm$^{-2}$]\\
\hline
{\bf CH$^+$}(1-0)\\
 835.1375 GHz\\
\\
NGC1333 I2A&$-$0.088&0.015&0.14&4.9&5.6&$-$1.9&3.8$^@$&9&1.1(13)\\
NGC1333 I4A&$-$0.071&0.016&0.21&5.6&5.8&$-$1.1&2.7$^@$&9&8.0(12)\\
NGC1333 I4B&$-$0.032&0.014&0.11&6.0&6.0&$-$1.0&1.3$^@$&9&2.9(12)\\
Ser SMM1&$-$0.096&0.013&0.25&7.3&4.1&$-$4.1&5.8$^@$&9&1.7(13)\\
L 1489&$>$-0.030&0.010&0.014&$-$&$-$&$-$&$<$6.3$^@$&9&$<$1.9(13)\\
NGC7129 FIRS2&$-$0.043&0.017&0.14&14.5&$-$5.4&4.5&2.4$^@$&9&7.0(12)\\

\\
W3 IRS5&$-$1.3&0.009&0.89&7.1&$-$38.5&$-$0.1&10.0$^@$&8$^a$&4.4(13)$^a$\\
&2.1&0.009&0.89&3.0&$-$34.5&3.9&9.8&38$^a$&9.6(12)$^a$\\
NGC6334 I&$-$3.6&0.030&3.7&5.9&$-$2.6&5.1&$>$48$^@$&7$^a$&2.6(14)$^a$ DC\\
&$-$2.2&0.030&3.7&4.5&$-$11&$-$3.3&1.6$^@$&12$^a$&5.7(13)$^a$\\
NGC6334 I(N)&$-$2.5&0.014&2.3&3.5&$-$1.9&2.6&$<$34$^@$&9&$\le$1.0(14)\\
&$-$1.5&0.014&2.3&7.1&6.5&11.0&4.9$^@$&9&1.5(13) DC\\
AFGL 2591&$-$0.28&0.010&0.47&10.3&$-$16.0&$-$10.5&8.7$^@$&3$^a$&1.8(14)$^a$\\
&0.33&0.010&0.47&13.5&$-$7.5&$-$2.0&0.86&43$^a$&8.5(12)$^a$\\
S 140&$-$0.30&0.030&$-$&2.7&$-$9.5&$-$2.4&2.1 - 20.1$^@$&9&0.6 - 5.9(13)\\
&0.27&0.030&$-$&1.9&$-$7.0&0.1&$<$0.38 - 2.4&38&0.16 - 1.0(12)\\
NGC7538 IRS1&0.63&0.014&0.91&13.5&$-$57.5&$-$0.1&9.3&44$^a$&4.1(12)$^a$\\
&$-$0.90&0.014&0.91&8.8&$-$51.8&5.6&8.7$^@$&3$^a$&1.1(14)$^a$\\
\hline
{\bf CH$^+$}(2-1)\\
1669.2813 GHz\\
\\
NGC1333 I2A&$<$0.17&0.058&&$-$&$-$&$-$&$<$0.48&&\\
NGC1333 I4A&$<$0.17&0.058&&$-$&$-$&$-$&$<$0.49&&\\
NGC1333 I4B&$<$0.18&0.061&&$-$&$-$&$-$&$<$0.51&&\\
Ser SMM1&$<$0.22&0.074&&$-$&$-$&$-$&$<$0.63&&\\
\\
W3 IRS5&1.1&0.045&4.1&7.7&$-$35.9&2.5&8.7&&\\
NGC6334 I&$-$2.8&0.12&11.4&3.9&$-$10.3&$-$2.6&1.15$^@$&&\\
&$-$0.92&0.16&11.4&4.5&$-$2.5&5.2&0.36$^@$&&\\
NGC6334 I(N)&$<$0.31&0.11&1.7&$-$&$-$&$-$&$<$0.87&&\\
AFGL 2591&0.85&0.077&2.3&8&$-$4.6&0.9&3.7&&\\
NGC7538 IRS1&0.41&0.12&2.9&10.7&$-$57.7&0.3&2.7&&\\
\hline
{\bf OH$^+$} (1-0)\\ 1033.1186 GHz\\
\\
NGC1333 I2A&$<$0.055&0.018&0.26&$-$&$-$&$-$&$<$-0.59$^@$&9&$<$5.2(12)\\
NGC1333 I4A&$-$0.040&0.020&0.25&1.4&4.1&$-$2.8&$\le$0.2$^@$&9&$\le$1.9(12)\\
NGC1333 I4B&$-$0.043&0.019&0.17&1.7&1.8&$-$5.2&$\le$0.63$^@$&9&$\le$5.7(12)\\
Ser SMM1&$-$0.20&0.065&0.43&9.7&4.8&$-$3.7&8.3$^@$&9&7.3(13)\\
L 1489&$<$0.50&0.017&0.037&$-$&$-$&$-$&$<$3.9$^@$&9&$<$3.4(13)\\
NGC7129 FIRS2&$-$0.024&0.021&0.22&7.4&$-$1.9&8.0&1.0$^@$&9&8.8(12)\\
\\
W3 IRS5&$-$0.2&0.015&1.7&5.9&$-$42&$-$3.6&0.99$^@$&9&8.8(12)\\
&0.1&0.015&1.7&9.4&$-$33&5.4&0.50&38&5.9(11)\\
NGC6334 I&$-$0.55&0.031&6.0&5.9&$-$10.0&$-$2.3&0.66$^@$&9&5.2(12)\\
&$-$2.1&0.031&6.0&6.4&$-$1.6&6.1&2.8$^@$&9&2.3(13) DC\\
NGC6334 I(N)&$-$0.99&0.021&3.0&4.9&$-$1.6&2.9&3.8$^@$&9&3.3(13)\\
&$-$1.35&0.021&3.0&3.5&3.2&7.7&3.7$^@$&9&3.3(13) DC\\
AFGL 2591&$-$0.14&0.016&0.90&11.8&$-$16.1&$-$10.6&1.0$^@$&3$^a$&1.6(13)$^a$\\
&$-$0.71&0.016&0.90&9.4&2.8&8.3&3.6$^@$&3$^a$&6.1(13)$^a$ DC\\
S 140&$-$0.30&0.024&0.96&4.5&$-$4.1&3.0&1.5$^@$&9&1.8(13)\\
&$-$0.18&0.024&0.96&6.7&3.5&10.6&1.3$^@$&9&1.5(13) DC\\
NGC7538 IRS1&$-$0.18&0.026&1.6&12.0&$-$48&9.4&1.6$^@$&9&1.4(13)\\
\hline
\end{tabular}
\end{center}
{\tiny {\bf Notes.} $T_{max}$ refers to the line peak intensity, negative values indicate absorption in [K] below background; $\Delta V$ to the FWHM line width; $V_{line}$ to the line mean velocity; $\delta V = V_{line}-V_{\mathrm{LSR}}$ to the line shift with respect to the systemic velocity of the object; rms to the root mean square noise level; $T_b$ to the background (continuum) intensity; the integrated line flux (line luminosity) is given in [K km s$^{-1}$]), $@$ indicates integrated optical depth [km s$^{-1}$]; $T_{\rm ex}$ is the assumed excitation temperature; and  $N_i$ is the total column density. Cases where the fitted line model had to be used for intensity or absorption instead of the observed data because of line blending are marked with an asterisk (*), and values marked (a) are from 1D slab model fitting \citep[][ Appendix B]{2010ApJ...720.1432B}. A minus sign (-) indicates observations but no detection. Absorption components marked with ``DC" are suspected to originate from diffuse interstellar clouds in the foreground. The error range in emission measure is up to factors of 0.8 - 2 for emission  lines, and factors of 0.8 - 1.5 for lines in absorption. Details and error margins of the column densities are given in Section 4 and Table 3.}

\label{table_{linelist}}
\end{table*}
\newpage
\pagebreak[4]

\begin{table*}[h]
\begin{center}
\caption{Continuation of Table B.1}
\begin{tabular}{lrrrrrrrrr}   
\hline \hline
\\
&$T_{max}$&rms&$T_b$&$\Delta V$&$V_{line}$&$\delta V$&$\int T_{MB} dV$&$T_{\rm ex}$&$N_i$\\
&[K]&[K]&[K]&[km s$^{-1}$]&[km s$^{-1}$]&[km s$^{-1}$]&$@=\int \tau dV$&[K]&[cm$^{-2}$]\\
\hline\\
{\bf H$_2$O$^+$}\\
1115.2041 GHz\\
\\
NGC1333 I2A&$<$0.069&0.023&0.34&$-$&$-$&$-$&$<$0.57$^@$&9&$<$4.9(12)\\
NGC1333 I4A&$<$0.060&0.020&0.29&$-$&$-$&$-$&$<$0.58$^@$&9&$<$4.9(12)\\
NGC1333 I4B&$<$0.069&0.023&0.20&$-$&$-$&$-$&$<$0.97$^@$&9&$<$8.5(12)\\
Ser SMM1&$<$0.071&0.024&0.52&5.5&5.0&$-$3.5&$<$0.52$^@$&9&$\leq$4.7(12)\\
L 1489&$<$0.065&0.022&0.033&$-$&$-$&$-$&$<$5.60$^@$&9&$<$5.1(13)\\
NGC7129 FIRS2&$<$0.060&0.020&0.25&$-$&$-$&$-$&$<$0.68$^@$&9&$<$6.0(12)\\
\\
W3 IRS5&$-$0.10&0.018&2.0&5.3&$-$42.7&$-$4.3&0.36$^@$&9&3.2(12)\\
NGC6334 I&$<$0.054&0.018&7.3&$-$&$-$&$-$&$<$0.070$^@$&9&$<$6.3(11)\\
NGC6334 I(N)&$<$0.070&0.023&3.1&$-$&$-$&$-$&$<$0.063$^@$&9&$<$1.9(12)\\
AFGL 2591&$-$0.27&0.017&1.1&13.5$^a$&$-$16.4$^a$&$-$10.9$^a$&0.24$^@$&8$^a$&5.4(12)$^a$\\
&$-$0.57&0.017&1.1&4.5$^a$&2.2$^a$&7.7$^a$&0.37$^@$&3$^a$&8.3(12)$^a$ DC\\
S 140&$<$0.096&0.032&1.0&$-$&$-$&$-$&$<$0.27$^@$&9&$<$1.8(12)\\
NGC7538 IRS1&$\le$0.042&0.018&1.9&4.5&$-$66.1&$-$9.9&$\le$0.057$^@$&9&$\le$5.1(11)\\
\hline\\
{\bf H$_3$O$^+$}\\
1031.2995 GHz\\
\\
NGC1333 I2A&$<$0.069&0.019&0.26&$-$&$-$&$-$&$<$0.18&225&$<$1.7(12)\\
NGC1333 I4A&$<$0.069&0.020&0.25&$-$&$-$&$-$&$<$0.19&225&$<$2.0(12)\\
NGC1333 I4B&$<$0.053&0.017&0.17&$-$&$-$&$-$&$<$0.14&225&$<$1.5(12)\\
Ser SMM1&$<$0.063&0.019&0.43&$-$&$-$&$-$&$<$0.17&225&$<$1.8(12)\\
L 1489&$<$0.066&0.017&0.036&$-$&$-$&$-$&$<$0.18&225&$<$1.9(12)\\
NGC7129 FIRS2&$<$0.054&0.017&0.23&$-$&$-$&$-$&$<$0.15&225&$<$1.5(12)\\
\\
W3 IRS5&0.61&0.017&1.7&5.7&$-$37.0&1.4&3.8*&225&3.9(13)\\
NGC6334 I&0.48&0.015&5.9&4.6&$-$4.8&2.9&2.4*&225&2.5(13)\\
NGC6334 I(N)&0.048&0.018&3.0&11&$-$5.2&$-$0.7&$\le$0.89*&225&$\le$9.2(12)\\
AFGL 2591&0.090&0.0095&0.90&7.4&$-$4.6&0.9&0.71*&225&7.4(12)\\
S 140&0.046&0.016&0.96&3.2&$-$5.5&1.6&$<$0.2*&225&$<$2.1(12)\\
NGC7538 IRS1&$<$0.061&0.020&1.5&$-$&$-$&$-$&$<$0.17&225&$<$1.8(12)\\
\hline\\
{\bf SH$^+$}\\
526.0479 GHz\\
\\
NGC1333 I2A&$<$0.011&0.0037&0.028&$-$&$-$&$-$&$<$0.031&38&$<$5.5(10)\\
NGC1333 I4A&$<$0.013&0.0044&0.082&$-$&$-$&$-$&$<$0.037&38&$<$6.5(10)\\
NGC1333 I4B&$<$0.014&0.0046&0.0051&$-$&$-$&$-$&$<$0.039&38&$<$6.9(10)\\
Ser SMM1&$<$0.015&0.0051&0.085&$-$&$-$&$-$&$<$0.043&38&$<$7.6(10)\\
L 1489&$<$0.016&0.0053&0.009&$-$&$-$&$-$&$<$0.045&38&$<$8.0(10)\\
NGC7129 FIRS2&$<$0.017&0.0057&0.036&$-$&$-$&$-$&$<$0.048&38&$<$8.5(10)\\
\\
W3 IRS5&0.062&0.0055&0.29&5.2&$-$38.9&$-$0.50&0.56*&38&9.5(11)\\
NGC6334 I&$<$0.026&0.0087&0.81&$-$&$-$&$-$&$<$0.073&38&$<$1.3(11)\\
NGC6334 I(N)&$<$0.015&0.0050&0.077&$-$&$-$&$-$&$<$0.042&38&$<$7.4(10)\\
AFGL 2591&$<$0.029&0.0097&0.11&$-$&$-$&$-$&$<$0.082&38&$<$1.5(11)\\
NGC7538 IRS1&$<$0.014&0.0048&0.26&$-$&$-$&$-$&$<$0.041&38&$<$7.3(10)\\
\hline
\\
\end{tabular}
\end{center}
\vskip-0.1cm
{\tiny {\bf Notes.} Values marked (a) are from 1D slab model fitting by \citet{2010A&A...521L..44B}. H$_2$O$^+$ column densities are total values assuming an ortho-to-para ratio of 3. For the symbols see Table B.1.}
\label{table_{linelist1}}
\end{table*}
\newpage
\pagebreak[4]

\begin{table*}[h]
\begin{center}
\caption{Continuation of Table B.1}
\begin{tabular}{lrrrrrrrrr}   
\hline \hline
\\
&$T_{max}$&rms&$T_b$&$\Delta V$&$V_{line}$&$\delta V$&$\int T_{MB} dV$&$T_{\rm ex}$&$N_i$\\
&[K]&[K]&[K]&[km s$^{-1}$]&[km s$^{-1}$]&[km s$^{-1}$]&$@=\int \tau dV$&[K]&[cm$^{-2}$]\\
\hline\\
{\bf HCO$^+$}\\
535.0615 GHz\\
\\
NGC1333 I2A&0.78&0.0038&0.028&1.7&7.5&0.0&2.3&38&1.5(12)\\
&0.20&0.0038&0.028&3.9&7.9&0.4&&\\
NGC1333 I4A&0.86&0.0050&0.078&2.1&6.9&0.0&3.6&38&2.3(12)\\
&0.15&0.0050&0.078&7.1&8.5&1.6&&\\
NGC1333 I4B&0.60&0.0052&0.050&2.4&6.9&$-$0.1&2.5&38&1.6(12)\\
&0.099&0.0052&0.050&8.1&7.3&0.3&&\\
Ser SMM1&1.1&0.0057&0.078&2.9&8.3&$-$0.2&6.0&38&3.8(12)\\
&0.15&0.0057&0.078&14&8.2&$-$0.3&&\\
L 1489&0.078&0.0033&0.0021&3.3&6.9&$-$0.3&0.45&38&3.0(11)\\
&0.024&0.0033&0.0021&9.4&6.2&$-$1.0&&\\
NGC7129 FIRS2&0.42&0.0055&0.036&2.5&$-$9.7&0.2&1.6&38&1.0(12)\\
&0.046&0.0055&0.036&8.3&$-$9.3&0.6&&\\
\\
W3 IRS5&6.1&0.0063&0.29&5.3&$-$38.1&0.3&44&38&2.8(13)\\
&0.48&0.0063&0.29&18&$-$38.9&$-$0.5&&\\
NGC6334 I&4.0&0.0055&0.80&6.1&$-$7.2&0.5&29&38&1.9(13)\\
&0.095&0.0055&0.80&30&$-$5.7&2.0&&\\
NGC6334 I(N)&2.3&0.0057&0.77&4.2&$-$5.5&$-$1.0&23&38&1.5(13)\\
&1.9&0.0057&0.77&5.3&$-$1.2&3.3&&\\
AFGL 2591&1.9&0.010&0.11&3.6&$-$5.8&$-$0.3&9.7&38&6.3(12)\\
&0.30&0.010&0.11&7.7&$-$7.0&$-$1.5&&\\
S 140&7.2&0.013&0.026&3.5&$-$6.9&0.2&33*&38&2.1(13)\\
&0.49&0.013&0.026&12&$-$6.2&0.9&&\\
NGC7538 IRS1&4.3&0.0041&0.26&3.7&$-$57.4&0.0&25&38&1.6(13)\\
&0.78&0.0041&0.26&8.7&$-$58.1&$-$0.7&&\\
\hline\\
{\bf C$^+$}\\
1900.5369 GHz\\
\\
NGC1333 I2A&$-$0.18&0.076&0.52&7.2&3.2&$-$4.3&2.8$^@$&38&4.7(17)\\
NGC1333 I4A&$-$0.040&0.059&0.70&5.8&1.1&$-$5.8&$\le$0.69$^@$&38&$\le$1.2(17)\\
NGC1333 I4B&$<$2.4&0.081&0.23&$-$&$-$&$-$&$<$2.8$^@$&38&$<$4.7(17)\\
Ser SMM1&$-$0.18&0.076&1.1&5.1&1.3&$-$7.2&0.93$^@$&38&1.6(17)\\
\\
W3 IRS5&42&0.055&4.3&5.9&$-$33.8&4.6&$>$310&38&$>$5.6(18)\\
&13&0.055&4.3&2.4&$-$43.7&$-$5.3&$-$&&\\
AFGL 2591&30&0.172&3.0&6.8&$-$4.2&1.3&$>$280&38&$>$5.0(18)\\
&12&0.172&3.0&5.9&$-$15.9&$-$10.4&$-$&&\\
S 140&12&0.78&12.5&4.3&$-$8.5&$-$1.4&$>$28&38&$\ge$5.0(17)\\
\hline\\
{\bf CH}\\
536.7611 GHz\\
\\
NGC1333 I2A&0.017&0.0024&0.027&1.8&7.1&$-$0.4&0.031&38&1.9(11)\\
NGC1333 I4A&0.040&0.0026&0.078&0.70&6.8&$-$0.1&0.068&38&4.3(11)\\
NGC1333 I4B&0.029&0.0025&0.053&0.59&6.8&$-$0.2&0.051&38&3.2(11)\\
Ser SMM1&0.028&0.0022&0.086&0.61&7.8&$-$0.7&0.049&38&3.1(11)\\
&0.012&0.0022&0.086&9.4&9.0&0.5&0.094&38&6.3(12)\\
L 1489&$<$0.017&0.0058&0.00099&$-$&$-$&$-$&$<$0.022&38&$<$1.4(11)\\
NGC7129 FIRS2&0.013&0.0025&0.040&7.2&$-$4.9&5.0&0.14&38&8.8(11)\\
\\
W3 IRS5&1.0&0.0060&0.28&4.7&$-$36.3&2.1&14&38&8.8(13)\\
NGC6334 I&0.28&0.014&0.80&2.6&$-$8.5&$-$0.8&2.1&38&1.3(13)\\
NGC6334 I(N)&$-$3.0&0.0059&0.7&3.3&$-$2.0&2.5&3.3$^@$&6&1.2(14)\\
AFGL 2591&0.26&0.0086&0.10&4.7&$-$5.5&0.0&2.9&38&1.8(13)\\
S 140&0.22&0.0061&0.021&3.5&$-$6.6&0.5&1.6&38&1.0(13)\\
NGC7538 IRS1&0.25&0.0059&0.26&4.8&$-$58.3&$-$0.9&4.0&38&2.5(13)\\\hline
\\
\end{tabular}
\end{center}
{\tiny {\bf Notes.} Column densities of HCO$^+$ (6-5) are lower limits because of high optical thickness (see Section 3). For the symbols see Table B.1.}
\end{table*}

\newpage
\pagebreak[4]
\clearpage

\section{Details of Correlation Analysis}

\begin{figure*}[]
\centering
\centering
\resizebox{17cm}{!}{\includegraphics{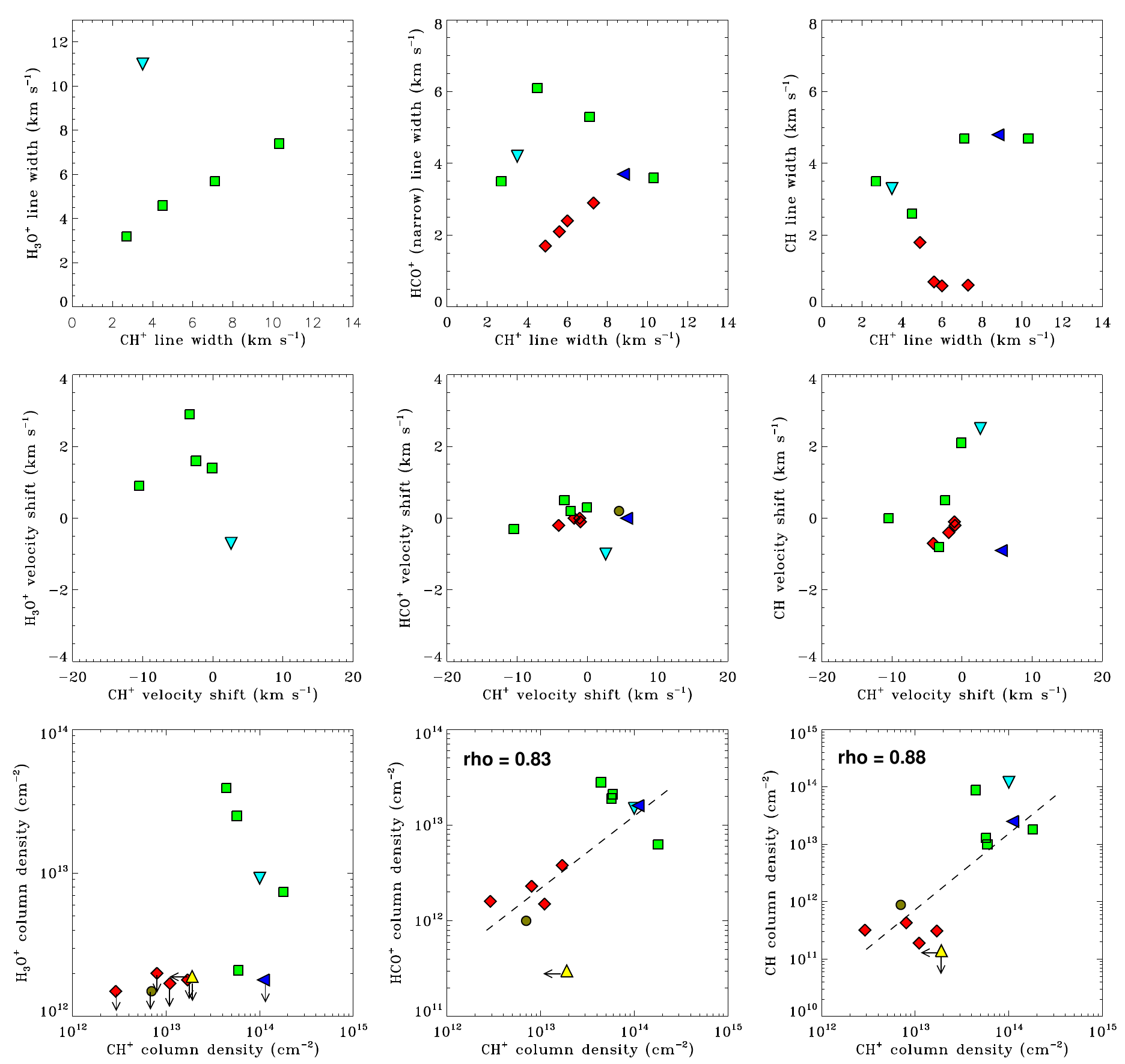}}
\caption{Observed line parameters (Tables B.2 and B.3) of H$_3$O$^+$, HCO$^+$ (narrow component), and CH vs. the observed line characteristics of CH$^+$. The Pearson correlation coefficient $\rho$ is given where statistically significant (see Section 4.2). The upper limit for L 1489 is not included in the analysis. The symbols mark different types of objects: Red diamonds for Class 0, yellow triangle up for Class I (L 1489), brown circle for intermediate mass, light blue triangle down for high-mass mid-IR quiet (NGC6334 I(N)), green square for high-mass mid-IR bright and hot molecular core, and dark blue triangle left for high-mass ultra-compact H{\tiny{\textsc II}} (NGC7538 IRS1).}
\label{H3Op-CHp}
\end{figure*}

\begin{figure*}[h]
\centering
\centering
\resizebox{17cm}{!}{\includegraphics{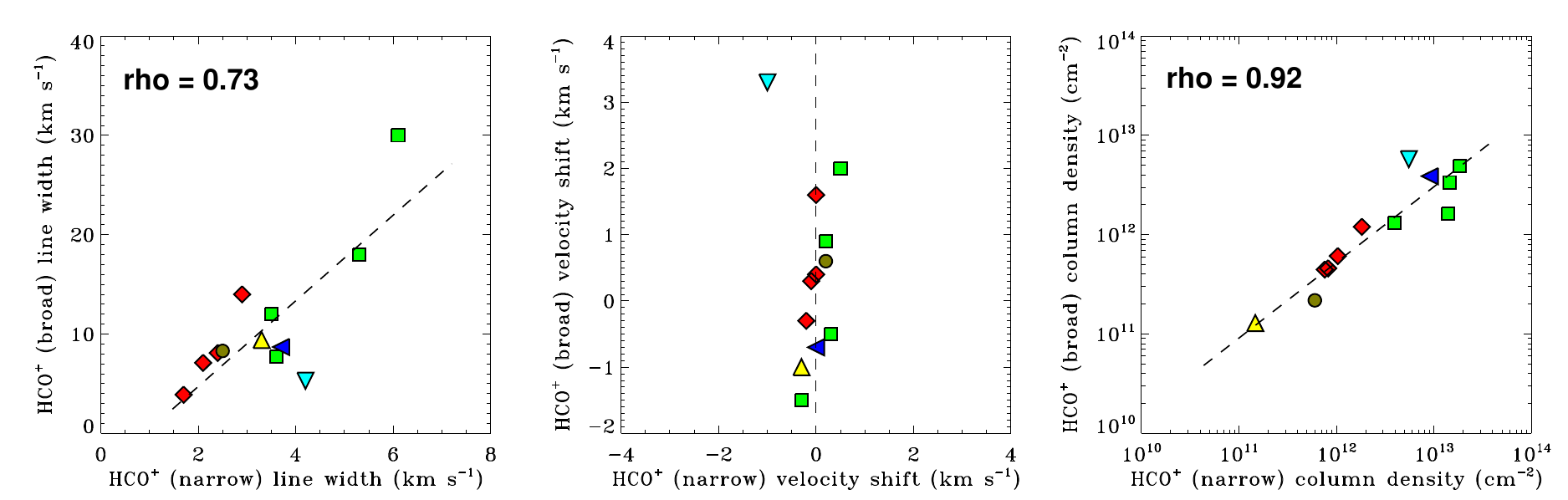}}
\caption{Observed line parameters of the narrow component vs. the broad component of HCO$^+$(6-5) as given in Table B.3. For the notation of the symbols see Fig. \ref{H3Op-CHp}.}
\label{HCOp1-HCOp2}
\end{figure*}

\begin{figure*}[h]
\centering
\centering
\resizebox{17cm}{!}{\includegraphics{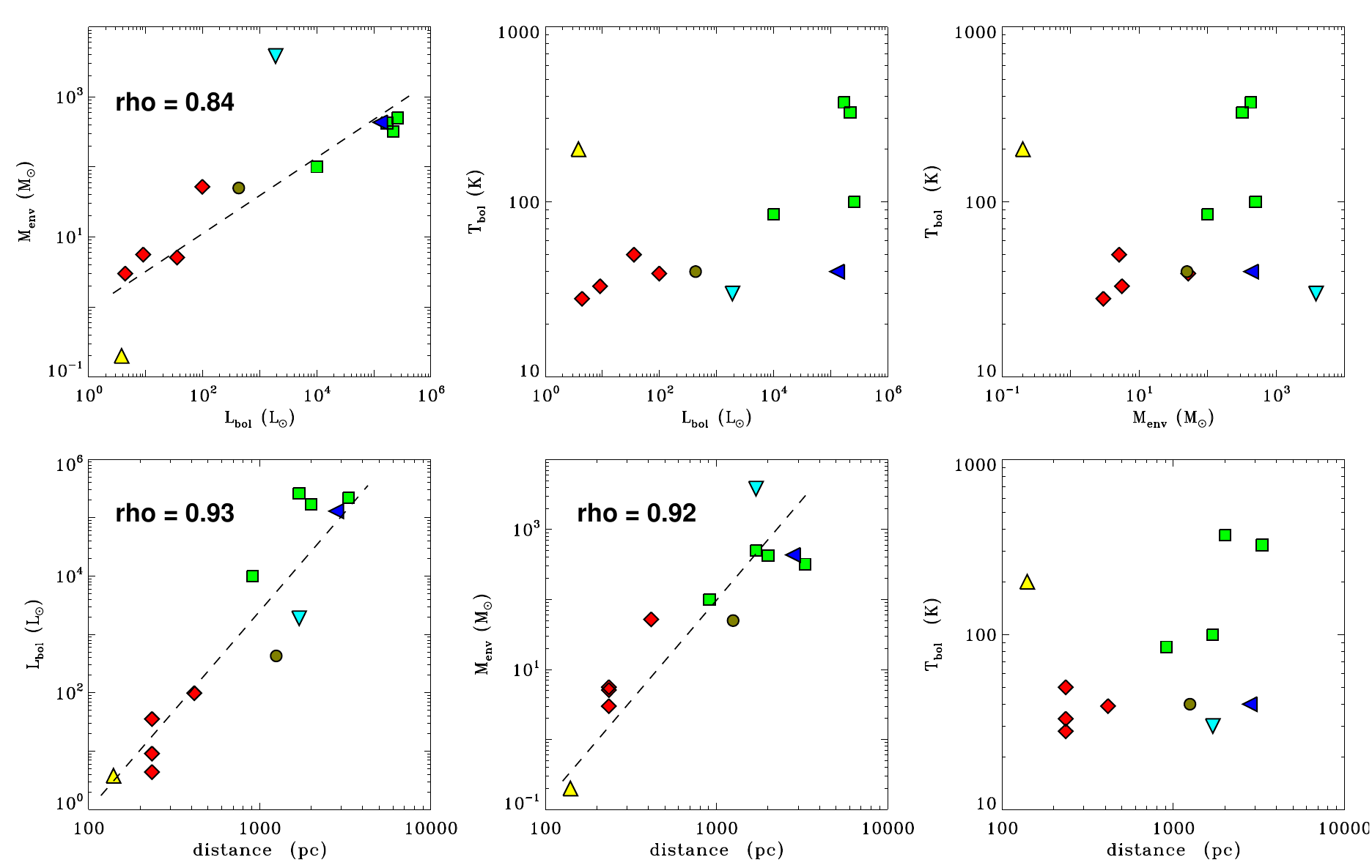}}
\caption{Correlations between object parameters as listed in Table 1.  The symbols mark different types of objects: Red diamonds for Class 0, yellow triangle up for Class I (L 1489), brown circle for intermediate mass, light blue triangle down for high-mass mid-IR quiet (NGC6334 I(N)), green square for high-mass mid-IR bright and hot molecular core, and dark blue triangle left for high-mass ultra-compact H{\tiny{\textsc II}} (NGC7538 IRS1).}
\label{D5}
\end{figure*}

\pagebreak[4]
\newpage

\begin{figure*}[h]
\centering
\centering
\resizebox{17cm}{!}{\includegraphics{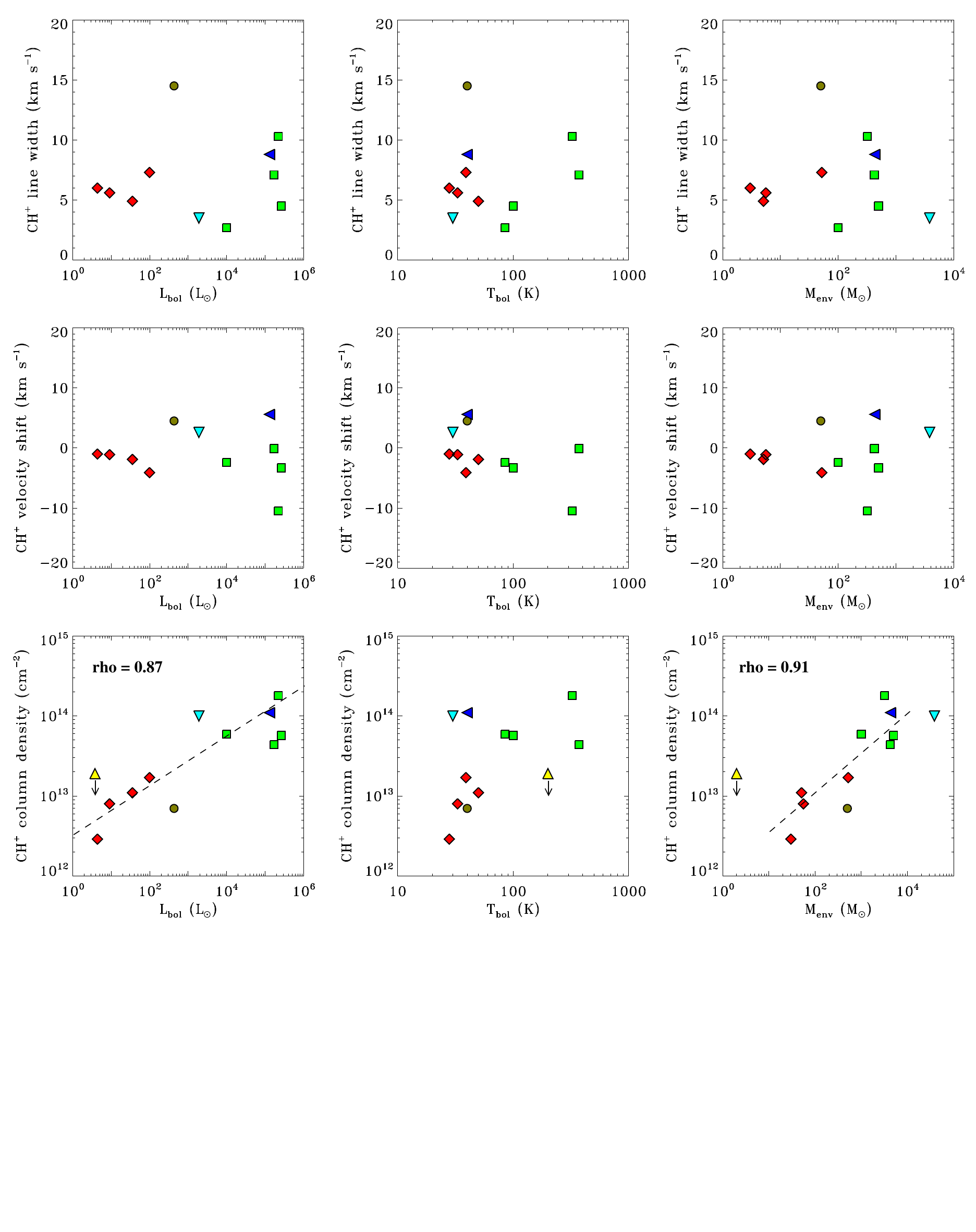}}
\caption{Observed CH$^+$(835 GHz) line characteristics (Table B.1) vs. object parameters given in Table 1. Only the line component in absorption is shown. L 1489 is not included in the correlation coefficient $\rho$, given where statistically significant.  The symbols mark different types of objects: Red diamonds for Class 0, yellow triangle up for Class I (L 1489), brown circle for intermediate mass, light blue triangle down for high-mass mid-IR quiet (NGC6334 I(N)), green square for high-mass mid-IR bright and hot molecular core, and dark blue triangle left for high-mass ultra-compact H{\tiny{\textsc II}} (NGC7538 IRS1).}
\label{CHp}
\end{figure*}

\pagebreak[4]
\newpage

\begin{figure*}[h]
\centering
\centering
\resizebox{17cm}{!}{\includegraphics{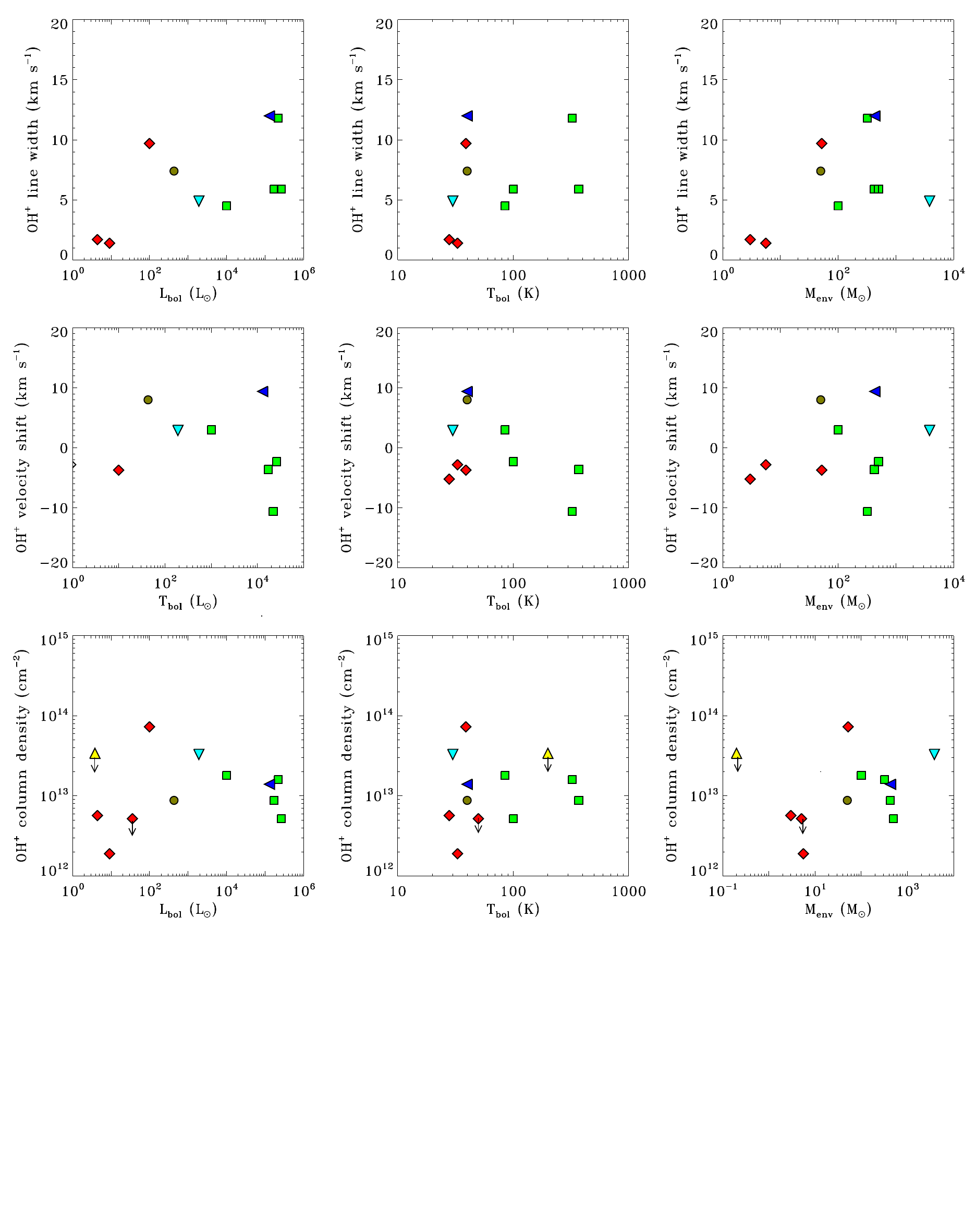}}
\caption{Observed OH$^+$(1033 GHz) line characteristics (Table B.1) vs. object parameters given in Table 1. Only the line component in absorption is considered. The correlations are not statistically significant.  The symbols mark different types of objects: Red diamonds for Class 0, yellow triangle up for Class I (L 1489), brown circle for intermediate mass, light blue triangle down for high-mass mid-IR quiet (NGC6334 I(N)), green square for high-mass mid-IR bright and hot molecular core, and dark blue triangle left for high-mass ultra-compact H{\tiny{\textsc II}} (NGC7538 IRS1).}
\label{OHp}
\end{figure*}
\pagebreak[4]
\newpage

\begin{figure*}[h]
\centering
\centering
\resizebox{17cm}{!}{\includegraphics{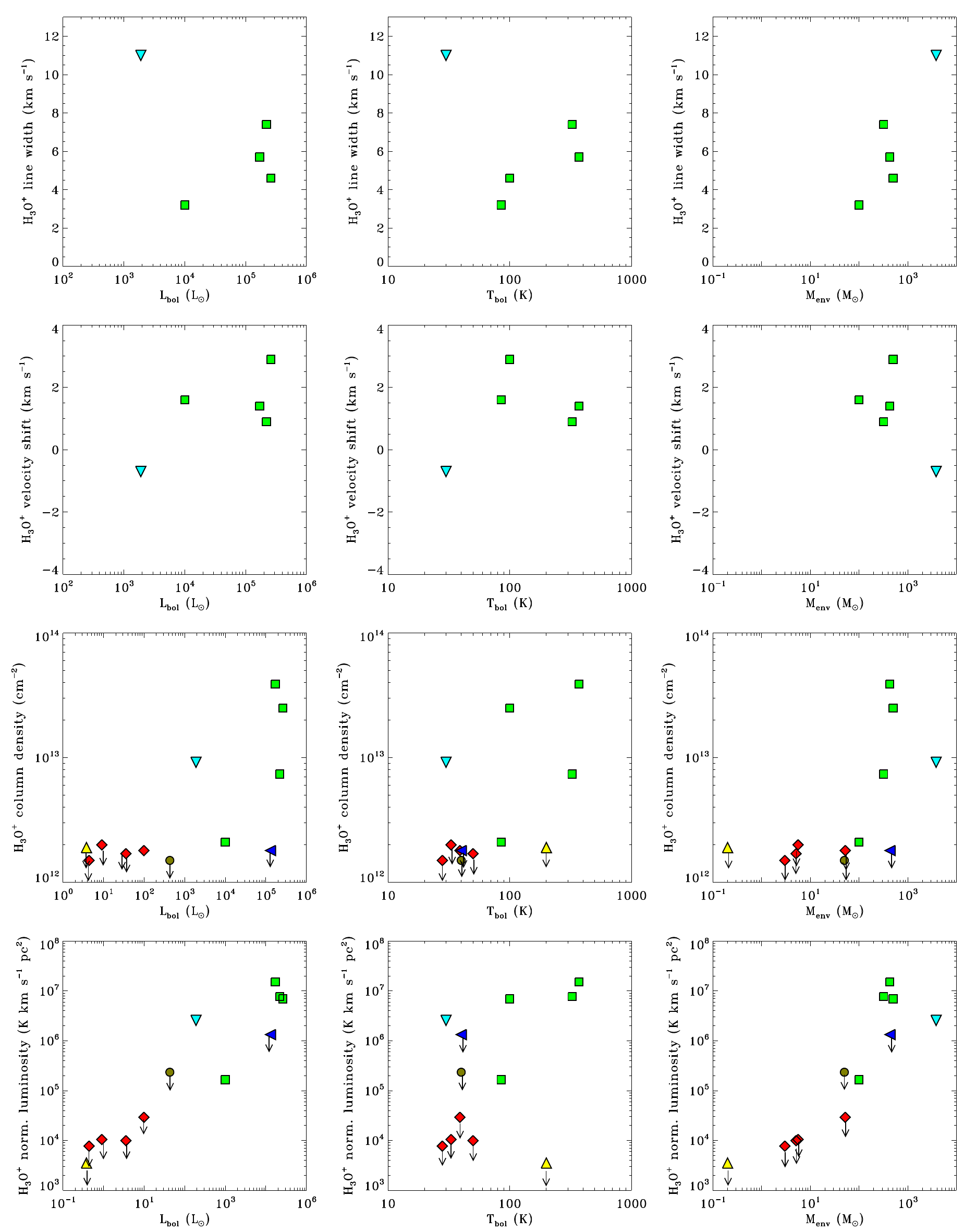}}
\caption{Observed H$_3$O$^+$(1031 GHz) line characteristics (Table B.2) vs. object parameters given in Table 1. The line is in emission. The bottom row shows the line luminosity multiplied with distance squared (normalized to 1 pc) to correct in case of point-source emission. The symbols mark different types of objects: Red diamonds for Class 0, yellow triangle up for Class I (L 1489), brown circle for intermediate mass, light blue triangle down for high-mass mid-IR quiet (NGC6334 I(N)), green square for high-mass mid-IR bright and hot molecular core, and dark blue triangle left for high-mass ultra-compact H{\tiny{\textsc II}} (NGC7538 IRS1).}
\label{H3Opa}
\end{figure*}
\pagebreak[4]
\newpage

\begin{figure*}[h]
\centering
\centering
\resizebox{17cm}{!}{\includegraphics{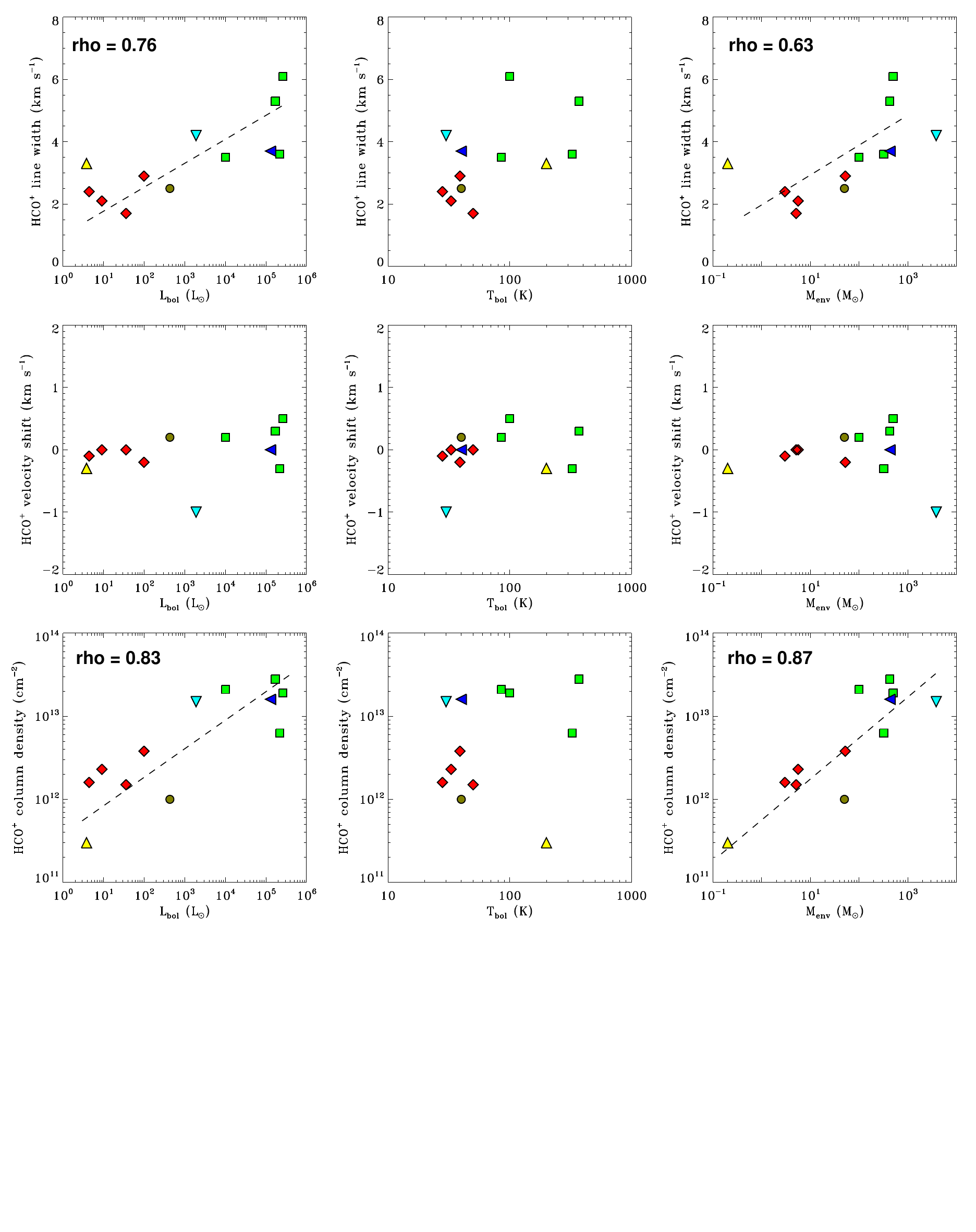}}
\caption{Observed line width and shift of the narrow component of HCO$^+$(6-5) and its total column density (Table B.2) vs. object parameters given in Table 1. The line is in emission. The correlations of intensity normalized in distance are displayed in Fig. \ref{HCOp_norm}. The symbols mark different types of objects: Red diamonds for Class 0, yellow triangle up for Class I (L 1489), brown circle for intermediate mass, light blue triangle down for high-mass mid-IR quiet (NGC6334 I(N)), green square for high-mass mid-IR bright and hot molecular core, and dark blue triangle left for high-mass ultra-compact H{\tiny{\textsc II}} (NGC7538 IRS1).}
\label{HCOp}
\end{figure*}
\pagebreak[4]
\newpage

\begin{figure*}[h]
\centering
\centering
\resizebox{17cm}{!}{\includegraphics{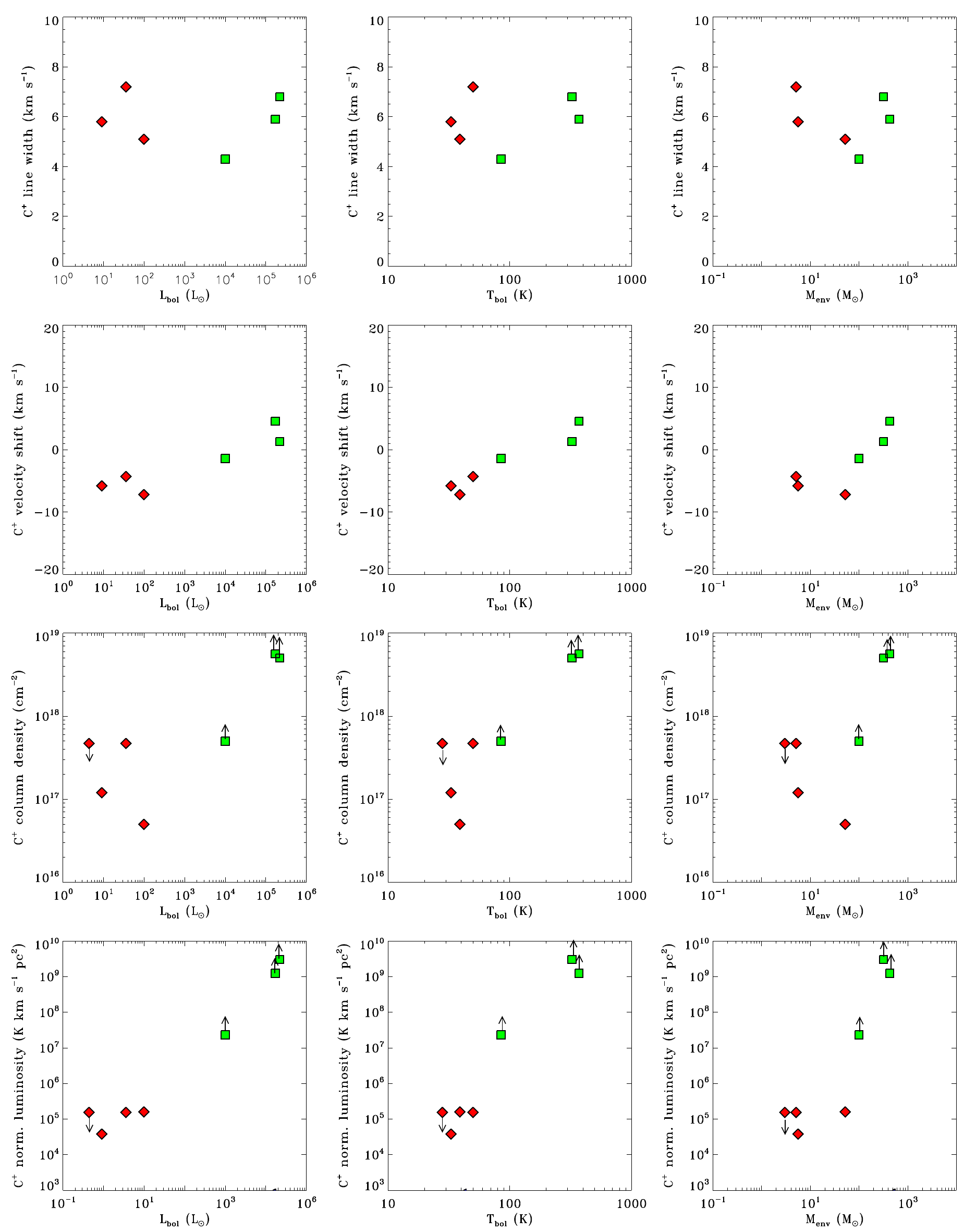}}
\caption{Observed C$^+$(1900 GHz) line characteristics (Table B.1) vs. object parameters given in Table 1. The line is in absorption toward the low-mass objects and in emission toward the high-mass objects. The bottom row shows the luminosity of the transition normalized for the case of point-source emission. The symbols mark different types of objects: Red diamonds for Class 0, yellow triangle up for Class I (L 1489), brown circle for intermediate mass, light blue triangle down for high-mass mid-IR quiet (NGC6334 I(N)), green square for high-mass mid-IR bright and hot molecular core, and dark blue triangle left for high-mass ultra-compact H{\tiny{\textsc II}} (NGC7538 IRS1).}
\label{CII}
\end{figure*}

\pagebreak[4]
\newpage

\begin{figure*}[h]
\centering
\centering
\resizebox{17cm}{!}{\includegraphics{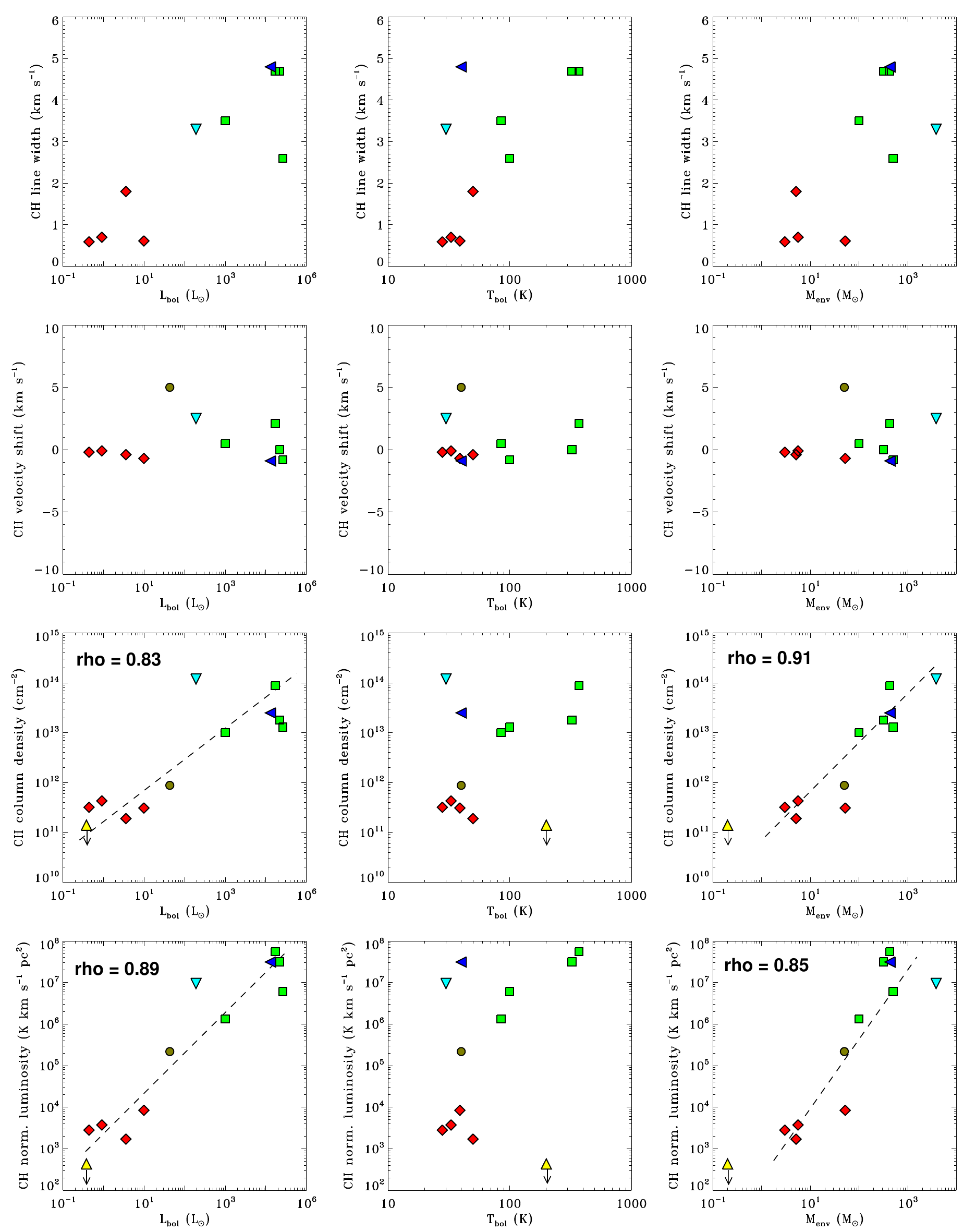}}
\caption{Observed CH (536 GHz) line characteristics (Table B.3) vs. object parameters given in Table 1. Only the line component in emission is considered; for Ser SMM1 only the narrow component. The bottom row shows the luminosity normalized to 1 pc for the case of point-source emission. L 1489 is not included in the correlation coefficient $\rho$, which is given where statistically significant. The symbols mark different types of objects: Red diamonds for Class 0, yellow triangle up for Class I (L 1489), brown circle for intermediate mass, light blue triangle down for high-mass mid-IR quiet (NGC6334 I(N)), green square for high-mass mid-IR bright and hot molecular core, and dark blue triangle left for high-mass ultra-compact H{\tiny{\textsc II}} (NGC7538 IRS1).}
\label{CH}
\end{figure*}

\newpage
\pagebreak[4]
\clearpage

\begin{figure*}[h]
\centering
\sidecaption
\resizebox{8cm}{!}{\includegraphics{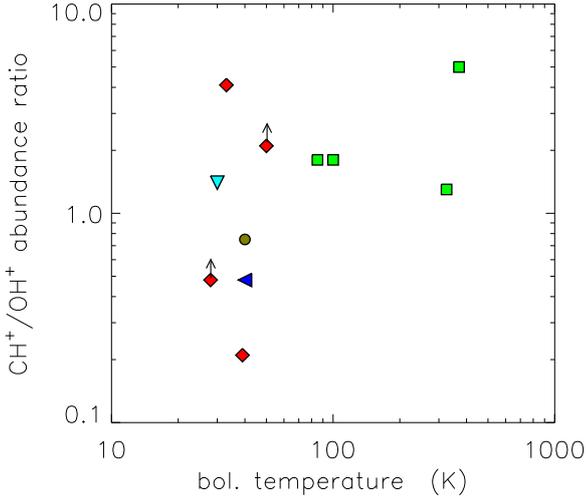}}
\caption{CH$^+$/OH$^+$ column density ratio of absorption components vs. bolometric temperature. L 1489 (Class I) has upper limits in both CH$^+$ and OH$^+$, and is not shown. The symbols mark different types of objects: Red diamonds for Class 0, yellow triangle up for Class I (L 1489), brown circle for intermediate mass, light blue triangle down for high-mass mid-IR quiet (NGC6334 I(N)), green square for high-mass mid-IR bright and hot molecular core, and dark blue triangle left for high-mass ultra-compact H{\tiny{\textsc II}} (NGC7538 IRS1). \vskip1cm}
\label{CHp_OHp_ratio_vsTbol}
\end{figure*}

\vskip4cm

\begin{table*}[h]
\begin{center}
\vskip4cm
\caption{List of observations according to the {\it Herschel} Science Archive (last 6 digits of ObsID, preceding 1342 omitted)}
\begin{tabular}{lcccccccccc}   
\hline \hline
Object &CH$^+$&CH$^+$&OH$^+$&H$_2$O$^+$&H$_3$O$^+$&SH$^+$&HCO$^+$&CH&C$^+$& \\
 &1 - 0&2 - 1&1$_1$ - 0$_1$&1$_{11}$ - 0$_{00}$&4$_{30}$ - 3$_{31}$&$1_2 - 0_1$&6 - 5&1$_{-1}$ - 1$_{1}$&$^2P_{\frac{3}{2}}$ - $^2P_{\frac{1}{2}}$ \\
\hline\\
NGC1333 I2A&203229&215966&203180&191657&203180&202024&202024&202024&201840\\
&&&&&&&&192206&\\
NGC1333 I4A&203230&203951&203181&191656&203181&202023&202023&202023&201841\\
&&&&&&&&192207&\\
NGC1333 I4B&203228&203952&203178&191655&203178&202022&202022&202022&201842\\
&&&&&&&&202033&\\
Ser SMM1&207620&207660&207656&207379&207656&207581&207581&207581&208575\\
&&&&&&&&194463&\\
L 1489&203227&&203159&203938&203159&203187&203187&\\
NGC7129 FIRS2&201704&&197969&191676&197969&198330&198330&192362&\\
&&&&&&&&&\\
NGC6334 I&214305&214454&204514&206385&204514&205281&205281&\\
&&&204515&&&&&&\\
S 140&196463&&197966&200762&197966&&195050&195050&190781\\
&&&197967&&&&&&\\
AFGL 2591&196471&196531&195021&196429&195021&194484&194484&194484&195118\\
&196472&&195022&196430&195022&&&&195119\\
W3 IRS5&201701&201763&191608&191661&191608&191501&191501&191501&191774\\
&201702&201764&191609&191662&191609&191502&191502&191502&191775\\
NGC6334 I(N)&214306&214455&204516&206383&204516&205280&205280&205280&\\
&214307&&204517&&204517&&&&\\
NGC7538 IRS1&201703&200758&197962&191663&197962&198331&198331&198331&\\
&&&197963&197976&197963&&197976&&\\
\hline
\end{tabular}
\end{center}
\label{table_obsIDs}
\end{table*}

\newpage
\pagebreak[4]
\clearpage

\section{HCO$^+$ Correlation with $L_{bol}$}

\begin{figure}[htb!]
\center
\includegraphics[width=1.0\hsize]{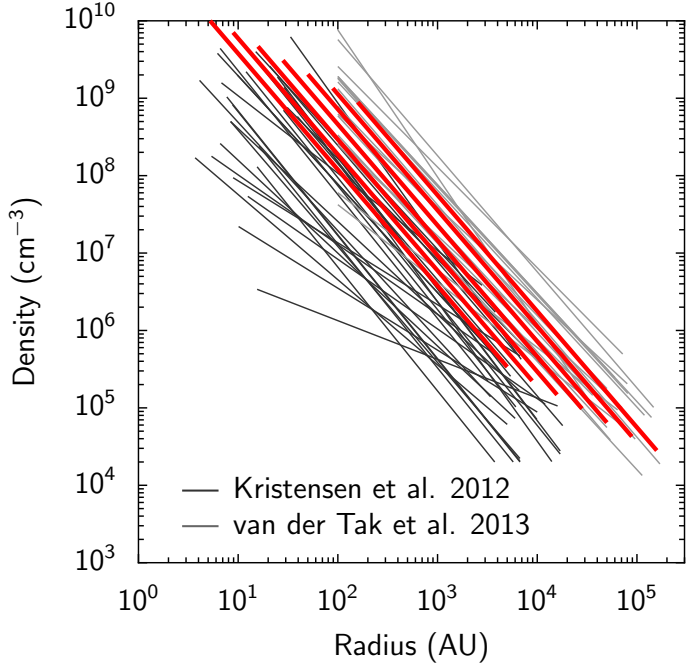}
\caption{Density profiles derived from observations (grey: high mass, black: low mass) compared to the model density profiles (red), which represent 7 bolometric luminosities of $L_{\rm bol} = \{ 1, 10, \ldots 10^6 \}$ $L_\odot$, increasing to the right. }
\label{fig:hcop_correl1}
\end{figure}

The correlation between the bolometric luminosity $L_{\rm bol}$ and the HCO$^+$($J=6-5$) distance-corrected integrated line intensity is excellent (Fig. \ref{HCOp_norm}). To explain the tight correlation, a set of radiative transfer models is run. We assume a spherically symmetric envelope with a power-law density profile
\begin{equation}
n(r) = n_0 \left(\frac{r}{r_{\rm out}} \right)^\alpha
\end{equation}
and a radial range between $r_{\rm in}$ and $r_{\rm out}$. To roughly approximate the envelope properties constrained by \citet{2012A&A...542A...8K} toward low-mass stars and \citet{2013A&A...554A..83V} toward high-mass stars, we set $\alpha =\ -1.5$, $r_{\rm in} = 5 \, L_{\rm bol}^{1/4}$ AU, and $r_{\rm out} / r_{\rm in} = 1000$. The parameter $n_0$ is adjusted to reproduce the $M_{\rm env} = 1.1 L_{\rm bol}^{0.54}$ relation presented in Fig. \ref{D5}, top left, and given in Eq. (\ref{lum_mass-corr}). Figure \ref{fig:hcop_correl1} shows the resulting density profiles, which lie within the range of profiles derived from observations. The largest deviation  at low density is for the (low envelope mass) Class I object.

\begin{figure}[htb!]
\center
\includegraphics[width=1.0\hsize]{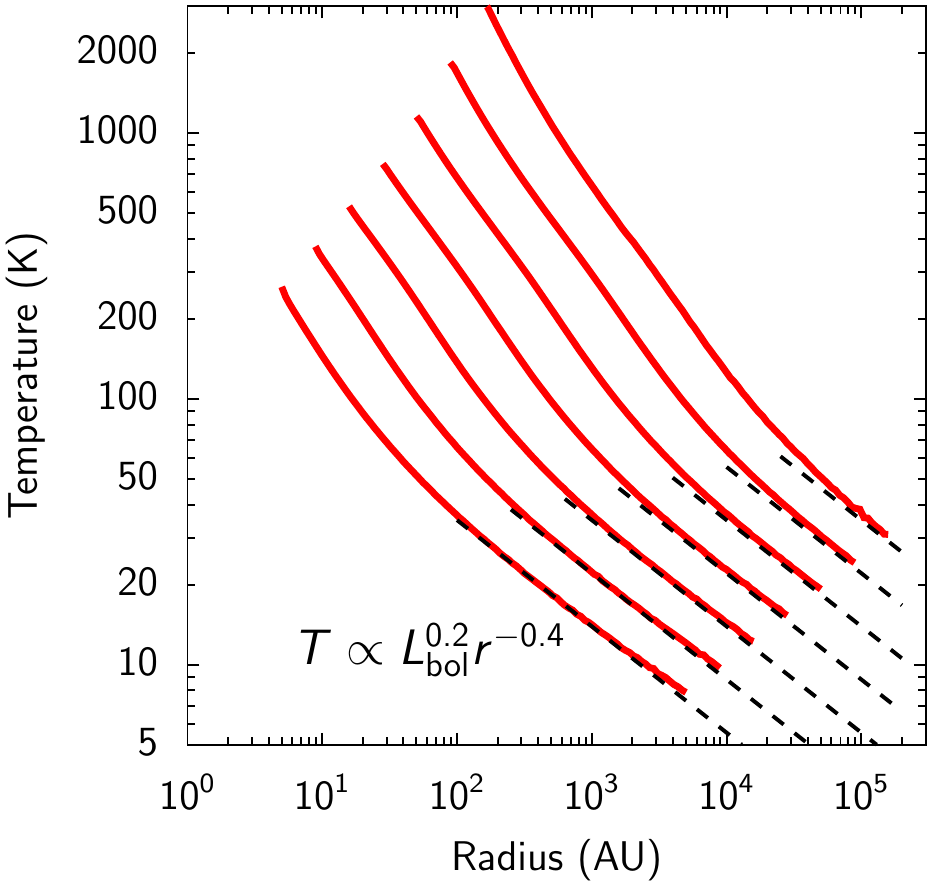}
\caption{Temperature profiles of the 7 models (red) given in Fig. \ref{fig:hcop_correl1} derived by a dust radiative transfer calculation. The fits used in the analytic approximation (Eq. \ref{temperature}) are indicated by dashed lines (black).}
\label{hcop_correl2}
\end{figure}

The model density profiles serve as input for a dust radiative transfer calculation to obtain the dust temperature. The code of \cite{2010ApJ...720.1432B}, benchmarked against DUSTY (\citealt{1997MNRAS.287..799I}) in \cite{2010PhDT..........1B}, Appendix D, is applied. We assume good coupling between dust and gas temperatures in the bulk mass of the envelope \citep{1997ApJ...489..122D,2002A&A...389..446D}, from where the HCO$^+$ emission originates, thus $T_{\rm gas} \approx T_{\rm dust}$ is an appropriate approximation. The result is shown in Fig. \ref{hcop_correl2}. In the outer layers, the temperature is a power law, which is used later in the analytical approach (Eq. \ref{temperature}).

\begin{figure}[htb!]
\center
\includegraphics[width=1.0\hsize]{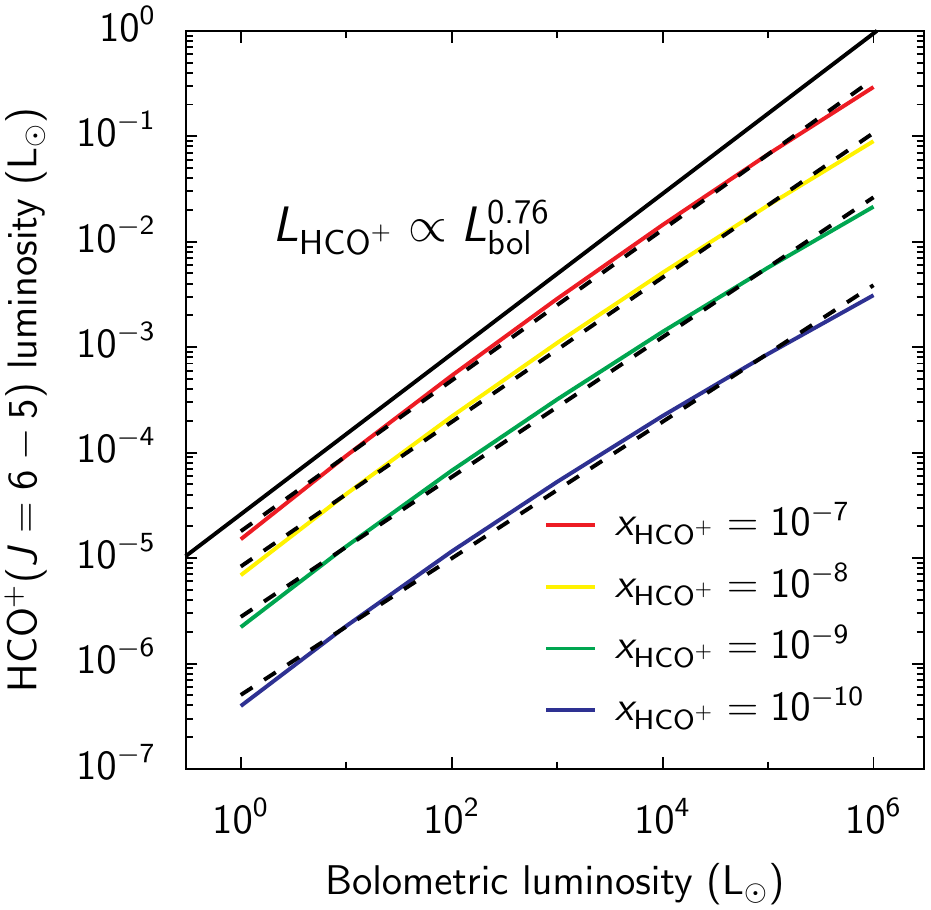}
\caption{$L_{\rm bol}$ vs. $L_{{\rm HCO}^+}$($J=6-5$) relation (full curves). The color indicates the 4 different HCO$^+$ abundances used. The model results are compared to the observed $L_{\rm bol}^{0.76\pm 0.08}$ power-law, indicated by a black solid line at the top (in arbitrary units) and power-law fits to the model relation (black dashed lines).}
\label{hcop_correl3}
\end{figure}

For the model calculations we use various constant fractional abundances of HCO$^+$ between $x_{{\rm HCO}^+} = 10^{-10}$ and $x_{{\rm HCO}^+} = 10^{-7}$.  The line radiative transfer problem is solved with the RATRAN code (\citealt{2000A&A...362..697H}). Molecular data from the LAMDA database (\citealt{2005A&A...432..369S}) are used. The results are presented in Fig. \ref{hcop_correl3}. From the fractional abundance $x_{{\rm HCO}^+}=10^{-10}$ to $x_{{\rm HCO}^+}=10^{-7}$, the power-law index increases from $0.64$ to $0.71$ (fits shown by dashed lines in the figure). We have also calculated models with a HCO$^+$ freeze-out taken into account, decreasing the HCO$^+$ abundance by a factor of 30 in regions with temperature below 20 K, but the results do not change significantly (power-law index between $0.66$ and $0.79$). Independent of the fractional abundance of HCO$^+$, the radiative transfer models can thus reproduce the observed relation having an exponent of 0.76$\pm 0.08$.

To understand the results of the numerical calculation, we roughly retrieve them from an analytical calculation. According to the model calculations, the HCO$^+$ ($J=6-5$) line is formed in the outer part of the envelope and optically thick; the high dust temperatures in the inner part of the models of the highest $L_{\rm bol}$ models do not affect the results. Thus we will use the fitted power-law temperatures for an analytical derivation below.

Since the HCO$^+$ lines are optically thick and the line luminosity scales as
\begin{equation} \label{eq:lumi}
L_{{\rm HCO}^+}\ \propto\ r_{\rm thick}^2 B_\nu(T(r_{\rm thick})) \ ,
\end{equation}
where $r_{\rm thick}$ is the radius at which the line gets optically thick ($\tau \approx 1$) and $B_\nu(T(r))$, the blackbody function (Rayleigh-Jeans approximation) for the temperature at the radius $r$. We assume that $r_{\rm thick}$ is such, that the column density, $N_{\rm thick}$, from this radius to the outer edge of the envelope is the same for all sources. This is a good approximation, since the line opacity is proportional to the column density, but neglects excitation effects. We thus solve
\begin{equation} \label{eq:col_eq}
N_{\rm thick} = \int_{r_{\rm thick}}^{r_{\rm out}} n(r) dr = {\rm const.}
\label{column_density}
\end{equation}
for $r_{\rm thick}$ in the two extreme cases $N_{\rm thick} (\alpha +1)/n_0 \ll {\rm and} \gg r_{\rm out}$. Using the model assumptions and Eq. (5),
\begin{equation} \label{eq:density}
n_0 \propto M_{\rm env} / r_{\rm out}^3 \propto L_{\rm bol}^{-0.21}\ \ .
\end{equation}
Therefore, Eq. (\ref{eq:col_eq}) yields a relation between $r_{\rm thick}$ and $L_{\rm bol}$
\begin{equation}
r_{\rm thick} \propto L_{\rm bol}^{0.25-0.33}\ \ ,
\label{r_ot}
\end{equation}
where the exponent is limited by the two extreme cases. The temperature profile of the outer envelope can be approximated by
\begin{equation}
T(r) \propto L_{\rm bol}^{0.2}\ r^{-0.4} \ ,
\label{temperature}
\end{equation}
as indicated in Fig. \ref{hcop_correl2}. This relation inserted into Eq. (\ref{eq:lumi}) yields
\begin{equation}
L_{{\rm HCO}^+}\ \propto\ r_{\rm thick}^2 L_{\rm bol}^{0.2}\ r_{\rm thick}^{-0.4} \ ,
\end{equation}
and with the range of power-law indices for $r_{\rm thick}$ given by Eq. (\ref{r_ot}) follows
\begin{equation}
L_{{\rm HCO}^+}\ \propto\ L_{\rm bol}^{0.60 - 0.73} \ \ \ .
\end{equation}
It is consistent with the model calculations (Fig. \ref{hcop_correl3}) and observations (Eq. \ref{HCOp_relation}).

The reason for the tight correlation between $L_{\rm bol}$ and HCO$^+$($J=6-5$) is the self-similarity of the dust radiative transfer problem \citep{1997MNRAS.287..799I}, which results in power-law relations for all relevant parameters. Most critical is the temperature. It can be approximated by single power-law in the HCO$^+$($J=6-5$) line photosphere (Fig. \ref{hcop_correl2}). The different power-law exponents of various YSOs in density, and the deviation from spherical symmetry have little influence on the relation between $L_{\rm bol}$ and HCO$^+$. The value of the power-law exponents depends mostly on the radius $r_{\rm thick}$ where the line gets optically thick, thus on the particular molecule and the transition.

\end{appendix}

\end{document}